\documentclass[10pt]{article}
\usepackage{amssymb}
\usepackage{amsmath}
\usepackage{amsfonts}

\makeatletter\@addtoreset{equation}{section}\makeatother

\newtheorem{theorem}{Theorem}[section]
\newtheorem{lemma}[theorem]{Lemma}
\newtheorem{proposition}[theorem]{Proposition}
\newtheorem{remark}[theorem]{Remark}
\newtheorem{corollary}[theorem]{Corollary}
\newtheorem{definition}[theorem]{Definition}

\numberwithin{equation}{section}

\title{Euclidean Gibbs Measures of Interacting Quantum Anharmonic
Oscillators}

\author{Yuri Kozitsky and Tatiana Pasurek }

\begin{document}

\maketitle

\begin{abstract}
A rigorous description of the equilibrium thermodynamic properties
of an infinite system of interacting $\nu$-dimensional quantum
anharmonic oscillators is given. The oscillators are indexed by the
elements of a countable set $\mathbf{L}\subset \mathbf{R}^d$,
possibly irregular; the anharmonic potentials vary from site to
site. The description is based on the representation of the Gibbs
states in terms of path measures -- the so called Euclidean Gibbs
measures. It is proven that: (a) the set of such measures
$\mathcal{G}^{\rm t}$ is non-void and compact; (b) every $\mu \in
\mathcal{G}^{\rm t}$ obeys an exponential integrability estimate,
the same for the whole set $\mathcal{G}^{\rm t}$; (c) every $\mu \in
\mathcal{G}^{\rm t}$ has a Lebowitz-Presutti type support; (d)
$\mathcal{G}^{\rm t}$ is a singleton at high temperatures. In the
case of attractive interaction and $\nu=1$ we prove that
$|\mathcal{G}^{\rm t}|>1$ at low temperatures. The uniqueness of
Gibbs measures due to quantum effects and at a nonzero external
field are also proven in this case. Thereby, a qualitative theory of
phase transitions and quantum effects, which interprets most
important experimental data known for the corresponding physical
objects, is developed. The mathematical result of the paper is a
complete description of the set $\mathcal{G}^{\rm t}$, which refines
and extends the results known for models of this type.
\end{abstract}

\section{Introduction}

The quantum anharmonic oscillator is a mathematical model of a
localized quantum particle moving in a potential field with possibly
multiple minima. Infinite systems of interacting quantum anharmonic
oscillators possess interesting properties connected with the
possibility of ordering caused by the interaction as well as with
quantum stabilization competing the ordering. Most of the systems of
this kind are related with solids, such as ionic crystals containing
localized light particles oscillating in the field created by heavy
ionic complexes, or quantum crystals consisting entirely of such
particles. For instance, a potential field with multiple minima is
seen by a helium atom located at the center of the crystal cell in
bcc helium \cite{Koehler}. The same situation exists in other
quantum crystals, ${\rm He}$, ${\rm H}_2$ and to some extent ${\rm
Ne}$. An example of the ionic crystal with localized quantum
particles moving in a double-well potential field is a ${\rm
KDP}$-type ferroelectric with hydrogen bounds, in which such
particles are protons or deuterons performing one-dimensional
oscillations along the bounds. In this case the particle carries
electric charge and its displacement produces dipole moment that
should be reflected in the choice of the interparticle interaction.
It is believed that structural phase transitions in such
ferroelectrics are triggered by the ordering of protons
\cite{BZ,St,Tokunaga,Vaks}. Another relevant physical object of this
kind is a system of light atoms, like ${\rm Li}$, doped into ionic
crystals, like ${\rm KCl}$. The particles in this system are not
necessarily regularly distributed. At last, quantum anharmonic
oscillators
 are used as parts of the models
describing interaction of vibrating quantum particles with a
radiation (photon) field \cite{HHS,OS} or strong electron-electron
correlations caused by the interaction of electrons with vibrating
ions responsible for such phenomena as superconductivity, charge
density waves, etc, see \cite{Freer,FreerL}. Thus, infinite systems
of interacting quantum anharmonic oscillators are quite important
models and their rigorous description is still
 a challenging mathematical task.

The model we consider has the following heuristic Hamiltonian
\begin{equation} \label{app1}
H = - \frac{1}{2} \sum_{\ell,\ell'} J_{\ell\ell'}\cdot(q_{\ell} ,
q_{\ell'}) + \sum_{\ell} H_{\ell},
\end{equation}
in which the interaction term is of dipole-dipole type. The sums run
through a countable set $\mathbf{L} \subset\mathbf{R}^d$, the
displacement $q_\ell$ is a $\nu$-dimensional vector. By $(\cdot,
\cdot)$ and $|\cdot|$ we denote the Euclidean scalar product and
norm in $\mathbf{R}^\nu$. The Hamiltonian
\begin{equation} \label{app2} H_\ell = H_\ell^{\rm har} + V_\ell (q_\ell
) \ \stackrel{\rm def}{=} \ \frac{1}{2m} |p_\ell|^2 + \frac{a}{2}
|q_\ell|^2 + V_\ell (q_\ell ), \quad a>0,
\end{equation}
 describes an isolated anharmonic oscillator of mass
$m$ and momentum $p_\ell$. Its part $H^{\rm har}_\ell$ corresponds
to a $\nu$-dimensional quantum harmonic oscillator of rigidity $a$.
The anharmonic potentials $V_\ell$, which may vary from site to
site, are supposed to obey certain uniform bounds responsible for
the stability of the whole system. We do not assume that the
interaction possesses special properties like translation invariance
 or has finite range. Therefore,
our model describes also systems with long-range interactions and
with spacial irregularities, e.g. caused by impurities, or random
components.

A complete description of the equilibrium thermodynamic properties
of infinite-particle systems may be made by constructing their
Gibbs states. Usually,  Gibbs states of quantum models  are
defined as positive normalized functionals on algebras of
observables, satisfying the Kubo-Martin-Schwinger (KMS) condition,
see \cite{BrR}. This condition is formulated in terms of the
limits $\Lambda \nearrow \mathbf{L}$ of the unitary operators
$\exp (\imath t H_\Lambda)$, $t\in \mathbf{R}$, which determine
the dynamics of the subsystem located in a finite $\Lambda\subset
\mathbf{L}$ and described by the Hamiltonian $H_\Lambda$. But for
our model, such a limit of $\exp (\imath t H_\Lambda)$ does not
exist and therefore
 the
KMS condition for the whole system cannot be formulated. Thus,
actually there is no canonical way to define Gibbs states, and
hence to give a complete description of the thermodynamic
properties of models like (\ref{app1}). The aim of this work is to
bridge this gap with the help of path integrals.

In \cite{AHK}, an approach based on the fact that the local
Hamiltonians $H_\Lambda$ generate stochastic processes has been
initiated. Here the description of the local Gibbs states employs
the properties of the semi-group $\exp(-\tau H_\Lambda)$,
$\tau>0$. This allows one to translate it into a ``probabilistic
language", that opens the possibility to apply here corresponding
concepts and techniques. In this language, our model
 is the system of interacting ``classical" spins $\omega_\ell$, $\ell\in \mathbf{L}$, which however are
infinite-dimensional --  they are continuous paths $\omega_\ell :[0,
\beta]\rightarrow \mathbf{R}^\nu$, $\omega_\ell (0) = \omega_\ell
(\beta)$, called also {temperature loops}. Each spin is described by
the path measure of the $\beta$-periodic Ornstein-Uhlenbeck process
corresponding to $H_\ell^{\rm har}$ multiplied by a density obtained
from the anharmonic potential with the help of the Feynman-Kac
formula. Afterwards, finite subsystems are associated with
conditional probability measures, which by the
Dobrushin-Lanford-Ruelle (DLR) equation determine the set of Gibbs
measures $\mathcal{G}^{\rm t}$. This approach is called Euclidean
due to its conceptual analogy with the Euclidean quantum field
theory. Its further development was conducted in the papers
\cite{AKK,AKKR01,AKKR,AKKR02b,AKPR01b,AKPR01c,AKRT,AKRT1,AKRT2,AKRT3,BK1,Kond,KK,Koz,Koz1,Ko,KoL,MPZ,MVZ00}.
Among the achievements  one has to mention the settlement in
\cite{AKKR01,AKKR02b,AKKRprl} of a long standing problem of the
influence of quantum effects on structural phase transitions in
quantum anharmonic crystals, which first was discussed in
\cite{Sch}, see also \cite{MVZ00,VZ1,VZ2}.

In the present article, we give a complete description of the set
$\mathcal{G}^{\rm t}$ for the model (\ref{app1}) and hence
essentially finalize the development of the Euclidean approach for
such models. Our results fall into two groups of theorems. The first
group describes the general case where $J_{\ell\ell'}$ and $V_\ell$
obey natural stability conditions only. We state that:
$\mathcal{G}^{\rm t}$ is non-void and compact (Theorem \ref{1tm});
the elements of $\mathcal{G}^{\rm t}$ obey certain exponential
moment estimates (Theorem \ref{2tm}) and have a Lebowitz-Presutti
type support (Theorem \ref{3tm}); $\mathcal{G}^{\rm t}$ is a
singleton at high temperatures (Theorem \ref{httm}). The second
group of theorems describes the case of $\nu=1$ and $J_{\ell \ell'}
\geq 0$. Here we employ the FKG order and show that the set
$\mathcal{G}^{\rm t}$ has maximal and minimal elements (Theorem
\ref{MAtm}). If the model is translation invariant, we prove that
the limiting pressure exists and is the same in all states (Theorem
\ref{pressuretm}). Then under natural additional conditions on
$V_\ell$ we show (Theorem \ref{phtm}) that the model undergoes a
phase transition (for $d\geq 3$) and, on the other hand,
$\mathcal{G}^{\rm t}$ is a singleton at all temperatures if a
quantum stabilization condition is satisfied (Theorem \ref{5tm}).
Finally, we describe a class of anharmonic potentials $V_\ell$, for
which $\mathcal{G}^{\rm t}$ is a singleton at a non-zero external
field (Theorem \ref{6tm}). Here we use a version of the Lee-Yang
theorem \cite{Koz1}, adapted to path measures. All these results are
novel both for the quantum model and its classical analogs.

The paper is organized as follows. In section \ref{2s} we describe
the model in detail (subsection \ref{2.1ss}) and present the basic
elements of the Euclidean approach (subsections \ref{ss2.2} and
\ref{ss2.3}). Afterwards, we introduce tempered configurations, a
local Gibbs specification, and tempered Euclidean Gibbs measures of
our model. In section \ref{3s} we give the results in the form of
the theorems discussed above. Comments, which in particular relate
these results with those known in the literature, conclude the
section. The remaining part of the article is dedicated to the proof
of the theorems and is quite technical. Here we first study in
detail the properties of the local Gibbs specification.

\section{Euclidean Gibbs Measures}
\label{2s}
\subsection{The model}
\label{2.1ss}
 The infinite system of quantum oscillators we consider is described
 by the formal Hamiltonian (\ref{app1}), (\ref{app2}),
  defined on the
set $\mathbf{L}\subset \mathbf{R}^d$, $d\in \mathbf{N}$. This set is
equipped with the Euclidean distance $|\ell - \ell'|$ inherited from
$\mathbf{R}^d$. We suppose that
\begin{equation} \label{april}
\sup_{\ell \in \mathbf{L}} \sum_{\ell' \in \mathbf{L}}
\frac{1}{\left( 1 + |\ell - \ell'|\right)^{d + \epsilon}} < \infty,
\end{equation} for every $\epsilon >0$. This is a kind of
regularity, which in particular means that big amounts of the
elements of $\mathbf{L}$ cannot concentrate in the subsets of
$\mathbf{R}^d$ of small volume. A regular case of $\mathbf{L}$ is a
lattice. In this case the model 
 is called \emph{the quantum anharmonic crystal.} For simplicity, we shall always
 assume that  $\mathbf{L}= \mathbf{Z}^d$ if $\mathbf{L}$ is  a lattice.

 Subsets of $\mathbf{L}$ are denoted by $\Lambda$. As
usual, $|\Lambda|$ stands for the cardinality of $\Lambda$ and
$\Lambda^c$ -- for its complement $\mathbf{L}\setminus \Lambda$.
We write $\Lambda \Subset \mathbf{L}$ if $\Lambda$ is non-void and
finite. By $\mathcal{L}$ we denote a cofinal (ordered by inclusion
and exhausting the lattice) sequence of finite subsets of
$\mathbf{L}$. Limits taken along such $\mathcal{L}$ are denoted by
$\lim_{\mathcal{L}}$. We write $\lim_{\Lambda \nearrow
\mathbf{L}}$ if the limit is taken along an unspecified sequence
of this type. If we say that something holds for all $\ell$, we
mean that it holds for all $\ell \in \mathbf{L}$; expressions like
$\sum_{\ell}$ mean $\sum_{\ell \in \mathbf{L}}$. By $(\cdot,
\cdot)$ and $|\cdot|$, we denote the  Euclidean scalar product and
norm in all spaces like $\mathbf{R}^\nu$, $\mathbf{R}^d$;
$\mathbf{N}_0$ will denote the set of nonnegative integers.

The Hamiltonian (\ref{app1}) has no direct mathematical meaning and
is ``represented" by the local Hamiltonians $H_\Lambda$,
$\Lambda\Subset \mathbf{L}$, which are
\begin{eqnarray} \label{sch}
H_\Lambda & = & \sum_{\ell \in \Lambda}\left[H_\ell^{\rm har}+
 V_\ell (q_\ell )\right] - \frac{1}{2} \sum_{\ell ,
\ell' \in \Lambda} J_{\ell \ell'} (q_l , q_{\ell'})
\\ & = &  \frac{1}{2m}\sum_{\ell \in \Lambda} |p_\ell|^2 + W_\Lambda
(q_\Lambda), \quad q_\Lambda = (q_\ell)_{\ell \in \Lambda}.
\nonumber
\end{eqnarray}
In the latter formula the first term is the kinetic energy; the
potential energy is
\begin{equation} \label{pe}
W_\Lambda (q_\Lambda ) = - \frac{1}{2}\sum_{\ell, \ell' \in \Lambda}
J_{\ell \ell'} (q_\ell , q_{\ell'}) + \sum_{\ell \in \Lambda}
\left[(a/2) |q_\ell|^2 + V_\ell (q_\ell)\right].
\end{equation}
The anharmonic potentials $V_\ell$ and the interaction intensities
$J_{\ell \ell'}$, such that
\begin{equation} \label{j}
J_{\ell \ell} = 0,\quad  J_{\ell \ell'} = J_{\ell' \ell} \in
\mathbf{R}, \  \ \ \ell, \ell' \in \mathbf{L},
\end{equation}
are subject to the following\\
{\bf Assumption (A)} 
All $V_\ell  : \mathbf{R}^\nu \rightarrow \mathbf{R}$ are continuous
and such that $V_\ell (0) = 0$; there exist $r>1$, $A_V
>0$, $B_V\in \mathbf{R}$, and a continuous function
$V:\mathbf{R}^\nu \rightarrow \mathbf{R}$, $V (0) = 0$,  such that
for all $\ell$ and $x\in \mathbf{R}^\nu$,
\begin{equation} \label{3}
 A_V |x|^{2r} + B_V \leq V_{\ell} (x) \leq V (x).
\end{equation}
We also assume that
\begin{equation} \label{6}
   \hat{J}_0  \ \stackrel{\rm def}{=} \sup_{\ell}
\sum_{\ell'} |J_{\ell \ell'}|< \infty.
\end{equation}

\noindent  The lower bound in (\ref{3}) is responsible for confining
each particle in the vicinity of its equilibrium position. The upper
bound is to guarantee that the oscillations of the particles located
far from the origin are not suppressed. An example of $V_\ell$ to
bear in mind is the polynomial
\begin{equation} \label{4}
V_\ell (x) = \sum_{s=1}^r b^{(s)}_\ell |x|^{2s} - (h, x), \quad
b^{(s)}_\ell \in \mathbf{R}, \ \ b^{(r)}_\ell >0, \ \ r\geq 2 ,
\end{equation}
in which $h\in \mathbf{R}^\nu$ is an external field and the
coefficients $b_\ell^{(s)}$ vary in certain intervals, such that
both estimates (\ref{3}) hold. Under Assumption (A)
 $H_\Lambda$ is a self-adjoint lower bounded operator in $L^2(\mathbf{R}^{\nu
|\Lambda|})$ having discrete spectrum. It generates a positivity
preserving semigroup such that
\begin{equation} \label{tr}
{\rm trace}[ \exp(-\tau H_\Lambda )] < \infty, \quad {\rm for} \
{\rm all} \ \tau >0.
\end{equation}
A part of our results describe the general case where the only
conditions are those set by (\ref{3}) and (\ref{6}). Another part
corresponds to more specific cases.
\begin{definition} \label{1df} The model is
ferroelectric\footnote{Usually such a model is called ferromagnetic;
we adopt the above terminology in view of the ferroelectric
interpretation mentioned in Introduction.} if $J_{\ell\ell'}\geq 0$
for all $\ell , \ell'$. The interaction has finite range if there
exists $R>0$ such that $J_{\ell\ell'} = 0$ whenever
$|\ell-\ell'|>R$. The model is translation invariant if $\mathbf{L}$
is a lattice, $V_\ell = V$ for all $\ell$, and the matrix $(J_{\ell
\ell'})_{\mathbf{L}\times \mathbf{L}}$ is invariant under
translations of $\mathbf{L}$. The model is rotation invariant if for
every orthogonal transformation $U\in O(\nu)$ and every $\ell$,
$V_\ell (U x) = V_\ell(x)$.
\end{definition}
\noindent If $V_\ell \equiv 0$ for all  $\ell$, one gets a system
of interacting quantum harmonic oscillators, a quantum harmonic
crystal if $\mathbf{L}$ is a lattice. It is stable if $\hat{J}_0 <
a$, see Remark \ref{apprm} below.

\subsection{Quantum Gibbs states in the Euclidean approach}

\label{ss2.2} Here we outline the basic elements of the Euclidean
approach in quantum statistical mechanics, its detailed presentation
may be found in \cite{AKKR,AKPR01c}.

For $\Lambda \Subset \mathbf{L}$, the Hamiltonian $H_\Lambda$,
defined by (\ref{sch}), acts in the physical Hilbert space
$\mathcal{H}_\Lambda \ \stackrel{\rm def}{=} \ L^2(\mathbf{R}^{\nu
|\Lambda|})$. In view of (\ref{tr}), one can introduce the local
Gibbs state
\begin{equation}
\mathfrak{C}_\Lambda \ni A\mapsto \varrho_{\Lambda }(A)\text{
}\overset{\mathrm{def}}{=}\text{ } \frac{\mathrm{trace}(Ae^{-\beta
H_{\Lambda }})}{\mathrm{trace}(e^{-\beta H_{\Lambda }})}, \label{8e}
\end{equation}%
which is a positive normalized  functional on the algebra
$\mathfrak{C}_\Lambda$ of all bounded linear operators (observables)
on $\mathcal{H}_\Lambda$. The mappings
\begin{equation}
\mathfrak{C}_\Lambda \ni A\mapsto
 \mathfrak{a}_{t}^{\Lambda }(A)\text{ }\overset{\mathrm{def}}{=}%
\text{ }e^{itH_{\Lambda }}Ae^{-itH_{\Lambda }},\ \ t\in \mathbf{R},
\label{9e}
\end{equation}%
constitute the group of time automorphisms which describes the
dynamics of the system in $\Lambda$. The state $\varrho_\Lambda$
satisfies the KMS (thermal equilibrium) condition relative to the
dynamics $\mathfrak{a}_t^{\Lambda}$, see Definition 1.1 in
\cite{KL}. Multiplication operators by bounded continuous functions
act as
\[
(F\psi)(x) = F(x) \cdot \psi(x), \quad \psi \in \mathcal{H}_\Lambda,
\ \ F \in C_{\mathrm{b}}(\mathbf{R}^{\nu |\Lambda |}).
\]
One can prove, see \cite{KoL}, that the linear span of the products
\begin{equation}
\mathfrak{a}_{t_{1}}^{\Lambda }(F_{1})\cdots
\mathfrak{a}_{t_{n}}^{\Lambda }(F_{n}),  \label{10e}
\end{equation}%
with all possible choices of $n\in \mathbf{N}$, $t_{1},\dots
,t_{n}\in \mathbf{R}$ and $F_{1},\dots ,F_{n}\in
C_{\mathrm{b}}(\mathbf{R}^{\nu |\Lambda |})$, is $\sigma $-weakly
dense in $\mathfrak{C}_{\Lambda }$. Therefore, as a $\sigma
$-weakly continuous functional (page 65 of the first volume of
\cite{BrR}), the state (\ref{8e}) is fully determined by its
values on (\ref{10e}), that is, by \emph{the Green functions}
\begin{equation}
G_{F_{1},\dots ,F_{n}}^{\Lambda }(t_{1},\dots ,t_{n})\text{ }\overset{%
\mathrm{def}}{=}\text{ }\varrho _{\Lambda }\left[ \mathfrak{a}%
_{t_{1}}^{\Lambda }(F_{1})\cdots \mathfrak{a}_{t_{n}}^{\Lambda }(F_{n})%
\right].   \label{11e}
\end{equation}%
They can be considered as restrictions of functions $G_{F_{1},\dots
,F_{n}}^{\Lambda }(z_{1},\dots ,z_{n})$, analytic in the domain
\begin{equation}
\mathcal{D}_{\beta }^{n}=\{(z_1 , \dots z_n) \in \mathbf{C}^n \ | \
0 < \Im (z_1) < \Im (z_2) < \cdots < \Im (z_n ) < \beta\},
\label{11z}
\end{equation}%
and continuous on its closure $\mathcal{\bar{D}}_{\beta
}^{n}\subset \mathbf{C}^{n}$.  Foe every $n \in \mathbf{N}$, the
``imaginary time" subset
\[
\{(z_1 , \dots , z_n)\in \mathcal{D}_\beta^n \ | \ \Re(z_1) = \cdots
= \Re(z_n) = 0\}
\]
is an inner set of uniqueness for functions analytic in
$\mathcal{D}_\beta^n$ (see pages 101 and 352 of \cite{Sch}).
Therefore, the Green functions (\ref{11e}), and hence the states
(\ref{8e}), are completely determined by {\it the Matsubara
functions}
\begin{eqnarray} \label{mats}
& & \quad  \mathit{\Gamma}^\Lambda_{F_1 , \dots , F_n}(\tau_1 ,
\dots , \tau_n) \ \stackrel{\rm def}{=} \ G^\Lambda_{F_1 , \dots ,
F_n}(\imath \tau_1 , \dots ,\imath \tau_n) \\
& & \qquad  = {\rm trace}[F_{1}e^{-(\tau _{2}-\tau _{1})H_{\Lambda
}}F_{2}e^{-(\tau _{3}-\tau _{2})H_{\Lambda }}\cdots F_{n}e^{-(
 \tau _{n+1}-\tau _{n})H_{\Lambda }}]/\mathrm{trace}[e^{-\beta
H_{\Lambda }}] \label{12e} \nonumber
\end{eqnarray}
taken at ordered arguments
$0 \leq \tau _{1}\leq \cdots \leq \tau _{n}\leq \tau _{1}+\beta \overset{\mathrm{def}%
}{=}\tau _{n+1}$, with all possible choices of $n \in \mathbf{N}$
and $F_1 , \dots , F_n \in C_{\rm b}(\mathbf{R}^{\nu|\Lambda|})$.
Their extensions to $[0, \beta]^n$ are
\[
\mathit{\Gamma}^\Lambda_{F_1 , \dots , F_n}(\tau_1 , \dots , \tau_n)
= \mathit{\Gamma}^\Lambda_{F_{\sigma(1)} , \dots ,
F_{\sigma(n)}}(\tau_{\sigma(1)} , \dots , \tau_{\sigma(n)}),
\]
where $\sigma$ is the permutation of $\{1, 2, \dots , n\}$ such that
$\tau_{\sigma(1)}\leq \tau_{\sigma(2)} \leq \cdots \leq
\tau_{\sigma(n)}$.  One can show that for every $\theta\in [0,
\beta]$,
\begin{equation} \label{mats1}
\mathit{\Gamma}^\Lambda_{F_1 , \dots , F_n}(\tau_1 + \theta, \dots ,
\tau_n + \theta) = \mathit{\Gamma}^\Lambda_{F_1 , \dots ,
F_n}(\tau_1 , \dots , \tau_n),
\end{equation}
where addition is modulo $\beta$. This periodicity along with the
analyticity of the Green functions is equivalent to the KMS property
of the state (\ref{8e}).

The central element of the Euclidean approach  is the representation
of the Matsubara functions (\ref{mats}) corresponding to $F_1 ,
\dots , F_n \in C_{\rm b}(\mathbf{R}^{\nu |\Lambda|})$ in the form
of
\begin{equation}\label{mul10}
\mathit{\Gamma }_{F_{1},\dots ,F_{n}}^{\Lambda }(\tau _{1},\dots
,\tau _{n})=\int_{\mathit{\Omega }_{\Lambda }}F_{1}(\omega _{\Lambda
}(\tau
_{1}))\ldots F_{n}(\omega _{\Lambda }(\tau _{n}))\mu _{\Lambda }(\mathrm{d}%
\omega _{\Lambda }),
\end{equation}
where $\mu_\Lambda$ is a certain probability measure on the space
$\mathit{\Omega}_\Lambda$, which we construct in the subsequent
part of this section. This measure is called \emph{a local
Euclidean Gibbs measure}. By standard arguments, it is uniquely
determined by the integrals (\ref{mul10}). In view of the fact
that the Matsubara functions $\mathit{\Gamma }_{F_{1},\dots
,F_{n}}^{\Lambda }$ uniquely determine the state
$\varrho_\Lambda$, the representation (\ref{mul10}) establishes a
one-to-one correspondence between the local Gibbs states
$\varrho_\Lambda$ and local Euclidean Gibbs measures
$\mu_\Lambda$.

 Thermodynamic properties of the
 model (\ref{app1}) are described by the Gibbs states
corresponding to the whole set $\mathbf{L}$. Such states should be
defined on the $C^*$-algebra of quasi-local observables
$\mathfrak{C}$, being the norm-completion of the algebra of local
observables $\cup_{\Lambda \Subset \mathbf{L}}
\mathfrak{C}_\Lambda$. Here each $\mathfrak{C}_\Lambda$ is
considered as a subalgebra of $\mathfrak{C}_{\Lambda'}$ for any
$\Lambda'$ containing $\Lambda$. The dynamics of the whole system
is to be defined by the limits $\Lambda \nearrow \mathbf{L}$ of
the time automorphisms (\ref{9e}), which would allow one to define
the Gibbs states on $\mathfrak{C}$ as KMS states. This
``algebraic" way can be realized for models described by bounded
local Hamiltonians $H_\Lambda$, e.g., quantum spin models, see
section 6.2 of \cite{BrR}. For the model considered here, such
limiting automorphisms do not exist and hence there is no
canonical way to define Gibbs states of the whole infinite system.
Therefore, the Euclidean approach based on the one-to-one
correspondence between the local states and measures arising from
the representation (\ref{mul10}) seems to be the only way of
developing a mathematical theory of the equilibrium thermodynamic
properties of such models. For some versions of quantum crystals,
a possibility of constructing the limiting states $\varrho =
\lim_{\Lambda \nearrow \mathbf{L}}\varrho_\Lambda$ in terms of the
limiting path measures $\mu = \lim_{\Lambda \nearrow
\mathbf{L}}\mu_\Lambda$ was discussed in \cite{Amour,MPZ,MVZ00}.
The set of Euclidean Gibbs measures $\mathcal{G}^{\rm t}$ we
construct and study in this article certainly includes all the
limiting points of this type. Furthermore, there exist axiomatic
methods, see \cite{Birke,Gelerak2}, analogous to the
Osterwalder-Schrader reconstruction theory \cite{GJ,Si74}, by
means of which KMS states are constructed on certain von Neumann
algebras from a complete set of Matsubara functions. In our case
such a set consists of the functions
\begin{equation} \label{kms}
\mathit{\Gamma}^{\mu}_{F_1 , \dots , F_n} (\tau_1 , \dots , \tau_n)
= \int_{\mathit{\Omega}} F_1 (\omega (\tau_1)) \cdots
F_n(\omega(\tau_n))\mu({\rm d}\omega), \quad \mu\in \mathcal{G}^{\rm
t},
\end{equation}
defined for all  bounded local multiplication operators $F_1,
\dots , F_n$. Therefore, the theory of  Euclidean Gibbs measures
 presented in this article can be further
developed  towards constructing such algebras and states, which we
leave as a task for the future.

\subsection{Path spaces and local Euclidean Gibbs measures}
\label{ss2.3} The local Euclidean Gibbs measures are defined on the
spaces of continuous paths. These are continuous functions defined
on the interval $[0, \beta]$, taking equal values at the endpoints
(temperature loops). Here $\beta^{-1} = T>0$ is absolute
temperature. One can consider the loops as functions on the circle
$S_\beta\cong[0, \beta]$ being a compact Riemannian manifold with
Lebesgue measure ${\rm d}\tau$ and distance
\begin{equation} \label{15}
|\tau - \tau'|_\beta \ \stackrel{\rm def}{=} \ \min\{|\tau - \tau'|
\ ; \ \beta - |\tau - \tau'| \}, \ \  \tau , \tau' \in S_\beta.
\end{equation}
As single-spin spaces  we use the standard Banach spaces
\[
C_\beta \ \stackrel{\rm def}{=} \ C(S_\beta \rightarrow
\mathbf{R}^\nu), \qquad C_\beta^\sigma \ \stackrel{\rm def}{=} \
C^\sigma(S_\beta \rightarrow \mathbf{R}^\nu), \ \ \sigma \in (0, 1),
\]
of all continuous and H\"older-continuous functions
$\omega_\ell:S_\beta \rightarrow \mathbf{R}^\nu$ respectively, which
are equipped with the supremum norm $|\omega_\ell |_{C_\beta}$ and
with the H\"older norm
\begin{equation} \label{16z}
|\omega_\ell |_{C^\sigma_\beta} = |\omega_\ell|_{C_\beta} +
\sup_{\tau, \tau' \in S_\beta, \ \tau \neq \tau'}\frac{|\omega_\ell
(\tau) - \omega_\ell (\tau')|}{|\tau - \tau'|^\sigma_\beta}.
\end{equation}
Along with them we use the real Hilbert space $ L^2_\beta = L^2
(S_\beta \rightarrow \mathbf{R}^\nu, {\rm d}\tau)$; its inner
product and norm  are denoted by $(\cdot , \cdot)_{L^2_\beta}$ and
$|\cdot|_{L^2_\beta}$ respectively. By
 $\mathcal{B}(C_\beta)$,
$\mathcal{B}(L^2_\beta)$ we denote the corresponding Borel
$\sigma$-algebras. Then one defines dense continuous embeddings
$C^\sigma_\beta \hookrightarrow C_\beta \hookrightarrow
L^2_\beta$, that by the Kuratowski theorem, page 499 of
\cite{Kurat}, yields
\begin{equation} \label{kt}
C_\beta \in \mathcal{B} (L_\beta^2) \quad {\rm and} \quad
\mathcal{B}({C}_\beta) = \mathcal{B}(L^2_\beta) \cap C_\beta.
\end{equation}
The space of H\"older-continuous functions $C_\beta^\sigma$ is not
separable, however, as a subset of $C_\beta$ or $L^2_\beta$, it is
measurable (page 278 of \cite{RS2}). Given $\Lambda \subseteq
\mathbf{L}$, we set
 \begin{equation} \label{17}
 \mathit{\Omega}_\Lambda  =  \{\omega_\Lambda = (\omega_\ell )_{\ell
 \in \Lambda} \ | \ \omega_\ell \in C_\beta\}, \ \ \mathit{\Omega}
 = \mathit{\Omega}_{\mathbf{L}} =
  \{\omega = (\omega_\ell )_{\ell
 \in \mathbf{L}} \ | \ \omega_\ell \in C_\beta\}. \quad
 \end{equation}
These spaces are equipped with the product topology and with the
Borel $\sigma$-algebras $\mathcal{B}(\mathit{\Omega}_\Lambda)$.
Thereby, each $\mathit{\Omega}_\Lambda$ is a Polish space; its
elements are called configurations in $\Lambda$. In particular,
$\mathit{\Omega}$ is the configuration space for the whole system.
 For $\Lambda \subset \Lambda'$, the decomposition $\omega_{\Lambda'} = \omega_{\Lambda} \times
\omega_{\Lambda' \setminus \Lambda}$ defines an embedding
$\mathit{\Omega}_\Lambda \hookrightarrow \mathit{\Omega}_{\Lambda'}$
 by identifying $\omega_{\Lambda} \in
\mathit{\Omega}_{\Lambda}$ with $\omega_{\Lambda} \times 0_{\Lambda'
\setminus \Lambda}\in \mathit{\Omega}_{\Lambda'}$. By
$\mathcal{P}(\mathit{\Omega}_\Lambda)$ and
$\mathcal{P}(\mathit{\Omega})$ we denote the sets of all probability
measures on $(\mathit{\Omega}_\Lambda,
\mathcal{B}(\mathit{\Omega}_\Lambda))$ and $(\mathit{\Omega},
\mathcal{B}(\mathit{\Omega}))$ respectively.

A $\nu$-dimensional quantum harmonic oscillator of mass $m>0$ and
rigidity $a>0$ is described by the Hamiltonian, c.f., (\ref{app2}),
\begin{equation} \label{16e}
H^{\rm har}_\ell = - \frac{1}{2m} \sum_{j=1}^\nu
\left(\frac{\partial}{\partial x^{(j)}_\ell} \right)^2 + \frac{a}{2}
|x_\ell|^2,
\end{equation}
acting in the complex Hilbert space $L^2 (\mathbf{R}^\nu)$. The
operator semigroup $\exp(- \tau H^{\rm har}_\ell)$, $\tau\in [0,
\beta]$, defines a Gaussian $\beta$-periodic Markov process -- the
periodic Ornstein-Uhlenbeck velocity process, see \cite{KL1}. In
quantum statistical mechanics it first appeared in R.
H{\o}egh-Krohn's paper \cite{H}. The canonical realization of this
process on $(C_\beta , \mathcal{B}(C_\beta))$ is described by the
path measure which one introduces as follows. In $L^2_\beta$ we
define the self-adjoint (Laplace-Beltrami type) operator
 \begin{equation} \label{16a}
 A = \left(- m \frac{{\rm d}^2 }{{\rm d }\tau^2} + a \right) \otimes
 \mathbf{I},
 \end{equation}
 where $\mathbf{I}$ is
 the identity operator in $\mathbf{R}^\nu$.
 Its spectrum consists of the eigenvalues
 \begin{equation} \label{16b}
 \lambda_k = m (2 \pi k/\beta)^2 + a, \quad k  \in \mathbf{Z}.
 \end{equation}
Thus, the inverse $A^{-1}$ is of trace class and the Fourier
transform
\begin{equation} \label{160}
\int_{L^2_\beta} \exp[\imath (\phi, \upsilon)_{L^2_\beta}]\chi({\rm
d}\upsilon) = \exp\left\{ - \frac{1}{2} (A^{-1} \phi,
\phi)_{L^2_\beta} \right\}, \ \ \phi \in L^2_\beta.
\end{equation}
defines a Gaussian measure $\chi$ on $(L^2_\beta,
\mathcal{B}(L^2_\beta))$. Employing the eigenvalues (\ref{16b}) one
can show (by Kolmogorov's lemma, page 43 of \cite{Si79}) that
\begin{equation} \label{16f}
\chi(C_\beta^\sigma) = 1, \quad {\rm for} \ {\rm  all}  \ \sigma\in
(0, 1/2).
\end{equation}
Then $\chi(C_\beta)=1$ and by (\ref{kt}) we redefine $\chi$ as a
probability measure on $(C_\beta , \mathcal{B}(C_\beta))$. An
account of the properties of $\chi$ may be found in \cite{AKKR}.
One of them, which plays a special role in our construction,
follows directly from  Fernique's theorem (Theorem 1.3.24 in
\cite{DS}).\begin{proposition} \label{1pn} For every $\sigma\in
(0, 1/2)$, there exists $\lambda_\sigma >0$ such that
\begin{equation} \label{16h}
\int_{L^2_\beta} \exp\left( \lambda_\sigma |\upsilon
|^2_{C_\beta^\sigma} \right)\chi({\rm d}\upsilon) < \infty.
\end{equation}
\end{proposition}
The measure $\chi$ is the local Euclidean Gibbs measure for a single
harmonic oscillator. The measure $\mu_\Lambda\in
\mathcal{P}(\mathit{\Omega}_\Lambda)$ which corresponds to the
system of interacting anharmonic oscillators located in $\Lambda
\Subset \mathbf{L}$ is associated with a stationary $\beta$-periodic
Markov process defined as follows. The marginal distributions of $\mu_\Lambda$ are
given by the integral kernels of the operators $\exp(-\tau H_\Lambda
)$, $\tau \in [0, \beta]$. This means that
\begin{eqnarray} \label{ea}
& & {\rm trace} [F_1 e^{-(\tau_2 - \tau_1)H_\Lambda } F_2
e^{-(\tau_3 - \tau_2)H_\Lambda }\cdots F_n e^{-(\tau_{n+1} -
\tau_n)H_\Lambda }]/ {\rm trace}[e^{-\beta H_\Lambda}]\\ & & \qquad
= \int_{\mathit{\Omega}_\Lambda }F_1 (\omega_\Lambda (\tau_1) \cdots
F_n (\omega_\Lambda (\tau_n))\mu_\Lambda ({\rm d} \omega_\Lambda),
\nonumber
\end{eqnarray}
for all $F_1 , \dots , F_n \in L^\infty (\mathbf{R}^{\nu
|\Lambda|})$, $n \in \mathbf{N}$ and $\tau_1 , \dots, \tau_n\in
S_\beta$, such that $\tau_1 \leq \cdots \leq \tau_n \leq \beta$,
$\tau_{n+1} = \tau_1 + \beta$. And vice verse, the representation
(\ref{ea}) uniquely, up to equivalence, defines $H_\Lambda$ (see
\cite{KL}). By means of the Feynman-Kac formula the measure
$\mu_\Lambda$ is obtained as a Gibbs modification
\begin{equation} \label{mul}
\mu_{\Lambda}({\rm d}\omega_\Lambda) =  \exp\left\{- I_\Lambda
(\omega_\Lambda) \right\}\chi_\Lambda ({\rm
d}\omega_\Lambda)/{Z_\Lambda},
\end{equation}
of the ``free measure"
\begin{equation} \label{18}
\chi_\Lambda ({\rm d}\omega_\Lambda) = \prod_{\ell \in \Lambda}
\chi({\rm d}\omega_\ell).
\end{equation}
Here
\begin{equation} \label{19}
I_\Lambda (\omega_\Lambda)  {=} - \frac{1}{2} \sum_{\ell , \ell' \in
\Lambda} J_{\ell \ell'} (\omega_\ell , \omega_{\ell'})_{L^2_\beta} +
\sum_{\ell \in \Lambda} \int_0^\beta V_\ell (\omega_\ell (\tau)){\rm
d} \tau
\end{equation}
is the energy functional describing the system of interacting paths
$\omega_\ell$, $\ell \in \Lambda$, whereas
\begin{equation} \label{mul1}
Z_\Lambda = \int_{\mathit{\Omega}_\Lambda} \exp\left\{- I_\Lambda
(\omega_\Lambda) \right\}\chi_\Lambda ({\rm d}\omega_\Lambda),
\end{equation}
is the partition function. As mentioned above, $\mu_\Lambda$ is
the local Gibbs measure, where {\em local} means corresponding to
a $\Lambda \Subset \mathbf{L}$.

\subsection{Tempered configurations}
\label{ss2.4} The next step is to construct the equilibrium states
of the whole infinite system (\ref{app1}), which we are going to
do in the DLR approach, which is standard for classical
(non-quantum) statistical mechanics, see  \cite{Ge,Pr}. In this
approach,  the Gibbs measures are constructed by means of local
conditional distributions.  In our case the single-spin spaces,
$C_\beta$, $C_\beta^\sigma$, are infinite-dimensional and hence
their topological properties are much richer, which makes the DLR
technique we develop here to be more sophisticated.

To go further we have to define functions on the spaces
$\mathit{\Omega}_\Lambda$ with infinite $\Lambda$, including
$\mathit{\Omega}$ itself. Among others, we will need the energy
functional $I_\Lambda (\cdot|\xi)$ describing the interaction with a
configuration $\xi\in \mathit{\Omega}$ fixed outside of $\Lambda$.
In accordance with (\ref{sch}) it is
\begin{equation} \label{23}
I_\Lambda (\omega | \xi) = I_\Lambda (\omega_\Lambda) - \sum_{\ell
\in \Lambda, \ \ell' \in \Lambda^c} J_{\ell \ell'} (\omega_\ell ,
\xi_{\ell'})_{L^2_{\beta}}, \quad \omega \in \mathit{\Omega},
\end{equation}
where $I_\Lambda$ is defined by (\ref{19}). Recall that $\omega =
\omega_\Lambda \times \omega_{\Lambda^c}$; hence,
\begin{equation} \label{23a}
I_\Lambda (\omega | \xi) = I_\Lambda (\omega_\Lambda \times
0_{\Lambda^c} | 0_\Lambda \times \xi_{\Lambda^c}).
\end{equation}
Clearly, the second term in (\ref{23}) makes sense for all $\xi \in
\mathit{\Omega}$ only if the interaction has finite range.
Otherwise, one has to restrict $\xi$ to a subset of
$\mathit{\Omega}$, naturally defined by the condition
\begin{equation} \label{24}
\forall{\ell}\in \mathbf{L}: \qquad \sum_{\ell'}
|J_{\ell\ell'}|\cdot |(\omega_\ell , \xi_{\ell'})_{L^2_\beta}| <
\infty,
\end{equation}
that can be rewritten in terms of growth restrictions imposed on
$\{|\xi_\ell |_{L^2_\beta}\}_{\ell \in \mathbf{L}}$, determined by
the decay of $J_{\ell \ell'}$ (c.f., (\ref{6})). Configurations
obeying such restrictions are called tempered. In one or another
way tempered configurations always appear in the theory of system
of unbounded spins, see \cite{BH,COPP,LP,PY}. To impose the
restrictions we use special mappings, which define the scale of
growth of $\{|\xi_\ell |_{L^2_\beta}\}_{\ell \in \mathbf{L}}$.
Such mappings, called weights, are introduced by the following
\begin{definition} \label{wpn}
Weights are the symmetric maps $w_\alpha : \mathbf{L}\times
\mathbf{L} \rightarrow (0, +\infty)$, indexed by
\begin{equation} \label{w}
\alpha \in \mathcal{I} {=} (\underline{\alpha} ,\overline{ \alpha}),
\quad 0\leq \underline{\alpha}< \overline{\alpha} \leq +\infty,
\end{equation}
which satisfy the conditions: \vskip.1cm
\begin{tabular}{ll}
(a) \quad &for any $\alpha \in \mathcal{I}$ and  $\ell$, $w_\alpha
(\ell , \ell) = 1$;\\[.2cm] (b) \quad &for any $\alpha \in \mathcal{I}$ and $\ell_1 ,
\ell_2 , \ell_3$,
\end{tabular}
\begin{equation} \label{te}
 w_\alpha (\ell_1 , \ell_2) \cdot w_\alpha (\ell_2 , \ell_3) \leq w_\alpha (\ell_1 ,
 \ell_3)
 \quad \textit{(triangle} \ \textit{ inequality)},
 \end{equation}
\vskip.1cm
\begin{tabular}{ll}
(c) \quad &for any $\alpha , \alpha' \in \mathcal{I}$, such that
$\alpha < \alpha'$, and arbitrary $\ell, \ell'$,
\end{tabular}
\begin{equation} \label{te1}
w_{\alpha'}(\ell , \ell') \leq w_\alpha (\ell , \ell'), \quad
\lim_{|\ell - \ell'|\rightarrow + \infty}w_{\alpha'}(\ell , \ell') /
w_\alpha (\ell , \ell') = 0.
\end{equation}
\end{definition}
\noindent The concrete choice of $\{w_\alpha\}_{\alpha \in
\mathcal{I}}$ depends on the decay of $J_{\ell \ell'}$, which thus
will be subject to the following\\ \noindent
{\bf Assumption (B)}
For all $\alpha \in \mathcal{I}$,
\begin{equation} \label{25}
 \sup_{ \ell} \sum_{\ell'}
\log(1 + |\ell - \ell'|) \cdot w_{ \alpha} (\ell, \ell') < \infty;
\end{equation}
\begin{equation} \label{26}
\hat{J}_\alpha  \  \stackrel{\rm def}{=} \ \sup_{\ell} \sum_{\ell'}
|J_{\ell \ell'}| \cdot \left[w_\alpha (\ell,  \ell') \right]^{-1} <
\infty.
\end{equation}
Given $\delta >0$, which is a parameter of the theory,  there exists
$\alpha\in \mathcal{I}$, such that
\begin{equation} \label{26a}
\hat{J}_\alpha - \hat{J}_0 < \delta.
\end{equation}

The choice of $\delta$, based on the parameters of the model, will
be done later. One observes that the conditions (\ref{25}) and
(\ref{26}) are competitive. One can easily find examples of
$J_{\ell\ell'}$, obeying (\ref{6}), for which (\ref{25}) and
(\ref{26}) cannot be satisfied simultaneously for any choice of
the weights.

Now we  give the basic examples which will be used in the sequel.
Suppose that
\begin{equation} \label{24a}
\sup_{\ell} \sum_{\ell'} |J_{\ell \ell'}|\cdot \exp\left(\alpha
|\ell - \ell'| \right) < \infty, \quad {\rm for} \ {\rm a} \ {\rm
certain} \ \alpha >0.
\end{equation}
The supremum of such $\alpha$ (possibly infinite) is denoted by
$\overline{\alpha}$. Then we set
\begin{equation} \label{24b}
\mathcal{I} = (0,\overline{\alpha}), \quad \ \   w_\alpha (\ell,
\ell') = \exp\left(- \alpha |\ell - \ell'| \right).
\end{equation}
If the condition (\ref{24a}) does not hold for any positive
$\alpha$, we assume that
\begin{equation} \label{24c}
\sup_{\ell} \sum_{\ell'} |J_{\ell \ell'}|\cdot \left( 1 + |\ell -
\ell'|\right)^{ \alpha d} < \infty,
\end{equation}
for a certain $\alpha>1$. Then $\overline{\alpha}$ is set to be the
supremum of $\alpha$ obeying (\ref{24c}) and
\begin{equation} \label{24d}
\mathcal{I} = (1 ,\overline{\alpha}), \quad \ \
 w_\alpha (\ell, \ell') =
\left(1 + \varepsilon |\ell - \ell'| \right)^{- \alpha d},
\end{equation}
where the parameter $\varepsilon>0$ will be chosen for (\ref{26a})
to be satisfied.

 Given $u = (u_\ell)_{\ell \in \mathbf{L}} \in
\mathbf{R}^\mathbf{L}$, $\ell_0$, and $\alpha\in \mathcal{I}$, we
set
\begin{eqnarray} \label{ww}
|u|_{l^1(w_\alpha)}  =  \sum_{\ell} |u_\ell| w_\alpha (\ell_0,
\ell),\qquad |u|_{l^{\infty}(w_\alpha)}  =  \sup_{\ell}
\left\{|u_\ell| w_\alpha (\ell_0, \ell)\right\}, \nonumber
\end{eqnarray}
and introduce the Banach spaces
\begin{equation} \label{27}
l^p (w_\alpha )  = \left\{ u \in \mathbf{R}^\mathbf{L} \
\left\vert \ |u|_{l^p (w_\alpha )}< \infty \right. \right\}, \quad
p = 1, +\infty.
\end{equation}
\begin{remark} \label{1rm}
By (\ref{te1}), for $\alpha < \alpha'$, the embedding $l^1
(w_\alpha ) \hookrightarrow l^1 (w_{\alpha'} )$ is compact. By
(\ref{26}), for every $\alpha \in \mathcal{I}$, the operator $u
\mapsto J u$, defined as $( J u )_{\ell} = \sum_{\ell'} J_{\ell
\ell'} u_{\ell'}$, is bounded in both spaces $l^p (w_\alpha )$, $p
= 1, +\infty$. Its norm does not exceed $\hat{J}_\alpha$.
\end{remark}
\noindent For $\alpha \in \mathcal{I}$, we introduce
\begin{equation} \label{29}
\mathit{\Omega}_\alpha {=}  \left\{ \omega \in \mathit{\Omega} \
\left\vert \ \|\omega \|_\alpha \ \stackrel{\rm def}{=} \ \right.
\left[\sum_{\ell} |\omega_{\ell}|^2_{L^2_\beta} w_\alpha
(\ell_0,\ell) \right]^{1/2} < \infty \right\},
\end{equation}
and endow this set with the metric
\begin{equation} \label{met}
\rho_\alpha (\omega , \omega') = \|\omega - \omega'\|_{\alpha} +
\sum_{\ell} 2^{-|\ell|} \cdot \frac{|\omega_\ell -
\omega'_\ell|_{C_\beta}}{1 +|\omega_\ell - \omega'_\ell|_{C_\beta}},
\end{equation}
which turns it into a Polish space.
\begin{remark} \label{augustrk}
The topology of each of the spaces $l^p(w_\alpha)$,
$\mathit{\Omega}_\alpha$ is independent of the particular choice
of $\ell_0$. This follows from the properties of the weights
$w_\alpha$ assumed in Definition \ref{wpn}.
\end{remark}

The set of tempered configurations is defined to be
\begin{equation} \label{30}
\mathit{\Omega}^{\rm t} = \bigcap_{\alpha \in \mathcal{I} }
\mathit{\Omega}_\alpha.
\end{equation}
Equipped with the projective limit topology $\mathit{\Omega}^{\rm
t}$ becomes a Polish space as well. For any $\alpha \in
\mathcal{I}$, we have continuous dense embeddings
$\mathit{\Omega}^{\rm t}\hookrightarrow \mathit{\Omega}_\alpha
\hookrightarrow \mathit{\Omega}$. Then by the Kuratowski theorem it
follows that $\mathit{\Omega}_\alpha, \mathit{\Omega}^{\rm t} \in
\mathcal{B}(\mathit{\Omega})$ and
 the Borel $\sigma$-algebras of all these Polish spaces
coincide with the ones induced on them by
$\mathcal{B}(\mathit{\Omega})$. Now we are at a position to complete
the definition of the function (\ref{23}).
\begin{lemma} \label{1lm}
For every $\alpha \in \mathcal{I}$ and $\Lambda \Subset \mathbf{L}$,
the map $\mathit{\Omega}_\alpha \times \mathit{\Omega}_\alpha \ni
(\omega , \xi)\mapsto I_\Lambda (\omega | \xi)$ is continuous.
Furthermore, for every ball $B_\alpha (R) = \{\omega \in
\mathit{\Omega}_\alpha \ | \ \rho_\alpha (0, \omega) < R\}$, $R>0$,
it follows that
\begin{equation} \label{31}
\inf_{\omega \in \mathit{\Omega}, \ \xi \in B_\alpha (R)} I_\Lambda
(\omega |\xi)
> - \infty, \quad  \sup_{\omega , \xi \in B_\alpha (R)}|I_\Lambda
(\omega |\xi)| < + \infty.
\end{equation}
\end{lemma}
{\bf Proof:}
As the functions $V_\ell: \mathbf{R}^\nu \rightarrow \mathbf{R}$
are continuous, the map $(\omega, \xi) \mapsto I_\Lambda
(\omega_\Lambda)$ is continuous and bounded on the balls $B_\alpha
(R)$. Furthermore,
\begin{eqnarray} \label{32}
& & \quad  \left\vert \sum_{\ell \in \Lambda , \ \ell' \in
\Lambda^c} J_{\ell \ell'} (\omega_\ell , \xi_{\ell'})_{L^2_\beta}
\right\vert \leq \sum_{\ell \in \Lambda , \ \ell' \in \Lambda^c}
\left\vert J_{\ell \ell'}\right\vert \cdot |\omega_\ell
|_{L^2_\beta}\cdot |\xi_{\ell'} |_{L^2_\beta} \nonumber \\ & & \quad
 = \sum_{\ell \in \Lambda}|\omega_\ell |_{L^2_\beta} [w_\alpha
(0,\ell)]^{-1/2} \nonumber \\ & & \quad \times \sum_{\ell' \in
\Lambda^c} |J_{\ell \ell'}| \left[w_\alpha (0,\ell)/w_\alpha
(0,\ell')\right]^{1/2}\cdot |\xi_{\ell'} |_{L^2_\beta} [w_\alpha
(0,\ell')]^{1/2} \nonumber \\ & & \quad \leq \sum_{\ell \in
\Lambda}|\omega_\ell |_{L^2_\beta} [w_\alpha
(0,\ell)]^{-1/2}\sum_{\ell' \in \Lambda^c} |J_{\ell \ell'}|\cdot
[w_\alpha (\ell, \ell')]^{-1/2} \cdot |\xi_{\ell'} |_{L^2_\beta}
[w_\alpha (0,\ell')]^{1/2}\nonumber \\ & & \quad \quad \leq
 \hat{J}_\alpha \|\omega\|_\alpha  \|\xi\|_\alpha  \sum_{\ell \in
\Lambda}[w_\alpha (0,\ell)]^{-1},
\end{eqnarray}
where we used the triangle inequality (\ref{te}). This yields the
continuity stated and the upper bound in (\ref{31}). To prove the
lower bound we employ the super-quadratic growth of $V_\ell$ assumed
in (\ref{3}). Then for any $\varkappa>0$ and $\alpha \in
\mathcal{I}$, one finds $C>0$ such that for any $\omega \in
\mathit{\Omega}$ and $\xi\in \mathit{\Omega}^{\rm t}$,
\begin{eqnarray} \label{llb}
\qquad I_\Lambda (\omega|\xi) & \geq & B_V \beta |\Lambda| + A_V
\beta^{1-r}\sum_{\ell \in \Lambda} |\omega_\ell |_{L^2_\beta}^{2r} -
\frac{1}{2}\sum_{\ell ,\ell' \in \Lambda} J_{\ell \ell'}
(\omega_\ell , \omega_{\ell'} )_{L^2_\beta}\\& - & \sum_{\ell \in
\Lambda , \ \ell' \in \Lambda^c} J_{\ell  \ell'} (\omega_\ell ,
\xi_{\ell'} )_{L^2_\beta} \geq - C |\Lambda| + \varkappa\sum_{\ell
\in \Lambda} |\omega_\ell|^2_{L^2_\beta} \nonumber\\& - &
\hat{J}_\alpha \|\xi \|^2_\alpha \sum_{\ell \in \Lambda}w_\alpha(0,
\ell). \nonumber
\end{eqnarray}
 To get the latter estimate we used the
Minkowski inequality.
$\blacksquare$
\vskip.1cm \noindent Now for $\Lambda \Subset \mathbf{L}$ and $\xi
\in \mathit{\Omega}^{\rm t}$, we introduce the partition function
(c.f., (\ref{23a}))
\begin{equation} \label{33}
Z_\Lambda (\xi) = \int_{\mathit{\Omega}_\Lambda} \exp\left[-
I_\Lambda(\omega_\Lambda \times 0_{\Lambda^c} |\xi)  \right]
\chi_\Lambda ({\rm d}\omega_\Lambda).
\end{equation}
An immediate corollary of the estimates (\ref{16h}) and (\ref{llb})
is the following
\begin{proposition} \label{2pn}
For every $\Lambda \Subset \mathbf{L}$, the function
$\mathit{\Omega}^{\rm t} \ni \xi \mapsto Z_\Lambda (\xi)\in (0, +
\infty)$ is continuous. Moreover, for any  $R>0$,
\begin{equation} \label{tania}
\inf_{\xi \in B_\alpha (R)} Z_\Lambda (\xi) > 0, \qquad \ \ \
\sup_{\xi \in B_\alpha (R)} Z_\Lambda (\xi)< \infty.
\end{equation}
\end{proposition}

\subsection{Local  specification and Euclidean Gibbs measures }
\label{ss2.5}

Note that the standard sources on the DLR approach are the books
\cite{Ge,Pr}.

 The local Gibbs specification is the
family $\{\pi_\Lambda \}_{\Lambda \Subset \mathbf{L}}$ of measure
kernels
\[
\mathcal{B}(\mathit{\Omega}) \times \mathit{\Omega} \ni (B,
\xi)\mapsto \pi_\Lambda (B|\xi) \in [0, 1]
\]
which we define as follows. For $\xi \in \mathit{\Omega}^{\rm t}$,
$\Lambda\Subset \mathbf{L}$, and $B \in \mathcal{B}(\Omega)$, we set
\begin{equation} \label{34}
\pi_\Lambda (B|\xi) = \frac{1}{Z_\Lambda(\xi)}
\int_{\mathit{\Omega}_\Lambda}\exp\left[- I_\Lambda(\omega_\Lambda
\times 0_{\Lambda^c} |\xi)  \right] \mathbf{I}_B (\omega_\Lambda
\times \xi_{\Lambda^c})\chi_\Lambda ({\rm d}\omega_\Lambda ),
\end{equation}
where $\mathbf{I}_B $ stands for the indicator of $B$. We also set
\begin{equation} \label{34a}
\pi_\Lambda (\cdot|\xi) \equiv 0, \quad {\rm for} \ \ \xi \in
\mathit{\Omega} \setminus \mathit{\Omega}^{\rm t}.
\end{equation}
To simplify notations we write $\pi_{\{\ell\}} = \pi_\ell$. From
these definitions one readily derives a consistency property
\begin{equation} \label{35}
\int_{\mathit{\Omega}} \pi_\Lambda (B|\omega) \pi_{\Lambda'} ({\rm
d} \omega |\xi) = \pi_{\Lambda'} (B |\xi), \quad \Lambda \subset
\Lambda',
\end{equation}
which holds for all $B\in \mathcal{B}(\mathit{\Omega})$ and $\xi \in
\mathit{\Omega}$. Furthermore, by (\ref{llb}) it follows that for
any $\xi\in \mathit{\Omega}$, $\sigma\in (0, 1/2)$, and
$\varkappa>0$,
\begin{equation} \label{llb1}
\int_{\mathit{\Omega}} \exp\left\{ \sum_{\ell \in \Lambda} \left(
\lambda_\sigma |\omega_\ell |^2_{C_\beta^\sigma} +
\varkappa|\omega_\ell|_{L^2_\beta }^2 \right) \right\} \pi_\Lambda
({\rm d} \omega|\xi) <\infty,
 \end{equation}
 where $\lambda_\sigma$ is the same as in Proposition \ref{1pn}.

By $C_{\rm b}(\mathit{\Omega}_\alpha)$ (respectively, $C_{\rm
b}(\mathit{\Omega}^{\rm t})$) we denote the Banach spaces of all
bounded continuous functions $f:\mathit{\Omega}_\alpha \rightarrow
\mathbf{R}$ (respectively, $f:\mathit{\Omega}^{\rm t} \rightarrow
\mathbf{R}$) equipped with the supremum norm. For every $\alpha \in
\mathcal{I}$, one has a natural embedding $C_{\rm
b}(\mathit{\Omega}_\alpha ) \hookrightarrow C_{\rm
b}(\mathit{\Omega}^{\rm t})$.
\begin{lemma} [Feller Property] \label{2lm}
For every $\alpha \in \mathcal{I}$, $\Lambda \Subset \mathbf{L}$,
and any $f \in C_{\rm b}(\mathit{\Omega}_{\alpha})$, the function
\begin{eqnarray} \label{f}
& & \mathit{\Omega}_\alpha \ni \xi \mapsto \pi_\Lambda (f | \xi) \\
& & \qquad \qquad  \stackrel{\rm def}{=} \
 \frac{1}{Z_\Lambda (\xi)}
\int_{\mathit{\Omega}_\Lambda}f (\omega_\Lambda \times
\xi_{\Lambda^c}) \exp\left[- I_\Lambda (\omega_\Lambda \times
0_{\Lambda^c}|\xi) \right]\chi_\Lambda ({\rm d}\omega_\Lambda),
\nonumber
\end{eqnarray}
belongs to $C_{\rm b}(\mathit{\Omega}_{\alpha})$. The linear
operator $f \mapsto \pi_\Lambda (f|\cdot)$ is a contraction on
$C_{\rm b}(\mathit{\Omega}_\alpha)$.
\end{lemma}
{\bf Proof:}
By Lemma \ref{1lm} and Proposition \ref{2pn} the integrand
\[
G^f_\Lambda (\omega_\Lambda|\xi) \ \stackrel{\rm def}{=} \ f
(\omega_\Lambda \times \xi_{\Lambda^c}) \exp\left[- I_\Lambda
(\omega_\Lambda \times 0_{\Lambda^c}|\xi) \right] / Z_\Lambda (\xi)
\]
is continuous in both variables. Moreover, by (\ref{31}) and
(\ref{tania}) the map
\[
\mathit{\Omega}_\alpha \ni \xi \mapsto \sup_{\omega_\Lambda \in
\mathit{\Omega}_\Lambda} |G^f_\Lambda (\omega_\Lambda|\xi)|
\]
is  bounded on every ball $B_\alpha (R)$. This allows one to apply
Lebesgue's dominated convergence theorem and obtain the continuity
stated. Obviously,
\begin{equation} \label{39}
\sup_{\xi \in \mathit{\Omega}_\alpha }\left\vert \pi_\Lambda (f
|\xi) \right\vert \leq \sup_{\xi \in \mathit{\Omega}_\alpha
}|f(\xi)|.
\end{equation}
$\blacksquare$
\noindent Note that by (\ref{34}), for $\xi \in \mathit{\Omega}^{\rm
t}$, $\alpha \in \mathcal{I}$, and $f \in C_{\rm
b}(\mathit{\Omega}_\alpha)$,
\begin{equation} \label{fp}
\pi_\Lambda (f|\xi) = \int_{\mathit{\Omega}}f(\omega)
\pi_\Lambda({\rm d}\omega|\xi).
\end{equation}
Recall that the particular cases of the model considered were
established by Definition \ref{1df}. For $B\in
\mathcal{B}(\mathit{\Omega})$ and $U\in O(\nu)$, we set
\[
  U \omega = (U
\omega_\ell)_{\ell \in \mathbf{L}} \qquad UB = \{ U \omega \  | \
\omega \in B\}.
\]
If $\mathbf{L}$ is a lattice, for a given $\ell_0$, we set
\[
t_{\ell_0} (\omega) = (\omega_{\ell - \ell_0})_{\ell \in
\mathbf{L}}, \qquad t_{\ell_0}( B) = \{t_{\ell_0}(\omega) \  | \
\omega \in B\}.
\]
Then if the model possesses the corresponding symmetry, one has
\begin{equation} \label{MA10}
\pi_\Lambda (U B |U\xi) = \pi_\Lambda (B| \xi), \qquad \pi_{\Lambda
+ \ell} (t_\ell (B)|t_\ell (\xi)) = \pi_\Lambda (B|\xi),
\end{equation}
which ought to hold for all $U$, $\ell$, $B$, and $\xi$.
\begin{definition} \label{3df}
A measure $\mu \in \mathcal{P}(\mathit{\Omega})$ is called a
tempered Euclidean Gibbs measure if it satisfies the
Dobrushin-Lanford-Ruelle (equilibrium) equation
\begin{equation} \label{40}
\int_{\mathit{\Omega}}\pi_\Lambda (B |\omega) \mu({\rm d}\omega) =
\mu(B), \quad { for} \ { all} \ \ \ \Lambda \Subset \mathbf{L} \ \ {
and} \ \ B \in \mathcal{B}(\mathit{\Omega}).
\end{equation}
\end{definition}
\noindent By $\mathcal{G}^{\rm t}$ we denote the set of all tempered
Euclidean Gibbs measures of our model. So far we do not know whether or not
$\mathcal{G}^{\rm t}$ is non-void; if it is, its elements are
supported by $\mathit{\Omega}^{\rm t}$. Indeed, by (\ref{34}) and
(\ref{34a}) $\pi_\Lambda (\mathit{\Omega} \setminus
\mathit{\Omega}^{\rm t} |\xi) = 0$ for every $\Lambda \Subset
\mathbf{L}$ and $\xi \in \mathit{\Omega}$. Then by (\ref{40}),
\begin{equation} \label{40a}
\mu (\mathit{\Omega} \setminus \mathit{\Omega}^{\rm t})  = 0 \ \
\Longrightarrow \ \ \mu(\mathit{\Omega}^{\rm t}) =1.
\end{equation}
Furthermore,
\begin{equation}
\mu \left( \left\{ \omega \in \mathit{\Omega }^{ \mathrm{t}} \ | \
\forall \ell \in \mathbf{L}: \  \omega_\ell \in C_{\beta }^{\sigma }
\right\} \right) =1,  \label{40b}
\end{equation}%
which follows from (\ref{llb1}). If the model is translation and/or
rotation invariant, then, for every $U\in O(\nu)$ and $\ell\in
\mathbf{L}$, the corresponding transformations preserve
$\mathcal{G}^{\rm t}$. That is, for any $\mu \in \mathcal{G}^{\rm
t}$,
\begin{equation} \label{MA11}
\Theta_U (\mu) \ \stackrel{\rm def}{=} \ \mu \circ U^{-1} \in
\mathcal{G}^{\rm t}, \qquad \theta_\ell (\mu) \ \stackrel{\rm
def}{=} \ \mu \circ t^{-1}_\ell \in \mathcal{G}^{\rm t}.
\end{equation}
In particular, if $\mathcal{G}^{\rm t}$ is a singleton, its unique
element should be invariant in the same sense as the model. One more
invariance of the Euclidean Gibbs measures is connected with the
dependence of their Matsubara functions on $\tau$'s.
\begin{definition} \label{shiftdf}
A measure $\mu \in \mathcal{G}^{\rm t}$ is called $\tau$-shift
invariant if its Matsubara functions (\ref{kms}) have the property
(\ref{mats1}).
\end{definition}
\noindent The $\tau$-shift invariance is crucial for reconstructing
quantum Gibbs states on von Neumann algebras, see
\cite{Birke,Gelerak2}. This means that only the elements of
$\mathcal{G}^{\rm t}$ which have this property are of physical
relevance.

Given $\alpha  \in \mathcal{I}$, by $\mathcal{W}_\alpha$ we denote
the usual weak topology on the set of all probability measures
$\mathcal{P}(\mathit{\Omega}_\alpha)$ defined by means of bounded
continuous functions on $\mathit{\Omega}_\alpha$. By
$\mathcal{W}^{\rm t}$ we denote the weak topology on
$\mathcal{P}(\mathit{\Omega}^{\rm t})$. With these topologies the
sets $\mathcal{P}(\mathit{\Omega}_\alpha)$ and
$\mathcal{P}(\mathit{\Omega}^{\rm t})$ become Polish spaces
(Theorem 6.5, page 46 of \cite{Pa}).

The proof of the existence of Euclidean Gibbs measures will be
based on the following statement.
\begin{lemma} \label{3lm}
For each $\alpha \in \mathcal{I}$, every
$\mathcal{W}_\alpha$-accumulation point $\mu \in
\mathcal{P}(\mathit{\Omega}^{\rm t})$ of the family $\{\pi_\Lambda
(\cdot |\xi) \ | \ \Lambda \Subset \mathbf{L}, \ \xi \in
\mathit{\Omega}^{\rm t}\}$ is an element of $\mathcal{G}^{\rm t}$.
\end{lemma}
{\bf Proof:}
For each $\alpha \in \mathcal{I}$, $C_{\rm b}
(\mathit{\Omega}_{\alpha})$ is a measure defining class for
$\mathcal{P}(\mathit{\Omega}^{\rm t})$. Then a measure $\mu \in
\mathcal{P}(\mathit{\Omega}^{\rm t})$ solves (\ref{40}) if and only
if for any $f \in C_{\rm b} (\mathit{\Omega}_{\alpha})$ and all
$\Lambda \Subset \mathbf{L}$,
\begin{equation} \label{42}
\int_{\mathit{\Omega}^{\rm t}} f (\omega)\mu ({\rm d} \omega) =
\int_{\mathit{\Omega}^{\rm t}} \pi_\Lambda (f |\omega) \mu ({\rm d}
\omega).
\end{equation}
Let $\{\pi_{\Lambda_k} (\cdot |\xi_k)\}_{k \in \mathbf{N}}$ converge
in $\mathcal{W}_{\alpha}$ to some $\mu\in
\mathcal{P}(\mathit{\Omega}^{\rm t})$. For every $\Lambda \Subset
\mathbf{L}$, one finds $k_\Lambda \in \mathbf{N}$ such that $\Lambda
\subset \Lambda_k$ for all $k > k_\Lambda$. Then by (\ref{35}), one
has
\[
\int_{\mathit{\Omega}^{\rm t}} f (\omega) \pi_{\Lambda_k }({\rm d}
\omega|\xi_k) = \int_{\mathit{\Omega}^{\rm t}} \pi_\Lambda (f
|\omega) \pi_{\Lambda_k }({\rm d} \omega|\xi_k).
\]
Now by Lemma \ref{2lm}, one can pass to the limit $k \rightarrow
+\infty$ and get (\ref{42}).
$\blacksquare$
\vskip.1cm Let us stress that in the lemma above we suppose that the
accumulation point is a probability measure on $\mathit{\Omega}^{\rm
t}$. In general, the convergence of $\{\mu_n\}_{n \in \mathbf{N}}
\subset \mathcal{P}({\mathit{\Omega}}^{\rm t})$ in every
$\mathcal{W}_\alpha$, $\alpha \in \mathcal{I}$, does not yet imply
its $\mathcal{W}^{\rm t}$-convergence. However, in Lemma \ref{tanlm}
and Corollary \ref{tanco} below we show that the topologies induced
by $\mathcal{W}_\alpha$ and $\mathcal{W}^{\rm t}$ on a certain
subset of $\mathcal{P}(\mathit{\Omega})$, which includes
$\mathcal{G}^{\rm t}$ and all $\pi_\Lambda (\cdot|\xi)$,   coincide.

\section{The Results}
\label{3s}

In the first subsection below we present the statements describing
the general case, whereas the second subsection is dedicated to the
case of $\nu=1$ and $J_{\ell \ell'}\geq 0$.

\subsection{Euclidean Gibbs measures in the general case}\label{ss3.1}

We begin by establishing existence of tempered Euclidean Gibbs
measures and compactness of their set $\mathcal{G}^{\rm t}$. For
models with non-compact spins, here they are even
infinite-dimensional, such a property is far from being evident.
\begin{theorem} \label{1tm}
For every $\beta>0$, the set of tempered Euclidean Gibbs measures
$\mathcal{G}^{\rm t}$ is non-void and $\mathcal{W}^{\rm t}$-
compact.
\end{theorem}
The next theorem gives an exponential moment estimate similar to
(\ref{16h}). Recall that the H\"{o}lder norm $|\cdot |_{C_{\beta
}^{\sigma }}$ was defined by (\ref{16z}).
\begin{theorem} \label{2tm}
For every $\sigma \in (0, 1/2)$ and $\varkappa >0$, there exists  a
positive constant $ C_{\ref{43}}$ such that, for any $\ell $ and for
all $\mu \in \mathcal{G}^{\rm t}$,
\begin{equation} \label{43}
\int_{\mathit{\Omega}} \exp\left(\lambda_\sigma |\omega_\ell
|_{C^\sigma_\beta}^2 + \varkappa |\omega_\ell |_{L^2_\beta}^2
\right)\mu({\rm d}\omega) \leq C_{\ref{43}},
\end{equation}
where $\lambda_\sigma$ is the same as in (\ref{16h}).
\end{theorem}
According to (\ref{43}), the one-site projections of each $\mu\in
\mathcal{G}^{\rm t}$ are sub-Gaussian. The bound $C_{\ref{43}}$ does
not depend on $\ell$ and is the same for all $\mu \in
\mathcal{G}^{\rm t}$, though it may depend on $\sigma $ and
$\varkappa$. The estimate (\ref{43}) plays a crucial role in the
theory of the set $\mathcal{G}^{\rm t}$. Such estimates are also
important in the study of the Dirichlet operators $H_\mu$ associated
with the measures $\mu\in \mathcal{G}^{\rm t}$, see \cite{AKR,AKR1}.

 The set of tempered configurations
$\mathit{\Omega}^{\rm t}$ was introduced in (\ref{29}), (\ref{30})
by means of rather slack restrictions (c.f. (\ref{24})) imposed on
the $L^2_\beta$-norms of $\omega_\ell$. By construction, the
elements of $\mathcal{G}^{\rm t}$ are supported by this set, see
(\ref{40a}). It turns out that they have a much smaller support (a
kind of the Lebowitz-Presutti one, see \cite{LP}). Given $b>0$ and
$\sigma\in (0, 1/2)$, we set
\begin{eqnarray} \label{47}
\mathit{\Xi} ( b , \sigma) &  = & \{ \xi \in \mathit{\Omega} \ |  \
(\forall \ell_0 \in \mathbf{L}) \ (\exists \Lambda_{\xi ,
\ell_0}\Subset
\mathbf{L}) \ (\forall \ell \in \Lambda_{\xi , \ell_0}^c): \\
& & \qquad \qquad \qquad \qquad  \qquad \qquad |\xi_\ell
|_{C^\sigma_\beta}^2 \leq b \log( 1 + |\ell - \ell_0|)\}, \nonumber
\end{eqnarray}
which in view of (\ref{25}) is a Borel subset of
$\mathit{\Omega}^{\rm t}$.
\begin{theorem}\label{3tm}
For every $\sigma \in (0, 1/2)$, there exists $b>0$, which depends
on $\sigma$ and on the parameters of the model only, such that for
all $\mu \in \mathcal{G}^{\rm t}$,
\begin{equation} \label{48}
\mu (\mathit{\Xi}(b , \sigma)) = 1.
\end{equation}
\end{theorem}
The last result in this group is a sufficient condition for
$\mathcal{G}^{\rm t}$ to be a singleton, which holds for high
temperatures (small $\beta$). It is obtained by controlling the
`non-convexity' of the potential energy (\ref{pe}). Let us decompose
\begin{equation}\label{decom}
V_{\ell} = V_{1, \ell} + V_{2, \ell},
\end{equation}
where $V_{1, \ell}\in C^2 (\mathbf{R}^\nu)$ is such that
\begin{equation} \label{dc1}
- a \leq b \ \stackrel{\rm def}{=} \ \inf_{\ell} \inf_{x, y \in
\mathbf{R}^\nu, \ y\neq 0}\left( V''_{1,\ell}(x)y, y \right)/|y|^2 <
\infty.
\end{equation}
As for the second term, we set
\begin{equation} \label{dc2}
0 \leq \delta \ \stackrel{\rm def}{=} \ \sup_{\ell} \left\{ \sup_{x
\in \mathbf{R}^\nu}V_{2, \ell}(x) - \inf_{x \in \mathbf{R}^\nu}V_{2,
\ell}(x) \right\} \leq \infty.
\end{equation}
Its role is to produce multiple minima of the potential energy
responsible for eventual phase transitions. Clearly, the
decomposition (\ref{decom}) is not unique; its optimal realizations
for certain types of $V_\ell$ are discussed in section 6 of
\cite{AKRT}.
\begin{theorem} \label{httm}
The set $\mathcal{G}^{\rm t}$ is a singleton if
\begin{equation}
\label{dc3} e^{\beta \delta} <(a + b)/\hat{J}_0.
\end{equation}
\end{theorem}
\begin{remark} \label{apprm}
The latter condition surely holds at all $\beta$ if
\begin{equation} \label{si}
\delta = 0 \quad {\rm and} \quad \hat{J}_0 < a + b.
\end{equation}
In this case the potential energy $W_\Lambda$ given by (\ref{pe}) is
convex. If the oscillators are harmonic, $\delta = b = 0$, which
yields the stability condition
\begin{equation} \label{si1}
\hat{J}_0 < a.
\end{equation}
The condition (\ref{dc3}) does not contain the particle mass $m$;
hence, the property stated holds also in the quasi-classical
limit\footnote{More details on this limit can be found in
\cite{AKKR}.} $m \rightarrow + \infty$.
\end{remark}

\subsection{Ferroelectric scalar models}
\label{sss3.2} Recall that here we consider the case where
$J_{\ell \ell'} \geq 0$ and $\nu =1$.

Let us introduce an order on the set $\mathcal{G}^{\rm t}$. As the
components of the configurations $\omega\in \mathit{\Omega}$ are
continuous functions $\omega_\ell :S_\beta \rightarrow
\mathbf{R}^\nu$, we can  set $\omega \leq \tilde{\omega}$ if
$\omega_\ell(\tau) \leq \tilde{\omega}_\ell(\tau)$ for all $\ell$
and  $\tau$. Thereby, we define
\begin{equation} \label{MA1}
K_+ (\mathit{\Omega}^{\rm t}) = \{ f\in C_{\rm
b}(\mathit{\Omega}^{\rm t}) \ | \ f(\omega) \leq
f(\tilde{\omega}), \quad {\rm if} \ \ \omega \leq
\tilde{\omega}\},
\end{equation}
which is a cone of bounded continuous functions.
\begin{lemma} \label{MAlm}
If for given $\mu, \tilde{\mu} \in \mathcal{G}^{\rm t}$, one has
\begin{equation} \label{MA1a}
\mu(f) = \tilde{\mu}(f), \qquad {\rm for} \ \ {\rm all} \ \ f \in
K_+ (\mathit{\Omega}^{\rm t}),
\end{equation}
then $\mu = \tilde{\mu}$.
\end{lemma}
The proof of this lemma will be given below in Section \ref{5s}.
We use it to establish the so called stochastic order on
$\mathcal{G}^{\rm t}$.
\begin{definition} \label{MAdf}
For $\mu, \tilde{\mu}\in\mathcal{G}^{\rm t}$, we say that $\mu\leq
\tilde{\mu}$, if
\begin{equation} \label{MAW}
\mu(f) \leq \tilde{\mu}(f), \qquad {\rm for} \ \ {\rm all} \ \ f \in
K_+(\mathit{\Omega}^{\rm t}).
\end{equation}
\end{definition}

 Our first result in this subsection is the following
\begin{theorem} \label{MAtm}
The set $\mathcal{G}^{\rm t}$ possesses maximal $\mu_{+}$ and
minimal $\mu_{-}$ elements in the sense of Definition \ref{MAdf}.
These elements are extreme and $\tau$-shift invariant; they are
also translation invariant if the model is translation invariant.
If $V_\ell (-x ) = V_\ell (x)$ for all $\ell $, then $\mu_{+} (B)
= \mu_{-} (- B)$ for all $B \in \mathcal{B}(\mathit{\Omega})$.
\end{theorem}
Now let the model be translation invariant. For this model, we are
going to study the limiting pressure which contains important
information about its thermodynamic properties. A particular
question here is the dependence of the pressure on the external
field $h$, c.f. (\ref{4}). The corresponding analytic properties are
then used in the study of phase transitions.

For $\Lambda \Subset \mathbf{L}$, we set
\begin{equation} \label{c1}
p_\Lambda (h, \xi) = \frac{1}{|\Lambda|} \log Z_\Lambda (\xi),
\quad \xi \in \mathit{\Omega}^{\rm t}.
\end{equation}
To simplify notations we write $p_\Lambda (h) = p_\Lambda (h, 0)$.
For $\mu \in \mathcal{G}^{\rm t}$, we set
\begin{equation} \label{c2}
p^\mu_\Lambda (h) = \int_{\mathit{\Omega}}p_\Lambda (h , \xi)
\mu({\rm d}\xi).
\end{equation}
If for a cofinal sequence $\mathcal{L}$, the limit
\begin{equation} \label{c3}
p^\mu (h) \ \stackrel{\rm def}{=} \
\lim_{\mathcal{L}}p^\mu_\Lambda (h),
\end{equation}
exists, we shall call it pressure in the state $\mu$. We shall also
consider
\begin{equation} \label{CC}
p(h) \  \stackrel{\rm def}{=} \ \lim_{\mathcal{L}} p_\Lambda (h).
\end{equation}
To obtain such limits we impose certain conditions on the sequences
$\mathcal{L}$. Given $l = (l_1 , \dots l_d)$, $l' = (l'_1 , \dots
l'_d)\in \mathbf{L}= \mathbf{Z}^d$, such that $l_j < l'_j$ for all
$j=1, \dots , d$, we set
\begin{equation} \label{de1}
\Gamma = \{ \ell \in \mathbf{L} \ | \ l_j \leq  \ell_j \leq l'_j,
\ \ {\rm for} \ {\rm all}\  j = 1 , \dots , d\}.
\end{equation}
For this parallelepiped, let $\mathfrak{G}(\Gamma)$ be the family
of all pair-wise disjoint translates of $\Gamma$ which cover
$\mathbf{L}$. Then for $\Lambda \Subset \mathbf{L}$, we set
$N_{-}(\Lambda|\Gamma)$ (respectively, $N_{+}(\Lambda|\Gamma)$) to
be the number of the elements of $\mathfrak{G}(\Gamma)$ which are
contained in $\Lambda$ (respectively, which have non-void
intersections with $\Lambda$). Then we introduce the following
notion, see \cite{Ru}.
\begin{definition} \label{rdf}
A cofinal sequence $\mathcal{L}$ is a van Hove sequence if for every
$\Gamma$,
\begin{equation} \label{de2}
(a) \ \  \lim_{\mathcal{L}} N_{-}(\Lambda |\Gamma) = +\infty;
\quad \quad (b) \ \ \lim_{\mathcal{L}}\left( N_{-}(\Lambda
|\Gamma)/ N_{+}(\Lambda |\Gamma)\right)= 1.
\end{equation}
\end{definition}
\begin{theorem} \label{pressuretm}
For every $h\in \mathbf{R}$ and any van Hove sequence
$\mathcal{L}$, it follows that the limits (\ref{c3}) and
(\ref{CC}) exist, do not depend on the particular choice of
$\mathcal{L}$, and are equal, that is $p(h) = p^\mu (h)$ for each
$\mu\in \mathcal{G}^{\rm t}$.
\end{theorem}

The following result, which will be proven in section \ref{sec7}
below, is a consequence of Theorems \ref{pressuretm} and
\ref{MAtm}.
\begin{corollary} \label{Mco}
If the pressure $p(h)$ is differentiable at a given $h$, then
$\mathcal{G}^{\rm t}$ is a singleton at this $h$.
\end{corollary}

Next we study the uniqueness/multiplicity problem for the
Euclidean Gibbs measures. In the DLR approach the multiplicity
corresponds to phase transitions.
 In physical systems phase transitions manifest themselves
in the macroscopic displacements of particles from their equilibrium
positions.
 For translation invariant ferroelectric
models with $V_\ell = V$ obeying certain conditions, the appearance
of such macroscopic displacements at low temperatures was proven in
\cite{BK1,DLP,Helf,Kond,Pastur}.  Thus, one can expect that
$|\mathcal{G}^{\rm t}|
> 1$ at big $\beta$, although the latter fact and the appearance
of macroscopic displacements are not equivalent. To avoid technical complications we
prove this for $\mathbf{L} = \mathbf{Z}^d$, $d\geq 3$,
however our scheme can be modified for certain types of irregular
$\mathbf{L} \subset \mathbf{R}^d$.

Let us impose further conditions  on $J_{\ell\ell'}$ and $V_\ell$.
The first one is
\begin{equation} \label{p1}
\inf_{\ell,\ell': \ |\ell - \ell'| =1} J_{\ell \ell'} \ \stackrel{\rm
def}{=} \ J>0.
\end{equation}
Next we suppose that $V_\ell$ are even continuous functions and
the upper bound in (\ref{3}) can be chosen in the form
\begin{equation} \label{ub}
V(x_\ell ) = \sum_{s=1}^r b^{(s)} x_\ell^{2s}; \quad 2 b^{(1)} < - a
; \ \ \ b^{(s)}\geq 0, \  s\geq 2,
\end{equation}
where $a$ is the same as in (\ref{16e}) or in (\ref{pe}), and $r\geq
2$ is either a positive integer or infinite. For $r=+\infty$, we
assume that the series
\begin{equation} \label{p3}
\mathit{\Phi}(t) = \sum_{s=2}^{+\infty}
\frac{(2s)!}{2^{s-1}(s-1)!}{b}^{(s)} t^{s-1},
\end{equation} converges at some $t>0$.
Since $2b^{(1)} + a <0$, the equation
\begin{equation} \label{p4}
a + 2b^{(1)} + \mathit{\Phi}(t) = 0,
\end{equation} has a unique solution $t_* >0$.
Finally, we suppose that for every $\ell$,
\begin{equation} \label{ub1}
V(x_\ell ) - V_\ell (x_\ell) \leq V(\tilde{x}_\ell )  - V_\ell
(\tilde{x}_\ell) , \quad {\rm whenever} \  \ x_\ell^2 \leq
\tilde{x}_\ell^2.
\end{equation}
By these assumptions all $V_\ell$ are `uniformly double-welled'. If
$V_\ell (x_\ell)= v_\ell (x_\ell^2)$ and $v_\ell$ are
differentiable, the condition (\ref{ub1}) may be formulated as an
upper bound for $v_\ell'$. For $d\geq 3$, we set
\begin{equation} \label{cj}
\theta_d = \frac{1}{(2\pi)^d}\int_{(- \pi, \pi]^d}\frac{{\rm
d}p}{{E(p)}}, \qquad \ E(p) = \sum_{j=1}^d [1-\cos p_j].
\end{equation}
Let also $f:[0,+\infty) \rightarrow [0,1)$ be the function defined
implicitly by
\begin{equation} \label{may1}
f(t \tanh t) = t^{-1} \cdot\tanh t, \ \ \ {\rm for} \ \  t>0, \ \
{\rm and} \ \ f(0) = 1.
\end{equation}
It is  convex and monotone decreasing on $(0,+\infty)$. For an
account of the properties of this function, see \cite{DLS}, where it
was introduced.

By (\ref{may1}) one readily proves that for every fixed $\alpha
>0$, the function
\begin{equation} \label{may2}
(0, +\infty) \ni t \mapsto \phi (t,\alpha) =  \alpha t f (t/\alpha),
\end{equation}
is monotone increasing to $\alpha^2$ as $t\rightarrow +\infty$.
\begin{theorem} \label{phtm}
Let $d\geq3$ and the above assumptions hold. Then under the
condition
\begin{equation} \label{cj1}
J > \theta_d / 8 m t_*^2,
\end{equation}
there exists $\beta_*>0$ such that $|\mathcal{G}^{\rm t}|
>1$ whenever $\beta> \beta_*$. The bound $\beta_*$ is the unique
solution of the equation
\begin{equation} \label{cja}
2 \theta_d m / J = \phi(\beta, 4 m t_*).
\end{equation}
\end{theorem}
As was shown in \cite{AKK,AKKRprl,Koz}, strong quantum effects,
occurring in particular at small values of the particle mass $m$,
can suppress abnormal fluctuations. Thus, one might expect that
such effects can cause $|\mathcal{G}^{\rm t}|=1$.  The strongest
result in this domain -- the uniqueness at all $\beta$ due to
quantum effects for the model with  nearest neighbor interaction
and a certain type of $V$ (so called BFS, see \cite{FFS}) -- was
proven in \cite{AKKR02b}. In the present paper we extend this
result in two directions. We prove it for a substantially larger
class of anharmonic potentials and make precise the bounds of the
uniqueness regime. Furthermore, unlike to the mentioned papers, we
do not suppose that the interaction has finite range and that
$\mathbf{L}$ is regular.

In Theorem \ref{5tm} below we suppose that the anharmonic
potentials $V_\ell$ are even and hence can be presented in the
form
\begin{equation} \label{MAWe1}
V_\ell(x) = v_\ell(x^2).
\end{equation}
Furthermore, we suppose that there exists the function $v:[0,
+\infty) \rightarrow \mathbf{R}$ which is convex and such that
\begin{equation} \label{49}
v_\ell (t ) -  v (t) \leq v_\ell (\theta ) - v (\theta) \quad {\rm
whenever} \ \ t < \theta.
\end{equation}
In typical cases of $V_\ell$, like (\ref{4}), as such a $v$ one can
take a convex polynomial of degree $r\geq 2$.

Next we introduce the following
 one-particle Hamiltonian (c.f. (\ref{16e}), (\ref{sch}))
\begin{equation} \label{2}
\tilde{H} = - \frac{1}{2m}\left(\frac{\partial }{\partial x
}\right)^2 + \frac{a}{2} x^2 + v (x^2), \quad x \in \mathbf{R} .
\end{equation}
It
 has purely discrete non-degenerate spectrum
$\{E_{n}\}_{n \in \mathbf{N}_0}$. Thus, one can define the parameter
\begin{equation} \label{51}
\mathit{\Delta} = \min_{n \in \mathbf{N}}\left(E_{n} - E_{n-1}
\right),
\end{equation}
which is positive and  depends on the model parameters $m$, $a$, and
on the choice of $v$. Recall, that $\hat{J}_0$ was defined by
(\ref{6}).
\begin{theorem} \label{5tm}
Let the anharmonic potentials $V_\ell$ be as above. Then the set of Euclidean Gibbs measures is a
singleton if
\begin{equation} \label{52}
m \mathit{\Delta}^2 > \hat{J}_0.
\end{equation}
\end{theorem}
\noindent Note that the above result is independent of $\beta>0$ and
that
 (\ref{52}) is a stability condition like (\ref{si}), where the
parameter $m \mathit{\Delta}^2$ appears as the oscillator rigidity.
If it holds, a stability-due-to-quantum-effects occurs, see
\cite{AKKRprl,KK,Koz,Ko}. If $v$ is a polynomial of degree $r \geq
2$, the rigidity $m \mathit{\Delta}^2$ is a continuous function of
the particle mass $m$; it gets small in the quasi-classical limit $m
\rightarrow +\infty$, see \cite{Ko}. At the same time, for $m
\rightarrow 0+$, one has $m \mathit{\Delta}^2 =
O(m^{-(r-1)/(r+1)})$, see \cite{AKK,Ko}. Hence, (\ref{52}) certainly
holds in the small mass limit, c.f., \cite{AKKR01,AKKR02b}. To
compare the latter result with Theorem \ref{phtm} let us assume that
 $\mathbf{L}=\mathbf{Z}^d$, $d\geq3$, $J_{\ell \ell'} = J$
iff $|\ell - \ell'|=1$, and all $V_\ell$ coincide with the
function given by (\ref{ub}). Then the parameter (\ref{51}) obeys
the estimate $\mathit{\Delta} < 1 /2 m t_*$, see \cite{Ko}, where
$t_*$ is the same as in (\ref{cj1}), (\ref{cja}). In this case the
condition (\ref{52}) can be rewritten as
\begin{equation} \label{cj2}
J < 1 / 8 d m t_*^2.
\end{equation}
One can show that $\theta_d > 1/d$ and $d \theta_d \rightarrow 1 $
as $d \rightarrow + \infty$; hence,  the estimates (\ref{cj1}) and
(\ref{cj2}), which give sufficient conditions for the phase
transition to occur or to be suppressed, become asymptotically
sharp.

 Now we consider a translation invariant version of our model, i.e.,
 $\mathbf{L}= \mathbf{Z}^d$. Set
\begin{equation} \label{m1}
\mathcal{F}_{\rm Laguerre} = \left\{ \varphi: \mathbf{R} \rightarrow
\mathbf{R} \ \left\vert \ \varphi(t) = \varphi_0 \exp( \gamma_0 t)
t^n \prod_{i=1}^{\infty} (1+ \gamma_i t) \right. \right\},
\end{equation}
where $\varphi_0 >0$, $n\in \mathbf{N}_0$, $\gamma_i \geq 0$ for all
$i \in \mathbf{N}_0$, and $\sum_{i=1}^\infty \gamma_i < \infty$.
Each $\varphi \in \mathcal{F}_{\rm Laguerre}$ can be extended to an
entire function $\varphi:\mathbf{C}\rightarrow \mathbf{C}$, which
has no zeros outside of $(-\infty, 0]$. These are Laguerre entire
functions, see \cite{Iliev,Koz1,KW}. In the next theorem the
parameter $a$ is the same as in (\ref{16e}).
\begin{theorem} \label{6tm}
Let the model we consider be translation invariant and
 the anharmonic potential  be of the form
\begin{equation} \label{m2}
V (x) = v (x^2) - h x, \quad h \in \mathbf{R},
\end{equation}
where $v(0) = 0$ and is such that for a certain $b\geq a/2$, the
derivative $v'$ obeys the condition $b + v' \in \mathcal{F}_{\rm
Laguerre}$. Then the set $\mathcal{G}^{\rm t}$ is a singleton if $h
\neq 0$.
\end{theorem}

\subsection{Comments} \label{ss3.2}
In what follows, we have developed a consistent rigorous theory of
the equilibrium thermodynamic properties of quantum models like
(\ref{app1}), based on a path measure representation of local
Gibbs states (\ref{8e}). In this theory, the model is considered
as a system of infinite-dimensional spins; its global properties
are described by the Euclidean Gibbs measures constructed with the
help of the DLR equation. As the spins are infinite-dimensional,
the methods employed are more involved and complicated than those
used for classical models. Additional complications arise from the
fact that we study a general case, where the model has no spacial
regularity and the interaction is of infinite range. In view of
the latter property, the only way to develop the theory is to
impose a priori restrictions on the support of the Gibbs measures,
which was done by means of the weights obeying the conditions
(\ref{te}) -- (\ref{26}). These conditions are competitive and, in
principle, can contradict each other if the interaction decays too
slowly. If they are satisfied, the set of tempered Gibbs measures
$\mathcal{G}^{\rm t}$ is non-void, Theorem \ref{1tm}. A
posteriori, by Theorem \ref{3tm} its elements have much smaller
support than $\mathit{\Omega}^{\rm t}$, which does not depend on
the particular choice of the weights. If the interaction has
finite range, the Gibbs measures can be defined with no support
restrictions. However, in this case the set of all such measures
may contain ``improper" elements, which have no physical meaning
and hence should be excluded from the theory. This can be done by
means of the weights obeying the same conditions, except for
 (\ref{26}) which now is satisfied automatically. Once this is done, the
 tempered Gibbs measures obtained have the support described by
 Theorem \ref{3tm}, independent of the weights.

Now let us compare our results with those known for similar
classical and quantum models.

 \vskip.1cm
\begin{itemize}
\item {\bf Theorem \ref{1tm}.} A standard tool for proving the
existence of Gibbs measures is the celebrated Dobrushin criterion,
see Theorem 1 in \cite{Do}. To apply it in our case one should
find a compact positive function $h$ defined on the single-spin
space ${C}_\beta$ such that for all $\ell$ and $\xi \in
\mathit{\Omega}$,
\begin{equation} \label{d1}
\int_{\mathit{\Omega}} h(\omega_\ell) \pi_\ell ({\rm d}\omega|\xi)
\leq A  + \sum_{\ell'} I_{\ell \ell'} h(\xi_{\ell'}),
\end{equation}
where
\[
 A >0 ; \ \quad I_{\ell \ell'} \geq 0 \quad  {\rm for} \ \ {\rm all} \ \ \ell ,
 \ell'
  , \ \quad {\rm and} \quad \  \sup_{\ell} \sum_{\ell'} I_{\ell \ell'}
< 1.
\]
 Then (\ref{d1})
would yield that for any $\xi \in \mathit{\Omega}$, such that
$\sup_{\ell} h(\xi_\ell ) < \infty$, the family
$\{\pi_\Lambda(\cdot|\xi)\}_{\Lambda \Subset \mathbf{L}}$ is
relatively compact in the weak topology on
$\mathcal{P}(\mathit{\Omega})$ (but not yet in
$\mathcal{W}_\alpha$, $\mathcal{W}^{\rm t}$). Next one would have
to show that any accumulation point of
$\{\pi_\Lambda(\cdot|\xi)\}_{\Lambda \Subset \mathbf{L}}$ is a
Gibbs measure, which is much stronger than the fact established by
our Lemma \ref{3lm}. Such a scheme was used in \cite{BH,COPP,Sin}
where the existence of Gibbs measures for lattice systems with the
single-spin space $\mathbf{R}$ was proven. In those papers the use
of the specific properties of the models, such as attractiveness
and translation invariance, was cricial. The direct extension of
this scheme to quantum models seems to be impossible. The scheme
we employ for proving Theorem \ref{1tm} is based on compactness
arguments in the topologies $\mathcal{W}_\alpha$,
$\mathcal{W}^{\rm t}$. After obvious modifications it can be
applied to models with more general inter-particle interactions.
Further comments on this item follow Corollary \ref{1co}.

\item {\bf Theorem \ref{2tm}} gives a uniform exponential moment
estimate for tempered Euclidean Gibbs measures in terms of model
parameters, which in principle can be proven before establishing the
existence. For systems of classical unbounded spins, the problem of
deriving such estimates was first posed in \cite{BH} (see the
discussion following Corollary \ref{1co}). For quantum anharmonic
systems, similar estimates were obtained in the so called analytic
approach, alternative to the traditional approach based on the DLR equation, see
\cite{AKPR01b,AKPR01c,AKRT2}. In this analytic approach
$\mathcal{G}^{\rm t}$ is defined as the set of probability measures
satisfying an integration-by-parts formula, determined by the model.
This gives additional tools for studying $\mathcal{G}^{\rm t}$ and
provides  a background for the stochastic dynamics method in which
the Gibbs measures are treated as invariant distributions for
certain infinite-dimensional stochastic evolution equations, see
\cite{AKRT3}. In both analytic and stochastic dynamics methods one
imposes a number of technical conditions on the interaction
potentials and uses advanced tools of stochastic analysis. The
method we employ for proving Theorem \ref{2tm} is much more
elementary. At the same time, Theorem \ref{2tm} gives an improvement
of the corresponding results of \cite{AKPR01c} because: (a) the
estimate (\ref{43}) gives a much stronger bound; (b) we do not
suppose that the functions $V_\ell$ are differentiable -- an
important assumption of the analytic approach.

\item {\bf Theorem \ref{3tm}.} As might be clear from the proof of
this theorem, every $\mu\in \mathcal{P}(\mathit{\Omega}^{\rm t})$
obeying the estimate (\ref{43}) possesses the support property
(\ref{48}). For Gibbs measures of classical lattice systems with
unbounded spins, a similar property  was first established in
\cite{LP}; hence, one can call $\mathit{\Xi} (b , \sigma)$ a
Lebowitz-Presutti type support. This result of \cite{LP} was
obtained by means of Ruelle's superstability estimates \cite{Ru1},
applicable to translation invariant models only. The generalization
to translation invariant quantum model was done in \cite{PY}, where
superstable Gibbs measures were specified by the following support
property
\[
\sup_{N\in \mathbf{N}} \left\{ (1 + 2N)^{-d}\sum_{\ell: |\ell|\leq
N} |\omega_\ell|^2_{L^2_\beta} \right\} \leq C(\omega) , \quad
\mu-{\rm a.s.}.
\]
Here we note that by the Birkhoff-Khinchine ergodic theorem, for any
translation invariant measure $\mu\in
\mathcal{P}(\mathit{\Omega}^{\rm t})$ obeying (\ref{43}), it follows a much stronger support property --
 for every $\sigma \in (0, 1/2)$, $\varkappa>0$, and
$\mu$-almost all $\omega$,
\[
\sup_{N\in \mathbf{N}} \left\{ (1 + 2N)^{-d}\sum_{\ell: |\ell|\leq
N} \exp\left(\lambda_\sigma |\omega_\ell|^2_{C_\beta^{\sigma}} +
\varkappa |\omega_\ell|^2_{L^2_\beta}\right)\right\} \leq C (\sigma
,\varkappa, \omega).
\]
In particular, every periodic Euclidean Gibbs measure constructed in
subsection \ref{sss5.2} below has the above property.

\item {\bf Theorem \ref{httm}} establishes a
sufficient uniqueness condition, holding in particular at
high-temperatures (small $\beta$). Here we follow the papers
\cite{AKRT,AKRT1}, where a similar uniqueness statement was proven
for translation invariant ferromagnetic scalar version of our model.
This was done by means of another renown Dobrushin result, Theorem 4
in \cite{Do}, which gives a sufficient condition for the uniqueness
of Gibbs measures. The main tool used in \cite{AKRT,AKRT1} for
estimating the elements of the Dobrushin matrix was the logarithmic
Sobolev inequality for the kernels $\pi_\ell$.

\item{\bf Theorem \ref{MAtm}.} For classical ferromagnetic spin
models, similar results were obtained in  \cite{BH,Pr} and
\cite{LML,LP}. The extreme elements $\mu_{\pm}$ play an important
role in proving Theorems \ref{phtm}, \ref{5tm}, and \ref{6tm}.

\item{\bf Theorem \ref{pressuretm}.} For classical ferromagnetic
spin models, a similar statement was proven in  \cite{BH,LP}.

\item {\bf Theorem \ref{phtm}.} For translation invariant lattice
models, phase transitions are established by showing the existence
of nonergodic (with respect to the group of lattice translations)
Gibbs measures. This mainly was being done by means of the infrared
estimates, see \cite{BK1,DLP,Helf,Kond,Pastur}. Here we use a
version of the technique developed in those papers and the
corresponding correlation inequalities which allow us to compare the
model considered with its translation invariant version (reference
model).

\item{\bf Theorem  \ref{5tm}.}
For translation invariant models with finite range interactions and
with the anharmonic potential being the polynomial (\ref{4}) with
all $b^{(s)}\geq 0$ except for $b^{(1)}$ (the so called EMN-class,
see \cite{FFS}), the uniqueness by quantum effects was proven in
\cite{AKKR02b} (see also \cite{AKKR01}). With the help of the
extreme elements $\mu_{\pm}\in \mathcal{G}^{\rm t}$ we essentially
extend the results of those papers. As in the case of Theorem
\ref{phtm}, we employ correlation inequalities to compare the model
considered with a proper reference model.

\item {\bf Theorem \ref{6tm}.} For classical lattice models, the
uniqueness at nonzero $h$  was proven in \cite{BH,LML, LP} under the
condition that the potential (\ref{m2}) possesses the property which
we establish below in Definition \ref{lydf}. The novelty of Theorem
\ref{6tm} is that it describes a quantum model and gives an explicit
sufficient condition for $V$ to possess such a
property\footnote{Examples follow Proposition \ref{lypn}.}. This
theorem is valid also in the quasi-classical limit $m \rightarrow
+\infty$, in which it covers all the cases considered in
\cite{BH,LML, LP}. For $(\phi^4)_2$ Euclidean quantum fields, a
similar statement was proven in \cite{Gelerak}.
\end{itemize}

\section{Properties of the Local Gibbs Specification}
\label{3as} Here we develop our main tools  based on the properties
of the kernels (\ref{34}).

\subsection{Moment estimates}
\label{ss4.1} Moment estimates for the kernels (\ref{34}) we are
going to derive will allow for proving the $\mathcal{W}^{\rm
t}$-relative compactness of the set
$\{\pi_\Lambda(\cdot|\xi)\}_{\Lambda \Subset \mathbf{L}}$, which by
Lemma \ref{3lm} will ensure that $\mathcal{G}^{\rm t} \neq
\emptyset$. Integrating them over $\xi\in\mathit{\Omega}^{\rm t}$ we
will get by the DLR equation (\ref{40}) the corresponding estimates
for the elements of $\mathcal{G}^{\rm t}$. Recall that $\pi_\ell$
stands for $\pi_{\{\ell\}}$.
\begin{lemma} \label{4lm}
For any $\varkappa$, $\vartheta >0$, and $\sigma \in (0, 1/2)$,
there exists $C_{\ref{53}}
>0$ such that for all $\ell \in \mathbf{L}$ and $\xi \in
\mathit{\Omega}^{\rm t}$,
\begin{equation} \label{53}
\int_{\mathit{\Omega}}\exp\left\{\lambda_\sigma |\omega_\ell
|^2_{C^\sigma_\beta} +  \varkappa |\omega_\ell |^2_{L^2_\beta}
\right\}\pi_\ell ({\rm d}\omega|\xi) \leq \exp\left\{ C_{\ref{53}} +
\vartheta \sum_{\ell'}
|J_{\ell\ell'}|\cdot|\xi_{\ell'}|_{L^2_\beta}^2\right\}.
\end{equation}
Here $\lambda_\sigma>0$ is the same as in (\ref{43}).
\end{lemma}
{\bf Proof:}
Note that by (\ref{llb1}) the left-hand side of (\ref{53}) is finite
and the second term in $\exp \{ \cdot\}$ on the right-hand side is
also finite since $\xi\in \mathit{\Omega}^{\rm t}$.

For any $\vartheta
>0$, one has (see (\ref{6}))
\begin{equation} \label{53z}
\left\vert \sum_{\ell'}J_{\ell \ell'} (\omega_\ell ,
\xi_{\ell'})_{L^2_\beta}\right\vert \leq \frac{ \hat{J}_0}{ 2
\vartheta}|\omega_\ell |^2_{L^2_\beta} + \frac{ \vartheta}{2}
\sum_{\ell'} |J_{\ell \ell'}|\cdot |\xi_{\ell'}|^2_{L^2_\beta},
\end{equation}
which holds for all $\omega, \xi \in \mathit{\Omega}^{\rm t}$.
 By these estimates and (\ref{19}), (\ref{23}), (\ref{33}),
 (\ref{34})
\begin{eqnarray} \label{54}
& & {\rm LHS}(\ref{53}) \leq [1/ Y_\ell( \vartheta)]\cdot
\exp\left\{ \vartheta \sum_{\ell'}
|J_{\ell\ell'}|\cdot |\xi_{\ell'}|_{L^2_\beta}^2\right\} \\
& & \times  \int_{\mathit{\Omega}}\exp\left\{ \lambda_\sigma
|\omega_\ell |^2_{C^\sigma_\beta} + \left( \varkappa  + \hat{J}_0 /
2 \vartheta \right)|\omega_\ell |^2_{L^2_\beta} - \int_0^\beta
V_\ell (\omega_\ell (\tau)){\rm d}\tau \right\}\chi ({\rm
d}\omega_\ell), \nonumber
\end{eqnarray}
where
\[
Y_\ell ( \vartheta) = \int_{\mathit{\Omega}}\exp\left\{ -
\frac{\hat{J}_0 } {2 \vartheta}\cdot |\omega_\ell |^2_{L^2_\beta} -
\int_0^\beta V_\ell (\omega_\ell (\tau)){\rm d}\tau \right\}\chi
({\rm d}\omega_\ell).
\]
Now we use the upper bound (\ref{3}) to estimate $\inf_{\ell} Y_\ell
(\vartheta)$, the lower bound (\ref{3}) to estimate the integrand in
(\ref{54}), take into account Proposition \ref{1pn}, and arrive at
(\ref{53}).
$\blacksquare$
\noindent By Jensen's inequality we readily get from (\ref{53}) the
following Dobrushin-like bound.
\begin{corollary} \label{1co}
For all $\ell$ and $\xi \in \mathit{\Omega}^{\rm t}$, the kernels
$\pi_\ell (\cdot|\xi)$,  obey the estimate
\begin{equation} \label{55}
\int_{\mathit{\Omega}} h(\omega_\ell) \pi_\ell ({\rm d}\omega |\xi)
\leq  C_{\ref{53}}  + (\vartheta/\varkappa) \sum_{\ell'}
|J_{\ell\ell'}|\cdot h(\xi_{\ell'}),
\end{equation}
with
\begin{equation} \label{56}
h(\omega_\ell ) = \lambda_\sigma |\omega_\ell |^2_{C^\sigma_\beta }
+ \varkappa |\omega_\ell |^2_{L^2_\beta },
\end{equation}
which is a compact function $h:C_\beta \rightarrow \mathbf{R}$.
\end{corollary}
\noindent For translation invariant lattice systems with the
single-spin space $\mathbf{R}$ and  ferromagnetic pair interactions,
integrability estimates like
\[
\log\left\{\int_{\mathbf{R}^{\mathbf{L}} }\exp (\lambda |x_\ell
|)\pi_\ell ({\rm d}x|y)\right\} < A + \sum_{\ell'} I_{\ell \ell'}
|y_{\ell'}|,
\]
were first obtained by J. Bellissard and R. H{\o}egh-Krohn, see
Proposition III.1 and Theorem III.2  in \cite{BH}. Dobrushin
 type estimates like (\ref{d1}) were also proven in \cite{COPP,Sin}.
The methods used there essentially employed the properties of the
model and hence cannot be of use in our situation. Our method of
getting such estimates is much simpler; at the same time, it is
applicable in both cases -- classical and quantum. Its peculiarities
are: (a) first we prove the exponential integrability (\ref{53}) and
then derive the Dobrushin bound (\ref{55}) rather than prove it
directly; (b) the function (\ref{56}) consists of two additive
terms, the first of which is to guarantee the compactness while the
second one controls the inter-particle interaction.

Now by means of (\ref{53}) we obtain the corresponding estimates
for the kernels $\pi_\Lambda$ with arbitrary $\Lambda \Subset
\mathbf{L}$. Let the parameters $\sigma$, $\varkappa$, and
$\lambda_\sigma$ be the same as in (\ref{53}). For $\ell \in
\Lambda \Subset \mathbf{L}$, we define
\begin{equation} \label{62}
n_\ell (\Lambda |\xi) = \log\left\{\int_{\mathit{\Omega}}\exp\left(
\lambda_\sigma |\omega_\ell |^2_{C^\sigma_\beta} + \varkappa
|\omega_\ell |^2_{L^2_\beta}\right) \pi_\Lambda ({\rm d}\omega|\xi)
\right\},
\end{equation}
which is finite by (\ref{llb1}).
\begin{lemma} \label{melm}
For every $\alpha \in \mathcal{I}$, there exists $C_{\ref{66a}}
(\alpha)
>0$ such that for all $\xi \in \mathit{\Omega }^{
\mathrm{t}}$,
\begin{equation} \label{66a}
\limsup_{\Lambda \nearrow \mathbf{L}} \sum_{\ell \in \Lambda
}n_{\ell }(\Lambda |\xi )w_{\alpha }(\ell _{0},\ell ) \leq
C_{\ref{66a}}(\alpha);
\end{equation}%
hence,
\begin{equation} \label{66c}
\limsup_{\Lambda \nearrow \mathbf{L}} n_{\ell_0} (\Lambda |\xi) \leq
C_{\ref{66a}}(\alpha), \quad {for} \ \ {any} \ \ \ell_0.
\end{equation}
Thereby,  there exists $C_{\ref{590}}(\ell, \xi)>0$ such that for
all $\Lambda \Subset \mathbf{L}$ containing $\ell$,
\begin{equation} \label{590}
n_{\ell} (\Lambda | \xi) \leq C_{\ref{590}}(\ell, \xi).
\end{equation}
\end{lemma}
{\bf Proof:}
Given $\varkappa >0$ and $\alpha \in \mathcal{I}$, we  fix
$\vartheta
>0$ such that
\begin{equation}
\vartheta \sum_{\ell ^{\prime }}|J_{\ell \ell ^{\prime }}|\leq \vartheta \hat{J}%
_{0}\leq \vartheta \hat{J}_{\alpha }<\varkappa .  \label{63b}
\end{equation}%
Then integrating both sides of the bound (\ref{53}) with respect to
the measure $\pi _{\Lambda }(\mathrm{d}\omega |\xi )$  we get
\begin{eqnarray}
\qquad \quad n_{\ell }(\Lambda |\xi ) &\leq &C_{\ref{53}} +\vartheta
\sum_{\ell ^{\prime }\in \Lambda ^{c}}|J_{\ell \ell ^{\prime
}}|\cdot |\xi _{\ell ^{\prime
}}|_{L_{\beta }^{2}}^{2}  \label{65} \\
&+&\log \left\{ \int_{\mathit{\Omega }}\exp \left( \vartheta
\sum_{\ell ^{\prime }\in \Lambda }|J_{\ell \ell ^{\prime }}|\cdot
 |\omega _{\ell ^{\prime }}|_{L_{\beta }^{2}}^{2}\right) \pi
_{\Lambda }(\mathrm{d}\omega
|\xi )\right\}  \notag \\
&\leq &C_{\ref{53}} +\vartheta \sum_{\ell ^{\prime }\in \Lambda
^{c}}|J_{\ell \ell ^{\prime }}|\cdot |\xi _{\ell ^{\prime
}}|_{L_{\beta }^{2}}^{2}  + \vartheta /\varkappa \sum_{\ell ^{\prime
}\in \Lambda }|J_{\ell \ell ^{\prime }}|\cdot n_{\ell ^{\prime
}}(\Lambda |\xi ).  \notag
\end{eqnarray}%
Here we have used (\ref{63b}) and the multiple H\"{o}lder inequality
\[
\int \left( \prod\nolimits_{i=1}^{n}\varphi _{i}^{\alpha
_{i}}\right)
\mathrm{d}\mu \leq \prod\nolimits_{i=1}^{n}\left( \int \varphi _{i}\mathrm{d}%
\mu \right) ^{\alpha _{i}},
\]
in which $\mu $ is a probability measure, $\varphi _{i}\geq 0$
(respectively,  $\alpha _{i}\geq 0$), $i=1 , \dots, n$, are
functions (respectively,  numbers such that $\sum_{i=1}^{n}\alpha
_{i}\leq 1$). Then (\ref{65}) yields
\begin{align}
& n_{\ell _{0}}(\Lambda |\xi )\leq \sum_{\ell \in \Lambda }n_{\ell
}(\Lambda
|\xi )w_{\alpha }(\ell _{0},\ell )  \label{66} \\
& \leq \frac{1}{1-\vartheta \hat{J}_{\alpha }/\varkappa }\left[
C_{\ref{53}} \sum_{\ell ^{\prime }\in \Lambda }w_{\alpha }(\ell
_{0},\ell ^{\prime })+\vartheta \hat{J}_{\alpha }\sum_{\ell ^{\prime
}\in \Lambda ^{c}}|\xi _{\ell
^{\prime }}|_{L_{\beta }^{2}}^{2}w_{\alpha }(\ell _{0},\ell ^{\prime })%
\right] .  \notag
\end{align}%
Therefrom, for all $\xi \in \mathit{\Omega }^{\mathrm{t}}$, we get
\begin{eqnarray}
& &  \limsup_{\Lambda \nearrow \mathbf{L}}n_{\ell _{0}}(\Lambda |\xi
)\leq \limsup_{\Lambda \nearrow \mathbf{L}} \sum_{\ell \in \Lambda
}n_{\ell
}(\Lambda |\xi )w_{\alpha }(\ell _{0},\ell )  \label{67d} \\
& &  \quad \qquad \leq \frac{C_{\ref{53}} }{1-\vartheta
\hat{J}_{\alpha }/\varkappa }\sum_{\ell }w_{\alpha }(\ell _{0},\ell
)\ \overset{\mathrm{def}}{=}\text{ }C_{\ref{66a}} (\alpha), \notag
\end{eqnarray}
which gives (\ref{66a}) and (\ref{66c}). The proof of (\ref{590}) is
straightforward.
$\blacksquare$
\vskip.1cm \noindent Recall that  the norm $\|\cdot\|_\alpha$ was
defined by (\ref{29}). Given $\alpha \in \mathcal{I}$ and $\sigma
\in (0, 1/2)$, we set, c.f. Remark \ref{augustrk},
\begin{equation} \label{59a}
\|\xi\|_{\alpha , \sigma} = \left[\sum_{\ell} |\xi_\ell
|^2_{C^\sigma_\beta}w_\alpha (\ell_0, \ell)\right]^{1/2}.
\end{equation}
\begin{lemma} \label{klm}
Let the assumptions of Lemma \ref{4lm} be satisfied. Then for every
 $\alpha \in \mathcal{I}$ and $\xi\in
 \mathit{\Omega}^{\rm t}$, one finds a positive
$C_{\ref{61q}}(\xi)$ such that for all  $\Lambda \Subset
\mathbf{L}$,
\begin{equation} \label{61q}
\int_{\mathit{\Omega}} \|\omega\|_{\alpha }^2 \pi_\Lambda ({\rm
d}\omega|\xi) \leq C_{\ref{61q}}(\xi).
\end{equation}
Furthermore, for every
 $\alpha \in \mathcal{I}$,
$\sigma \in (0, 1/2)$, and $\xi\in
 \mathit{\Omega}^{\rm t}$ for which the norm (\ref{59a}) is finite,
 one finds a $C_{\ref{61a}}(\xi)>0$ such that for
 all $\Lambda \Subset \mathbf{L}$,
 \begin{equation} \label{61a}
\int_{\mathit{\Omega}} \|\omega\|_{\alpha , \sigma}^2 \pi_\Lambda
({\rm d}\omega|\xi) \leq C_{\ref{61a}}(\xi).
\end{equation}
\end{lemma}
{\bf Proof:}
For any fixed $\xi \in\mathit{\Omega}^{\rm t}$, by the Jensen
inequality and (\ref{66}) one has
\begin{eqnarray} \label{59s}
& & \lim\sup_{\Lambda \nearrow \mathbf{L}}\int_{\mathit{\Omega}}
\|\omega\|^2_{\alpha } \pi_\Lambda ({\rm d}\omega|\xi) \\ & & \quad
\leq \lim\sup_{\Lambda \nearrow
\mathbf{L}}\left[\frac{1}{\varkappa}\sum_{\ell \in \Lambda } n_\ell
(\Lambda|\xi) w_\alpha (0, \ell) + \sum_{\ell \in
\Lambda^c}|\xi_\ell |_{L_\beta^2}^2 w_\alpha (0, \ell) \right] \nonumber \\
& & \quad \leq C_{\ref{66a}}(\alpha )/ \varkappa . \nonumber
\end{eqnarray}
Hence, the set consisting of the left-hand sides of (\ref{61q})
indexed by $\Lambda \Subset \mathbf{L}$ is bounded. The proof of
(\ref{61a}) is analogous.
$\blacksquare$

\subsection{Weak convergence of tempered measures}
\label{wcss}

Recall that $f:\mathit{\Omega}\rightarrow \mathbf{R}$ is a local
function if it is measurable with respect to
$\mathcal{B}(\mathit{\Omega}_\Lambda)$ for a certain $\Lambda\Subset
\mathbf{L}$.
\begin{lemma} \label{tanlm}
Let a sequence $\{\mu_n\}_{n \in \mathbf{N}}\subset
\mathcal{P}(\mathit{\Omega}^{\rm t})$ have the following properties:
(a) for every $\alpha \in \mathcal{I}$, each its element obeys the
estimate
\begin{equation} \label{ta1}
\int_{\mathit{\Omega}^{\rm t}} \|\omega\|_\alpha^2  \mu_n ({\rm
d}\omega) \leq C_{\ref{ta1}} (\alpha),
\end{equation}
with one and the same $C_{\ref{ta1}}(\alpha)$; (b) for every local
$f\in C_{\rm b}(\mathit{\Omega}^{\rm t})$, $\{\mu_n (f)\}_{n \in
\mathbf{N}}\subset \mathbf{R}$ is a Cauchy sequence. Then
$\{\mu_n\}_{n \in \mathbf{N}}$ converges in $\mathcal{W}^{\rm t}$ to
a certain $\mu \in \mathcal{P}(\mathit{\Omega}^{\rm t})$.
\end{lemma}
{\bf Proof:}
The topology of the Polish space $\mathit{\Omega}^{\rm t}$ is
consistent with the following metric (c.f. (\ref{met}))
\begin{eqnarray} \label{tan2}
\rho (\omega , \tilde{\omega}) = \sum_{k=1}^\infty
2^{-k}\frac{\|\omega - \tilde{\omega}\|_{\alpha_k} }{1 + \|\omega
- \tilde{\omega}\|_{\alpha_k} } + \sum_{\ell} 2^{-|\ell_0 - \ell|}
\frac{|\omega_\ell - \tilde{\omega}_\ell|_{C_\beta}}{1 +
|\omega_\ell - \tilde{\omega}_\ell|_{C_\beta}},
\end{eqnarray}
where $\{\alpha_k\}_{k \in \mathbf{N}}\subset
\mathcal{I}=(\underline{\alpha}, \overline{\alpha})$ is a monotone
strictly decreasing sequence converging to $\underline{\alpha}$.
Let us denote by $C^{\rm u}_{\rm b}(\mathit{\Omega}^{\rm t};
\rho)$ the set of all bounded functions $f: \mathit{\Omega}^{\rm
t} \rightarrow \mathbf{R}$, which are uniformly continuous with
respect to (\ref{tan2}). Thus, in accord with a known fact, see
e.g. Theorem 2.1.1, page 19 of \cite{Borkar}, to prove the lemma
it suffices to show that under its conditions $\{\mu_n (f)\}_{n
\in \mathbf{N}}$ is a Cauchy sequence for every $f \in C^{\rm
u}_{\rm b}(\mathit{\Omega}^{\rm t}; \rho)$. Given $\delta
>0$, we choose $\Lambda_\delta \Subset \mathbf{L}$ and $k_\delta
\in \mathbf{N}$ such that
\begin{equation} \label{tan3}
\sum_{\ell \in \Lambda_\delta^c} 2^{-|\ell_0 -\ell|} < \delta/3,
\qquad \ \sum_{k =k_\delta}^\infty 2^{-k} = 2^{- k_\delta + 1} <
\delta/3.
\end{equation}
For this $\delta$ and a certain $R>0$, we choose $\Lambda_\delta
(R)\Subset \mathbf{L}$ such that
\begin{equation} \label{tan4}
\sup_{\ell \in \mathbf{L} \setminus \Lambda_\delta (R)}\left\{
w_{\alpha_{k_\delta
-1}}(\ell_0,\ell)/w_{\alpha_{k_\delta}}(\ell_0,\ell)\right\} <
\frac{\delta}{3 R^2},
\end{equation}
which is possible in view of (\ref{te1}). Finally, for $R>0$, we set
\begin{equation} \label{tan40}
B_R = \{ \omega \in \mathit{\Omega}^{\rm t} \ | \
\|\omega\|_{\alpha_{k_\delta}} \leq R\}.
\end{equation}
By (\ref{ta1}) and the Chebyshev inequality, one has that for all $n
\in \mathbf{N}$,
\begin{equation} \label{tan5}
\mu_n \left(\mathit{\Omega}^{\rm t} \setminus B_R \right) \leq
C_{\ref{ta1}}(\alpha_{k_\delta})/ R^2 .
\end{equation}
Now for $f \in C^{\rm u}_{\rm b}(\mathit{\Omega}^{\rm t}; \rho)$,
$\Lambda \Subset \mathbf{L}$, and $n , m\in \mathbf{N}$, we have
\begin{eqnarray} \label{tan6}
\left\vert \mu_n (f) - \mu_m (f)  \right\vert & \leq & \left\vert
\mu_n (f_\Lambda ) - \mu_m (f_\Lambda)  \right\vert \\ & + & 2
\max\{\mu_n ( |f - f_\Lambda|); \mu_m ( |f - f_\Lambda|) \},
\nonumber
\end{eqnarray}
where we set $f_\Lambda (\omega ) = f(\omega_\Lambda \times
0_{\Lambda^c})$. By (\ref{tan5}),
\begin{eqnarray} \label{tan7}
\mu_n ( |f - f_\Lambda|) & \leq & 2 C_{\ref{ta1}}(\alpha_{k_\delta})
\|f\|_\infty
/ R^2\\
& + & \int_{B_R} \left\vert f(\omega) - f(\omega_\Lambda \times
0_{\Lambda^c} )\right\vert \mu_n ({\rm d} \omega). \nonumber
\end{eqnarray}
For chosen $f \in C^{\rm u}_{\rm b}(\mathit{\Omega}^{\rm t}; \rho)$
and $\varepsilon >0$, one finds $\delta >0$ such that for all
$\omega, \tilde{\omega} \in \mathit{\Omega}^{\rm t}$,
\[
\left\vert f (\omega) - f(\tilde{\omega})\right\vert <
\varepsilon/6, \quad {\rm whenever} \ \  \rho(\omega ,
\tilde{\omega}) < \delta.
\]
For these $f$, $\varepsilon$, and $\delta$, one picks up
$R(\varepsilon, \delta)>0$ such that
\begin{equation} \label{tan8}
C_{\ref{ta1}}(\alpha_{k_\delta})\|f\|_{\infty}/ \left[R(\varepsilon,
\delta)\right]^2 < {\varepsilon}/{12}.
\end{equation}
Now one takes $\Lambda \Subset \mathbf{L}$, which contains both
$\Lambda_\delta$ and $\Lambda_\delta [R(\varepsilon, \delta)]$
defined by (\ref{tan3}), (\ref{tan4}). For this $\Lambda$, $\omega
\in B_{R(\varepsilon, \delta)}$, and $k = 1 , 2 , \dots , k_\delta
-1$, one has
\begin{eqnarray} \label{tan9}
\qquad \|\omega - \omega_\Lambda \times
0_{\Lambda^c}\|_{\alpha_k}^2 & = & \sum_{\ell \in \Lambda^c}
|\omega_\ell|_{L^2_\beta}^2
w_{\alpha_{k_\delta}}(\ell_0,\ell)\left[w_{\alpha_{k}}(\ell_0,\ell)/w_{\alpha_{k_\delta}}(\ell_0,\ell)
\right]\\ & \leq & \frac{\delta}{3\left[R(\varepsilon,
\delta)\right]^2} \sum_{\ell \in
\Lambda^c}|\omega_\ell|_{L^2_\beta}^2
w_{\alpha_{k_\delta}}(\ell_0,\ell) < \frac{\delta}{3}, \nonumber
\end{eqnarray}
where (\ref{tan4}), (\ref{tan40}) have been used. Then by
(\ref{tan2}), (\ref{tan3}), it follows that
\begin{equation} \label{tan10}
\forall \omega\in B_{R(\varepsilon, \delta)}: \quad \rho (\omega ,
\omega_\Lambda \times 0_{\Lambda^c}) < \delta,
\end{equation}
which together with (\ref{tan8}) yields in (\ref{tan7})
\[
\mu_n ( |f - f_\Lambda|) < \frac{\varepsilon}{6} +
\frac{\varepsilon}{6}\mu_n \left(B_{R(\varepsilon, \delta)}
\right)\leq \frac{\varepsilon}{3}.
\]
By assumption (b) of the lemma, one finds $N_\varepsilon$ such that
for all $n , m >N_\varepsilon$,
\[
\left\vert \mu_n (f_\Lambda ) - \mu_m (f_\Lambda ) \right\vert <
\frac{\varepsilon}{3}.
\]
Applying the latter two estimates in (\ref{tan6}) we get that
$\{\mu_n\}_{n\in \mathbf{N}}$ is a Cauchy sequence in the topology
$\mathcal{W}^{\rm t}$ in which $\mathcal{P}(\mathit{\Omega}^{\rm
t})$ is complete.
$\blacksquare$

\section{Proof of Theorems \ref{1tm} -- \ref{httm}}
\label{4s}

The existence of Euclidean Gibbs measures and the estimate
(\ref{43}) can be proven independently. To establish the compactness
of $\mathcal{G}^{\rm t}$ we will need (\ref{43}), thus, we first
prove  Theorem \ref{2tm}. \noindent

 \textbf{Proof of Theorem
\ref{2tm}:} \ Let us show that every $\mu \in \mathcal{P}(\Omega)$
which solves the DLR equation (\ref{40}) ought to obey (\ref{43})
with one and the same $C_{\ref{43}}$. To this end we apply the
bounds for the kernels $\pi_\Lambda (\cdot|\xi)$ obtained above.
Consider the functions
\[
G_N (\omega_\ell) \ \stackrel{\rm def}{=} \ \exp\left(\min\left\{
\lambda_\sigma |\omega_\ell|_{C_\beta^\sigma}^2 + \varkappa
|\omega_\ell|_{L_\beta^2}^2; N\right\} \right), \quad N\in
\mathbf{N}.
\]
By (\ref{40}), Fatou's lemma, and the estimate (\ref{66c}) with an
arbitrarily chosen
 $\alpha \in \mathcal{I}$, we get
\begin{eqnarray*}
& & \int_{\mathit{\Omega}} G_N (\omega_\ell)\mu({\rm d}\omega)
=\limsup_{\Lambda \nearrow \mathbf{L}}\int_{\mathit{\Omega}} \left[
\int_{\mathit{\Omega}} G_N (\omega_\ell)\pi_\Lambda ({\rm
d}\omega|\xi)\right]\mu({\rm d}\xi) \\ & & \quad \leq
\limsup_{\Lambda \nearrow \mathbf{L}}\int_{\mathit{\Omega}}
\left[\int_{\mathit{\Omega}}\exp\left( \lambda_\sigma
|\omega_\ell|_{C_\beta^\sigma}^2 + \varkappa
|\omega_\ell|_{L_\beta^2}^2 \right)\pi_\Lambda ({\rm
d}\omega|\xi)\right]\mu({\rm d}\xi) \\ & & \quad \leq
\int_{\mathit{\Omega}}\left[ \limsup_{\Lambda \nearrow
\mathbf{L}}\int_{\mathit{\Omega}}\exp\left( \lambda_\sigma
|\omega_\ell|_{C_\beta^\sigma}^2 + \varkappa
|\omega_\ell|_{L_\beta^2}^2 \right)\pi_\Lambda ({\rm
d}\omega|\xi)\right]\mu({\rm d}\xi)\\ & & \quad \leq \exp
C_{\ref{66a}}(\alpha)  \ \stackrel{\rm def}{=} \ C_{\ref{43}} .
\end{eqnarray*}
In view of the support property (\ref{40b}) of any measure solving
the equation (\ref{40}) we can pass here to the limit $N \rightarrow
+\infty$ and get (\ref{43}).
 $\square$
\begin{corollary} \label{tanco}
For every $\alpha \in \mathcal{I}$, the topologies induced on
$\mathcal{G}^{\rm t}$ by $\mathcal{W}_\alpha$ and $\mathcal{W}^{\rm
t}$ coincide.
\end{corollary}
{\bf Proof:}
Follows immediately from Lemma \ref{tanlm} and the estimate
(\ref{43}).
$\blacksquare$
\vskip.1cm \noindent
 \textbf{Proof of Theorem
\ref{1tm}:} \ Let us introduce the next scale of Banach spaces
(c.f. (\ref{29}))
\begin{equation} \label{57}
\mathit{\Omega}_{\alpha, \sigma} = \left\{ \omega \in
\mathit{\Omega} \ \left\vert \ \|\omega\|_{\alpha , \sigma}  <
\infty \right. \right\}, \quad \sigma \in (0, 1/2), \ \ \alpha \in
\mathcal{I},
\end{equation}
where the norm $\|\cdot\|_{\alpha , \sigma}$ was defined by
(\ref{59a}).  For any pair $\alpha , \alpha'\in \mathcal{I}$ such
that $\alpha < \alpha'$, the embedding $\mathit{\Omega}_{\alpha ,
\sigma} \hookrightarrow \mathit{\Omega}_{\alpha'}$ is compact, see
Remark \ref{1rm}. This fact and the estimate (\ref{61a}), which
holds for any $\xi \in \mathit{\Omega}_{\alpha, \sigma}$, imply by
Prokhorov's criterion the relative compactness of the set
$\{\pi_{\Lambda}(\cdot|\xi)\}_{\Lambda \Subset \mathbf{L}}$ in
$\mathcal{W}_{\alpha'}$. Therefore, the sequence
$\{\pi_{\Lambda}(\cdot|0)\}_{\Lambda \Subset \mathbf{L}}$ is
relatively compact in every $\mathcal{W}_\alpha$, $\alpha \in
\mathcal{I}$. Then Lemma \ref{3lm} yields that $\mathcal{G}^{\rm t}
\neq \emptyset$. By the same Prokhorov criterion and the estimate
(\ref{43}), we get the $\mathcal{W}_\alpha$-relative compactness of
$\mathcal{G}^{\rm t}$. Then in view of the Feller property (Lemma
\ref{2lm}), the set $\mathcal{G}^{\rm t}$ is closed and hence
compact in every $\mathcal{W}_\alpha$, $\alpha\in \mathcal{I}$,
which by Corollary \ref{tanco} completes the proof. $\square$
\vskip.1cm
 \noindent
 \noindent
 \textbf{Proof of Theorem \ref{3tm}:} \
To some extent we shall  follow the line of arguments used in the
proof of Lemma 3.1 in \cite{LP}. Given $\ell, \ell_0$, $b>0$,
$\sigma \in (0, 1/2)$, and $\Lambda \subset \mathbf{L}$, we
introduce
\begin{eqnarray} \label{44}
\qquad \mathit{\Xi}_\ell (\ell_0 , b , \sigma) & = & \{ \xi \in
\mathit{\Omega} \ | \ |\xi_\ell |_{C^\sigma_\beta}^2 \leq b \log ( 1
+ |\ell - \ell_0|) \}, \\ \qquad \mathit{\Xi}_\Lambda (\ell_0 , b ,
\sigma) & = & \bigcap_{\ell \in \Lambda} \Xi_\ell (\ell_0 , b ,
\sigma) . \nonumber
\end{eqnarray}
For a cofinal sequence $\mathcal{L}$, we set
\begin{equation} \label{45}
\mathit{\Xi} (\ell_0 , b , \sigma)  = \bigcup_{\Lambda \in
\mathcal{L}}\mathit{\Xi}_{\Lambda^c} (\ell_0 , b , \sigma), \quad
\mathit{\Xi}( b , \sigma)  = \bigcap_{\ell_0 \in
\mathbf{L}}\mathit{\Xi} (\ell_0 , b , \sigma).
\end{equation}
The latter $\mathit{\Xi} (b, \sigma)$ is a subset of
$\mathit{\Omega}^{\rm t}$ and is the same as the one given by
(\ref{47}). To prove the theorem let us show that for any $\sigma
\in (0, 1/2)$, there exists $b>0$ such that for all $\ell_0$ and
$\mu \in \mathcal{G}^{\rm t}$,
\begin{equation} \label{69}
\mu \left( \mathit{\Omega} \setminus \mathit{\Xi}(\ell_0 , b ,
\sigma)\right) = 0.
\end{equation}
By (\ref{44}) we have
\begin{eqnarray} \label{mar}
\qquad \quad \mathit{\Omega} \setminus \mathit{\Xi}_{\Lambda^c}
(\ell_0 , b, \sigma)  & = & \{ \xi \in \mathit{\Omega} \ | \
(\exists \ell \in \Lambda^c):\ \ |\xi_\ell |_{C^\sigma_\beta}^2 > b
\log (1 + |\ell - \ell_0|)\} \\ &  \subset &  \{ \xi \in
\mathit{\Omega} \ | \ (\exists \ell \in \Delta^c):\ \ |\xi_\ell
|_{C^\sigma_\beta}^2 > b \log (1 + |\ell - \ell_0|)\}, \nonumber
\end{eqnarray}
for any $\Delta \subset \Lambda$. Therefore,
\begin{equation} \label{70}
\mu \left(\bigcap_{\Lambda \in \mathcal{L}} \left[  \mathit{\Omega}
\setminus \mathit{\Xi}_{\Lambda^c} (\ell_0 , b, \sigma)\right]
\right) = \lim_{\mathcal{L}}\mu \left(  \mathit{\Omega} \setminus
\mathit{\Xi}_{\Lambda^c} (\ell_0 , b, \sigma)\right),
\end{equation}
which holds for any cofinal sequence $\mathcal{L}$. By (\ref{mar}),
\begin{eqnarray*}
\mu \left(  \mathit{\Omega} \setminus \mathit{\Xi}_{\Lambda^c}
(\ell_0 , b, \sigma) \right) & = & \mu \left(\bigcup_{\ell \in
\Lambda^c} \left[\mathit{\Omega} \setminus \mathit{\Xi}_\ell
(\ell_0 , b , \sigma) \right] \right)\\ & \leq &  \sum_{\ell \in
\Lambda^c} \mu\left( \left\{ \xi \ | \ \exp\left( \lambda_\sigma
|\xi_\ell|^2_{C^\sigma_\beta} \right) > (1 + |\ell - \ell_0|)^{b
\lambda_\sigma} \right\} \right).
\end{eqnarray*}
Applying here the Chebyshev inequality and the estimate (\ref{43})
we get
\[
\mu \left( \mathit{\Omega} \setminus \mathit{\Xi}_{\Lambda^c}
(\ell_0 , b, \sigma) \right) \leq C_{\ref{43}} \sum_{\ell \in
\Lambda^c} ( 1 + |\ell - \ell_0|)^{ - b \lambda_\sigma}.
\]
In view of (\ref{april}) the latter series converges for any $b >d
/\lambda_\sigma$. In this case by (\ref{70})
\begin{eqnarray*}
\mu \left(  \mathit{\Omega} \setminus \mathit{\Xi} (\ell_0 , b,
\sigma) \right)  =  \lim_{\mathcal{L}}\mu \left(\left[
\mathit{\Omega} \setminus \mathit{\Xi}_{\Lambda^c} (\ell_0 , b,
\sigma)\right] \right) = 0,
\end{eqnarray*}
which yields (\ref{69}). $\square$ \vskip.1cm \noindent Let
$\mathcal{E}$ be the set of all
 continuous local functions $f:\mathit{\Omega}^{\rm t} \rightarrow \mathbf{R}$,
 for which there exist $\sigma\in (0,1/2)$,
$\Delta_f \Subset \mathbf{L}$, and $D_f >0$, such that
\begin{equation} \label{loc}
|f(\omega) |^2 \leq D_f \sum_{\ell \in \Delta_f} \exp\left(
\lambda_\sigma |\omega_\ell|_{C^\sigma_\beta}^2 \right), \quad {\rm
for \ all} \ \  \omega\in \mathit{\Omega}^{\rm t},
\end{equation}
where $\lambda_\sigma$ is the same as in (\ref{16h}) and (\ref{43}).
Let also ${\rm ex}(\mathcal{G}^{\rm t})$ stand for the set of all
extreme elements of $\mathcal{G}^{\rm t}$.
\begin{lemma} \label{locco}
For every $\mu\in{\rm ex}(\mathcal{G}^{\rm t})$ and any cofinal
sequence $\mathcal{L}$, it follows that:
 (a) the sequence $\{\pi_\Lambda
(\cdot|\xi)\}_{\Lambda \in \mathcal{L}}$ converges in
$\mathcal{W}^{\rm t}$ to this $\mu$ for $\mu$-almost all $\xi \in
\mathit{\Omega}^{\rm t}$; (b) for every $f \in \mathcal{E}$, one has
$\lim_{\mathcal{L}}\pi_\Lambda (f|\xi)= \mu(f)$ for $\mu$-almost all
$\xi \in \mathit{\Omega}^{\rm t}$.
\end{lemma}
{\bf Proof:}
Claim (c) of Theorem 7.12, page 122 in \cite{Ge}, implies that for
any local $f\in C_{\rm b}(\mathit{\Omega}^{\rm t})$,
\begin{equation} \label{a1}
\lim_{\mathcal{L}} \pi_\Lambda (f|\xi) = \mu(f), \quad {\rm for} \
\mu{\rm -almost} \ {\rm all} \ \xi\in\mathit{\Omega}^{\rm t}.
\end{equation}
Then the convergence stated in claim (a) follows from Lemmas
\ref{klm} and \ref{tanlm}. Given $f \in \mathcal{E}$ and $N \in
\mathbf{N}$, we set $\mathit{\Omega}_N = \{\omega \in
\mathit{\Omega} \ | \ |f(\omega)|> N\}$ and
\[
f_N (\omega) = \left\{ \begin{array}{ll} f(\omega)  \ \ &{\rm if} \
\ |f(\omega)| \leq N; \\ N f(\omega)/|f(\omega)| \ \ &{\rm
otherwise}.
\end{array} \right.
\]
Each $f_N$ belongs to $C_{\rm b}(\mathit{\Omega}^{\rm t})$ and $f_N
\rightarrow f$ point-wise as $N \rightarrow +\infty$. Then by
(\ref{a1}) there exists  a Borel set $\mathit{\Xi}_\mu \subset
\mathit{\Omega}^{\rm t}$, such that $\mu(\mathit{\Xi}_\mu) = 1$ and
for every $N \in \mathbf{N}$,
\begin{equation} \label{1aa}
\lim_{\mathcal{L}} \pi_\Lambda (f_N |\xi) = \mu(f_N), \quad {\rm for
} \ {\rm all} \ \xi \in \mathit{\Xi}_\mu.
\end{equation}
Note that by (\ref{62}), (\ref{590}), and (\ref{loc}), for any $\xi
\in \mathit{\Xi}_\mu$ one finds a positive $C_{\ref{591}}(f, \xi)$
such that for all $\Lambda \Subset \mathbf{L}$, which contain
$\Delta_{ f}$, it follows that
\begin{equation} \label{591}
\int_{\mathit{\Omega}} |f(\omega)|^2 \pi_\Lambda ({\rm d}\omega|\xi)
\leq {C}_{\ref{591}} (f , \xi).
\end{equation}
Hence
 \begin{eqnarray*}
& & |\pi_\Lambda (f |\xi) - \pi_\Lambda (f_N|\xi)| \leq 2
\int_{\mathit{\Omega}_N}  |f(\omega)|\pi_\Lambda ({\rm
d}\omega|\xi)\\
& & \qquad \leq \frac{2}{N} \cdot\int_{\mathit{\Omega}}|f(\omega)|^2
\pi_\Lambda ({\rm d}\omega|\xi) \leq \frac{2}{N} \cdot
{C}_{\ref{591}} (f , \xi).
\end{eqnarray*}
 Similarly, by means of (\ref{loc}) and Theorem
\ref{2tm}, one gets
\[
|\mu (f) - \mu (f_N)| \leq \frac{2}{N}\cdot  D_f {C}_{\ref{43}}.
\]
The latter two inequalities  and (\ref{1aa}) allow us to estimate
$|\pi_\Lambda (f|\xi) - \mu(f)|$ and thereby to complete the proof.
$\blacksquare$
\vskip.1cm \noindent \textbf{Proof of Theorem \ref{httm}:} For the
scalar translation invariant version of the model considered here,
the high-temperature uniqueness was proven in \cite{AKRT,AKRT1} by
means of Dobrushin's criterium. The proof given below is a
modification of the arguments used there.

The main idea of the method of Dobrushin is to control the
Wasserstein distance  $R[\pi_\ell (\cdot|\xi); \pi_\ell
(\cdot|\xi')]$ between the measures $\pi_\ell (\cdot|\xi)$ and
$\pi_\ell (\cdot|\xi')$ with  $\xi\neq  \xi'$. In our context, its
appropriate choice may be made as follows. For given $\ell$ and
$\xi, \xi' \in \mathit{\Omega}^{\rm t}$,  we set
\begin{equation} \label{WD}
   R[\pi_\ell (\cdot|\xi); \pi_\ell (\cdot|\xi')]
 =  \sup_{f \in {\rm Lip}_1 (L^2_\beta)} \left\vert
\int_{\mathit{\Omega}} f(\omega_\ell)\pi_\ell({\rm d}\omega|\xi) -
\int_{\mathit{\Omega}} f(\omega_\ell)\pi_\ell({\rm
d}\omega|\xi')\right\vert,
\end{equation}
where ${\rm Lip}_1 (L^2_\beta)$ stands for the set of
Lipschitz-continuous functions $f:L^2_\beta\rightarrow \mathbf{R}$
 with the Lipschitz constant equal one. The
 Dobrushin criterium (see Theorem 4 in \cite{Do})
employs the matrix
\begin{equation} \label{ta21}
C_{\ell\ell'} = \sup\left\{ \frac{R[\pi_\ell(\cdot|\xi);
\pi_\ell(\cdot|\xi')]}{|\xi_\ell -
\xi_{\ell'}|_{L^2_\beta}}\right\}, \quad \ell \neq \ell', \ \ \ell,
\ell'\in \mathbf{L},
\end{equation}
where the supremum is taken over all $\xi, \xi' \in
\mathit{\Omega}^{\rm t}$ which differ only at $\ell'$. According to
this criterium the uniqueness stated will follow from the fact
\begin{equation} \label{ta22}
\sup_{\ell} \sum_{\ell' \in \mathbf{L}\setminus\{\ell\}} C_{\ell
\ell'} < 1.
\end{equation}
In view of (\ref{llb1}) the map
\begin{equation} \label{ta23}
L^2_\beta \ni \xi_{\ell'} \mapsto \mathit{\Upsilon}(\xi_{\ell'}) \
\stackrel{\rm def}{=} \ \int_{\mathit{\Omega}} f(\omega_\ell)
\pi_\ell ({\rm d}\omega |\xi)
\end{equation}
has the following derivative in direction $\zeta \in L^2_\beta$
\[
\left(\nabla \mathit{\Upsilon} (\xi_{\ell'}) ,
\zeta\right)_{L^2_\beta} =  - J_{\ell \ell'} \left[\pi_\ell \left(f
\cdot (\omega_\ell , \zeta)_{L^2_\beta} \left\vert \xi
\right.\right) - \pi_\ell \left(f|\xi\right) \cdot \pi_\ell \left(
(\omega_\ell , \zeta)_{L^2_\beta}\left\vert \xi \right.
\right)\right].
\]
By Theorem 5.1 of \cite{AKRT} the measures $\pi_\ell (\cdot|\xi)$
obey the logarithmic Sobolev inequality with the constant
\begin{equation}\label{ta25}
C_{\rm LS} = e^{\beta \delta}/ (a+b),
\end{equation}
which is independent of $\xi$. By standard arguments this yields the
estimate
\begin{equation}\label{ta26}
\left\vert \left(\nabla \mathit{\Upsilon} (\xi_{\ell'}) ,
\zeta\right)_{L^2_\beta}\right\vert \leq C_{\rm LS} |J_{\ell
\ell'}|\cdot |\zeta|^2_{L^2_\beta}.
\end{equation}
Then with the help of the mean value theorem from (\ref{ta21}) and
(\ref{ta25}) we get
\[
C_{\ell \ell'} \leq |J_{\ell \ell'}| \cdot  e^{\beta \delta}/ (a+b).
\]
Thereby, the validity of the uniqueness condition (\ref{ta22}) is
ensured by (\ref{dc3}). $\square$

\section{Proof of Theorems  \ref{MAtm} and \ref{pressuretm}}
\label{5s}

\subsection{Stochastic order and  the proof of Theorem \ref{MAtm}}
First we prove that the cone $K_+(\mathit{\Omega}^{\rm t})$ may be
used to establish an order on $\mathcal{G}^{\rm t}$, that is it
has the property: if $\mu(f) \leq \tilde{\mu}(f)$ and
$\tilde{\mu}(f) \leq{\mu}(f)$ for all $f \in
K_+(\mathit{\Omega}^{\rm t})$, then $\mu=\tilde{\mu}$.

\noindent
 {\bf Proof of Lemma \ref{MAlm}:} Let us show that
the cone $K_+ (\mathit{\Omega}^{\rm t})$ contains a defining class
for $\mathcal{G}^{\rm t}$. Usually, measure defining classes of
functions are established by means of monotone class theorems, see
e.g., \cite{[Patrick]}, pages 36 - 39. In our situation, a
sufficient condition for a measure defining class of bounded
continuous functions may be formulated as follows: (a) to contain
constant functions; (b) to be closed under multiplication; (c) to
separate points of $\mathit{\Omega}^{\rm t}$. The class
(\ref{MA1}) does not meet (b); hence, to prove the stated one has
to use additional arguments.

 A continuous function $f:\mathit{\Omega}^{\rm
t} \rightarrow \mathbf{R}$ is called a cylinder function if it
possesses the representation
\begin{equation} \label{MA2}
 f(\omega ) = \phi
(\omega_{\ell_1} (\tau_1), \dots , \omega_{\ell_n} (\tau_n)),
\end{equation}
with certain $n\in \mathbf{N}$, $\ell_1 , \dots , \ell_n$, $\tau_1,
\dots , \tau_n$,   and a continuous $\phi:\mathbf{R}^n \rightarrow
\mathbf{R}$. By $K^{\rm cyl}_+ (\mathit{\Omega}^{\rm t})$ we denote
the subset of $ K_+ (\mathit{\Omega}^{\rm t})$ consisting of
cylinder functions. Suppose that the equality (\ref{MA1a}) holds for
all $f \in K^{\rm cyl}_+ (\mathit{\Omega}^{\rm t})$. Then
\begin{equation} \label{MA3}
\int_{\mathit{\Omega}^{\rm t}} \omega_\ell (\tau) \mu({\rm d}\omega)
= \int_{\mathit{\Omega}^{\rm t}} \omega_\ell (\tau) \tilde{\mu}({\rm
d}\omega), \qquad {\rm for} \ \ {\rm all} \ \ \ell, \tau, j.
\end{equation}
For fixed $\ell_1 , \dots , \ell_n$ and $\tau_1, \dots , \tau_n$,
 let $P$ and $\tilde{P}$ be the projections
of the measures $\mu$ and $\tilde{\mu}$ on $\mathbf{R}^n$. That is,
each of $P$ and $\tilde{P}$ obeys
\[
\int_{\mathit{\Omega}^{\rm t}}f( \omega) \mu({\rm d}\omega) =
\int_{\mathbf{R}^n} \phi (x_1 , \dots , x_n) P({\rm d}x),
\]
for $f$ and $\phi$ as in (\ref{MA2}). Then by (\ref{MA1a}), it
follows that
\begin{equation} \label{MA4}
\int_{\mathbf{R}^n} \phi (x_1 , \dots , x_n) P({\rm d}x) \leq
\int_{\mathbf{R}^n} \phi (x_1 , \dots , x_n) \tilde{P}({\rm d}x),
\end{equation}
for all increasing $\phi$. Let $\widehat{P}$ be a probability
measure on $\mathbf{R}^{2n}$, such that
\[
P({\rm d}x) = \int_{\mathbf{R}^n}  \widehat{P}({\rm d}x, {\rm
d}\tilde{x}), \qquad \tilde{P}({\rm d}\tilde{x})=
\int_{\mathbf{R}^n} \widehat{ P}({\rm d}x, {\rm d}\tilde{x}).
\]
Thus, $\widehat{P}$ is a \emph{coupling} of $P$ and $\tilde{P}$. Of
course, the above equalities do not determine $\widehat{P}$
uniquely. By the Kantorovich-Rubinstein duality theorem, the
Wasserstein distance, c.f. (\ref{WD}), between the measures $P$ and
$\tilde{P}$ which have first moments, can be defined as follows, see
\cite{Du},
\begin{equation} \label{MA4a}
R (P, \tilde{P}) = \inf\int_{\mathbf{R}^{2n}}|x -
\tilde{x}|\widehat{P}({\rm d}x, {\rm d}\tilde{x}),
\end{equation}
where infimum is taken over all couplings of $P$ and $\tilde{P}$.
It is  a metric, and the convergence of a sequence of measures in
this metric is equivalent to its weak convergence combined with
the convergence of the first moments.
 Consider
\[
M = \{ (x, \tilde{x})\in \mathbf{R}^{2n} \ | \ x_i \leq \tilde{x}_i,
\ \quad {\rm for} \ {\rm all} \ i = 1, \dots , n\}.
\]
This set is closed in $\mathbf{R}^{2n}$. Then from (\ref{MA4}) by
Strassen's theorem, see page 129 of \cite{Li}, it follows that there
exists a coupling $\widehat{P}_*$ such that
\begin{equation} \label{MA5}
\widehat{P}_*\left(M \right) = 1.
\end{equation}
Thereby,
\begin{eqnarray*}
R (P, \tilde{P})  & \leq & \int_{M} |x -
\tilde{x}|\widehat{P}_*({\rm d}x, {\rm d}\tilde{x}) \\ & \leq &
\sum_{i=1}^n \int_{\mathbf{R}^{2n}} (\tilde{x}_i -
x_i)\widehat{P}_*({\rm d}x, {\rm d}\tilde{x}) \\ & = & \sum_{i=1}^n
\int_{\mathbf{R}^n} x_i \left[\tilde{P}({\rm d}x) - P({\rm d}x)
\right] =0.
\end{eqnarray*}
The latter equality follows from (\ref{MA3}). Since the subset of
$C_{\rm b}(\mathit{\Omega}^{\rm t})$ consisting of all cylinder
functions (\ref{MA2}) is a defining class for
$\mathcal{P}(\mathit{\Omega}^{\rm t})$, the equality of all the
projections of $\mu$ and $\tilde{\mu}$ yields $\mu= \tilde{\mu}$.
$\square$

Observe that for (\ref{MA4}) to hold, it was enough to have $\mu\leq
\tilde{\mu}$, c.f., (\ref{MA1}). Thus, we have one more important
fact arising from the proof of the above lemma.
\begin{corollary} \label{MAcor}
If for any $\mu, \tilde{\mu}\in \mathcal{G}^{\rm t}$, such that $\mu
\leq \tilde{\mu}$,  all their first moments coincide, i.e.,
(\ref{MA3}) holds, then $\mu = \tilde{\mu}$.
\end{corollary}
\begin{remark} \label{MAWrk}
For every $\ell$, $t_\ell (\omega) \leq t_\ell (\tilde{\omega})$ if
$\omega \leq \tilde{\omega}$. This means that the transformation
$\theta_\ell$ defined in (\ref{MA11}) is order preserving.
\end{remark}
\noindent \textbf{Proof of Theorem \ref{MAtm}:} \ In establishing
the existence of the elements $\mu_{\pm}$ the main point was to
prove Lemma \ref{MAlm}. Thereby, the existence of $\mu_{\pm}$ can be
proven by literal repetition of the arguments used in \cite{BH} for
proving Theorem IV.3. They are unique by definition. Indeed, for two
maximal elements, say $\mu_+$ and $\tilde{\mu}_+$, one would have
$\mu_+ \leq \tilde{\mu}_+$ and $\tilde{\mu}_+ \leq \mu_+$ at the
same time. Thus,  $\mu_+ = \tilde{\mu}_+$. The proof of the
extremeness (respectively, the symmetry properties) of $\mu_{\pm}$
can be done by following the proof of Proposition V.1 (respectively,
Proposition V.3) in \cite{BH}. Some additional properties of
$\mu_{\pm}$  will be described in the subsequent section. $\square$

The result just proven and Corollary \ref{MAcor} yield the following
\begin{lemma} \label{bhpn}
Suppose that, for all $\ell$,
\begin{equation} \label{July}
\mu_{+} (\omega_\ell (0) ) = \mu_{-} (\omega_\ell (0) ).
\end{equation}
Then $\mathcal{G}^{\rm t}$ is a singleton. If the model is
symmetric, then (\ref{July}) turns into the condition
\[
\mu_{+} (\omega_\ell (0) ) = \mu_{-} (\omega_\ell (0) )=0.
\]
\end{lemma}

\subsection{Existence of the pressure}
 Given
$R>0$ and $\Lambda \Subset \mathbf{L}$, let $\partial^{+}_R
\Lambda$ be the set of all $\ell \in \Lambda^c$, such that ${\rm
dist}(\ell, \Lambda) \leq R$. Then for a van Hove sequence
$\mathcal{L}$ and any $R>0$, one has $\lim_{\mathcal{L}}
|\partial^{+}_R \Lambda|/|\Lambda| = 0$, yielding
\begin{equation} \label{c5}
\lim_{\mathcal{L}} \frac{1}{|\Lambda|}\sum_{\ell \in \Lambda,
\ell' \in \Lambda^c} J_{\ell \ell'} = 0.
\end{equation}
 The existence of van Hove sequences means the
amenability of the graph $(\mathbf{L}, E)$, $E$ being the set of
all pairs $\ell, \ell'$, such that $|\ell - \ell'|=1$. For
nonamenable graphs, phase transitions with $h\neq 0$ are possible;
hence, statements like Theorem \ref{6tm} do not hold, see
\cite{JS,Ly}.

Let us prove first the existence of the pressure corresponding to
the zero boundary conditions.
\begin{lemma} \label{conlm}
For every $h\in \mathbf{R}$, the limiting pressure $p(h)=
\lim_{\mathcal{L}}p_\Lambda (h)$ exists for every van Hove
sequence $\mathcal{L}$. It is independent of the particular choice
of $\mathcal{L}$.
\end{lemma}
{\bf Proof:}
For $t \geq 0$, $\xi \in \mathit{\Omega}^{\rm t}$, and $\Delta
\subset \Lambda$, let $ \varpi_{\Lambda, \Delta}^{(t)}$,
$Y_{\Lambda, \Delta}(t)$ be defined by (\ref{w4z}) with the
potentials $V_\ell = V$ having the form (\ref{m2}). Then we define
\begin{equation} \label{c6}
f_{\Lambda, \Delta} (t) = \frac{1}{|\Lambda|} \log Y_{\Lambda,
\Delta} (t), \quad t\geq 0.
\end{equation}
This function is differentiable and
\begin{eqnarray} \label{c7}
g_{\Lambda, \Delta} (t) \ \stackrel{\rm def}{=} \ f'_{\Lambda,
\Delta} (t) & = & \frac{1}{2|\Lambda|}\sum_{\ell ,\ell' \in
\Delta} J_{\ell \ell'} \varpi^{(t)}_{\Lambda ,
\Delta}[(\omega_\ell , \omega_{\ell'})_{L^2_\beta} ] \\ & + &
\frac{1}{|\Lambda|}\sum_{\ell \in \Delta ,\ell' \in \Lambda
\setminus  \Delta} J_{\ell \ell'} \varpi^{(t)}_{\Lambda, \Delta}
[(\omega_\ell , \omega_{\ell'})_{L^2_\beta}]\geq 0. \nonumber
\end{eqnarray}
Here we used that $ \varpi^{(t)}_{\Lambda, \Delta} [(\omega_\ell ,
\omega_{\ell'})_{L^2_\beta}] \geq 0,$ which follows from the GKS
inequality (\ref{w3b}). The function $g_{\Lambda, \Delta}$ is also
differentiable and
\begin{equation} \label{c8}
  g'_{\Lambda, \Delta} (t) \geq 0,
\end{equation}
which may be proven similarly by means of the GKS inequality
(\ref{w3c}). Therefore,
\begin{equation} \label{c8a}
f_{\Lambda, \Delta} (0) \leq f_{\Lambda, \Delta} (1) \leq
g_{\Lambda, \Delta} (1).
\end{equation}
Now we take here $\Delta = \Lambda$ and obtain that $p_\Lambda$ is
a convex function of $h$. Furthermore, by (\ref{61q}), for any
$\alpha \in \mathcal{I}$,
\begin{equation} \label{w32}
\log Y_{\{\ell\}, \{\ell\}} (0) \leq p_\Lambda (h) \leq
 \hat{J}_0  C_{\ref{61q}} (0)/2.
\end{equation}
By the translation invariance the lower bound in (\ref{w32}) is
independent of $\ell$. Therefore, the set $\{p_\Lambda
(h)\}_{\Lambda \Subset \mathbf{L}}$ has accumulation points. For
one of them, $p (h)$, let $\{\Gamma_n\}_{n \in \mathbf{N}}$ be the
sequence of parallelepipeds such that $p_{\Gamma_n} (h)
\rightarrow p(h)$ as $n \rightarrow +\infty$. Let also
$\mathcal{L}$ be a van Hove sequence. Given $n \in \mathbf{N}$ and
$\Lambda \in \mathcal{L}$, let $\mathfrak{L}_n^{-}
(\Lambda)\subset \mathfrak{G}(\Gamma_n)$ (respectively,
$\mathfrak{L}_n^{+} (\Lambda)\subset \mathfrak{G}(\Gamma_n)$)
consist of the translates of $\Gamma_n$ which are contained in
$\Lambda$ (respectively, which have non-void intersections with
$\Lambda$). Let also
\begin{equation} \label{c8z}
\Lambda_n^{\pm} = \bigcup_{\Gamma \in \mathfrak{L}_n^{\pm}}
\Gamma.
\end{equation}
Now we take in (\ref{c6}) first $\Delta = \Lambda_n^{-}$, then
$\Delta = \Lambda$, $\Lambda = \Lambda_n^{+}$ and obtain by
(\ref{c8a})
\begin{equation} \label{c8w}
\frac{|\Lambda_n^{-}|}{|\Lambda|} p_{\Lambda_n^{-}} (h) \leq
p_\Lambda (h) \leq \frac{|\Lambda_n^{+}|}{|\Lambda|}
p_{\Lambda_n^{+}} (h).
\end{equation}
Let us estimate $p_{\Lambda_n^{\pm}}(h) - p_{\Gamma_n}(h)$. To
this end we introduce for $t\geq 0$, c.f., (\ref{w4z}),
\begin{eqnarray} \label{c8v}
 X_{\Lambda_n^{-}} (t) & = & \int_{\mathit{\Omega}_{\Lambda_n^{-}}} \exp \left\{
\frac{1}{2}\sum_{\Gamma \in \mathfrak{L}_n^{-}} \sum_{\ell, \ell'
\in \Gamma} J_{\ell \ell'}
(\omega_{\ell},\omega_{\ell'})_{L_\beta^2} \right.\\  \quad & + &
\left. t \sum_{\Gamma , \Gamma' \in \mathfrak{L}_n^{-}, \ \Gamma
\neq \Gamma'} \sum_{\ell \in \Gamma} \sum_{\ell' \in \Gamma'}
J_{\ell \ell'} (\omega_{\ell},\omega_{\ell'})_{L_\beta^2}\right.
\nonumber \\
& + & \left. \sum_{\ell \in \Lambda_n^{-}}\int_0^\beta\left[ h
\omega_\ell (\tau) - v([\omega_\ell(\tau)]^2) \right]{\rm d}\tau
\right\}\chi_{\Lambda_n^{-}} ({\rm d}\omega), \nonumber
\end{eqnarray}
and
\begin{equation} \label{c8t}
f_{\Lambda_n^-}(t) = \frac{1}{|\Lambda_n^{-}|}\log
X_{\Lambda_n^{-}} (t).
\end{equation}
Then
\begin{equation} \label{c8q}
f_{\Lambda_n^-}(1) = p_{\Lambda_n^{-}} (h), \quad
f_{\Lambda_n^-}(0) = \frac{|\Gamma_n|}{|\Lambda_n^{-}|}
\sum_{\Gamma \in \mathfrak{L}_n^{-}} p_{\Gamma}(h) = p_{\Gamma_n}
(h).
\end{equation}
Observe that $p_{\Gamma}(h) = p_{\Gamma_n} (h)$ for all $\Gamma
\in \mathfrak{G}(\Gamma_n)$, which follows from the translation
invariance of the model. Thereby,
\begin{eqnarray} \label{c8s}
0 & \leq & p_{\Lambda_n^{-}} (h) - p_{\Gamma_n} (h) \leq
f'_{\Lambda_n^{-}} (1) \\ & = & \frac{1}{|\Lambda_n^{-}|}
\sum_{\Gamma , \Gamma' \in \mathfrak{L}_n^{-}, \ \Gamma \neq
\Gamma'} \sum_{\ell \in \Gamma} \sum_{\ell' \in \Gamma'} J_{\ell
\ell'} \pi_{\Lambda_n^{-}} \left(
(\omega_{\ell},\omega_{\ell'})_{L_\beta^2}|0\right) \nonumber \\ &
\leq & \frac{1}{|\Lambda_n^{-}|} \sum_{\Gamma  \in
\mathfrak{L}_n^{-}} \sum_{\ell \in \Gamma} \sum_{\ell' \in
\Gamma^c} J_{\ell \ell'} \pi_{\Lambda_n^{-}} \left(
(\omega_{\ell},\omega_{\ell'})_{L_\beta^2}|0\right)\nonumber
\\
& & \leq \hat{J}(\Gamma_n)C_{\ref{61q}} (0) , \nonumber
\end{eqnarray}
where we used the estimate (\ref{61q}) and set
\begin{equation} \label{c8o}
\hat{J}(\Gamma_n) = \frac{1}{|\Gamma_n|} \sum_{\ell \in \Gamma_n }
\sum_{\ell' \in \Gamma_n^c} J_{\ell \ell'} =
\frac{1}{|\Gamma|}\sum_{\ell \in \Gamma } \sum_{\ell' \in
\Gamma^c} J_{\ell \ell'}, \quad {\rm for} \ {\rm every} \ \Gamma
\in \mathfrak{G}(\Gamma_n).
\end{equation}
In deriving (\ref{c8s}) we took into account that the function
(\ref{c8t}) has positive first and second derivatives, c.f.,
(\ref{c7}) and (\ref{c8}). By literal repetition one proves that
both estimates from (\ref{c8s}) hold also for $p_{\Lambda_n^{+}}
(h) - p_{\Gamma_n} (h)$. In view of (\ref{c5}) the above
$\hat{J}(\Gamma_n)$ may be made arbitrarily small by taking big
enough $\Gamma_n$. Thereby, for any $\varepsilon>0$, one can
choose $n\in \mathbf{N}$ such that the following estimates hold
(recall that $p_{\Gamma_n} \rightarrow p$ as $n \rightarrow
+\infty$)
\begin{equation} \label{c80}
|p_{\Gamma_n} (h) - p(h)| < \varepsilon/3, \quad 0 \leq
p_{\Lambda_n^-} (h) - p_{\Gamma_n} (h) \leq p_{\Lambda_n^+} (h) -
p_{\Gamma_n} (h) < \varepsilon/3.
\end{equation}
As $\mathcal{L}$ is a van Hove sequence, one can pick up $\Lambda
\in \mathcal{L}$ such that
\[
\max\left\{\left(\frac{|\Lambda_n^{+}|}{|\Lambda|}- 1 \right)
p_{\Lambda_n^{+}} (h); \left(1 - \frac{
|\Lambda_n^{-}|}{|\Lambda|} \right) p_{\Lambda_n^{+}} (h) \right\}
< \varepsilon/3,
\]
which is possible in view of (\ref{w32}). Then for the chosen $n$
and $\Lambda \in \mathcal{L}$, one has
\begin{eqnarray*}
& & |p_{\Lambda} (h) - p(h) | \leq |p_{\Gamma_n} (h) - p(h) | +
p_{\Lambda_n^+} (h)- p_{\Gamma_n} (h) \\ & & \quad  +
\max\left\{\left(\frac{|\Lambda_n^{+}|}{|\Lambda|}- 1 \right)
p_{\Lambda_n^{+}} (h); \left(1 - \frac{
|\Lambda_n^{-}|}{|\Lambda|} \right) p_{\Lambda_n^{+}} (h) \right\}
< \varepsilon,
\end{eqnarray*}
which obviously holds also for all $\Lambda'\in \mathcal{L}$ such
that $\Lambda \subset \Lambda'$.$\blacksquare$ \vskip.1cm \noindent
{\bf Proof of Theorem \ref{pressuretm}:} The proof will be done if
we show that, for every $\mu \in \mathcal{G}^{\rm t}$ and any van
Hove sequence $\mathcal{L}$,
\[
\lim_{\mathcal{L}}p^{\mu}_\Lambda (h) = p (h).
\]
By the Jensen inequality one obtains for $t_1 , t_2 \in
\mathbf{R}$, $\xi \in \mathit{\Omega}^{\rm t}$,
\begin{eqnarray*}
Z_{\Lambda}((t_1 + t_2)\xi) \geq Z_{\Lambda}(t_1 \xi) \exp\left\{
t_2 \sum_{\ell \in \Lambda , \ell' \in \Lambda^c} J_{\ell \ell'}
\pi_\Lambda \left[(\omega_\ell , \xi_{\ell'})_{L^2_\beta}
\left\vert t_1 \xi \right. \right]\right\}.
\end{eqnarray*}
We set here first $t_1 = 0, \ t_2 = 1$, then $t_1 = - t_2 = 1$,
and obtain after taking logarithm and dividing by $|\Lambda|$
\begin{eqnarray} \label{c9}
 p_\Lambda (h) & + & \frac{1}{|\Lambda|} \sum_{\ell \in \Lambda,
\ell' \in \Lambda^c} J_{\ell \ell'} \pi_\Lambda \left[(\omega_\ell
,
\xi_{\ell'})_{L^2_\beta}| 0 \right] \leq p_\Lambda (h , \xi) \\
& & \quad \qquad \quad \leq p_\Lambda (h) + \frac{1}{|\Lambda|}
\sum_{\ell \in \Lambda, \ell' \in \Lambda^c} J_{\ell \ell'}
\pi_\Lambda \left[(\omega_\ell , \omega_{\ell'})_{L^2_\beta}| \xi
\right], \nonumber
\end{eqnarray}
where we used that $\pi_\Lambda \left[(\omega_\ell ,
\omega_{\ell'})_{L^2_\beta}| \xi \right] = \pi_\Lambda
\left[(\omega_\ell , \xi_{\ell'})_{L^2_\beta}| \xi \right]$, see
(\ref{34}). Thereby, we integrate (\ref{c9}) with respect to
$\mu\in \mathcal{G}^{\rm t}$, take into account (\ref{40}), and
obtain after some calculations the following
\begin{eqnarray} \label{c10}
 p_\Lambda (h) & - & \frac{1}{|\Lambda|} \sum_{\ell \in \Lambda,
\ell' \in \Lambda^c} J_{\ell \ell'}\pi_\Lambda
\left(|\omega_\ell|_{L^2_\beta} \left\vert 0 \right.
\right)\mu\left(
|\xi_{\ell'}|_{L^2_\beta}\right) \leq p_\Lambda^\mu \\
& & \quad \qquad \quad \leq p_\Lambda (h) + \frac{1}{|\Lambda|}
\sum_{\ell \in \Lambda, \ell' \in \Lambda^c} J_{\ell \ell'} \mu
\left((\omega_\ell, \omega_{\ell'})_{L^2_\beta} \right). \nonumber
\end{eqnarray}
By means of Theorem \ref{2tm} (respectively, Lemma \ref{klm}), one
estimates $\mu \left((\omega_\ell, \omega_{\ell'})_{L^2_\beta}
\right)$, $\mu\left( |\xi_{\ell'}|_{L^2_\beta}\right)$
(respectively, $\pi_\Lambda (|\omega_\ell|_{L^2_\beta}|0)$) by
positive constants independent of $\ell, \ell'$. Thereby, the
property stated follows from (\ref{c5}) and Lemma \ref{conlm}.
$\square$ \vskip.1cm \noindent \vskip.1cm \noindent{\bf Proof of
Corollary \ref{Mco}:} By (\ref{c1}),
\[
\frac{\partial }{\partial h} p_\Lambda (h, \xi) =
\frac{1}{|\Lambda|} \sum_{\ell \in \Lambda} \int_0^\beta
\pi_\Lambda(\omega_\ell (\tau)| \xi){\rm d }\tau.
\]
Then, for every $\mu \in \mathcal{G}^{\rm t}$ and $\Lambda \Subset
\mathbf{L}$, one has
\begin{eqnarray} \label{X1}
\frac{\partial }{\partial h} p^\mu_\Lambda (h)& = &
\int_{\mathit{\Omega}} \frac{\partial }{\partial
h}\left(p^\mu_\Lambda (h, \xi) \right)\mu({\rm d}\xi) \\ & = &
\frac{1}{|\Lambda|} \sum_{\ell \in \Lambda}\int_0^\beta
\int_{\mathit{\Omega}} \pi_\Lambda \left[ \omega_\ell
(\tau)|\xi\right]\mu({\rm d}\xi) { \rm d}\tau  \nonumber \\ & = &
\frac{1}{|\Lambda|} \sum_{\ell \in \Lambda}\int_0^\beta \mu\left[
\omega_\ell (\tau)\right]{\rm d}\tau \nonumber
\end{eqnarray}
By Theorem \ref{pressuretm}, it follows that
\begin{equation} \label{X}
\frac{\partial }{\partial h} p^{\mu_+} (h) = \frac{\partial
}{\partial h} p^{\mu_-} (h).
\end{equation}
Both extreme measures $\mu_{\pm}$ are translation and shift
invariant. Then combining (\ref{X}) and (\ref{X1}) one obtains
$\mu_{+} (\omega_\ell (0)) = \mu_{-} (\omega_\ell (0))$ for any $h
\neq 0$. By Lemma \ref{bhpn} this gives the proof. $\square$

\section{Proof of Theorems \ref{phtm} and {\ref{5tm}}}
\label{sec7}

 We prove these theorem by comparing the model considered
with a certain model, for which the property desired is being
proven directly. The comparison is based on correlation
inequalities, which we present in the next subsections. They were
proven in the framework of the lattice approximation technique,
analogous to that of Euclidean quantum fields \cite{Si74}.

Recall that Theorems \ref{phtm} -- \ref{6tm} describe the model with
$\nu=1$ and $J_{\ell \ell'} \geq 0$, which will tacitely be assumed
in the statements below.

\subsection{Correlation inequalities}

\label{ss5.1}

We begin with the FKG inequality, Theorem 6.1 in \cite{AKKR}. Recall
that the family of functions $K_+ (\mathit{\Omega})$ and $K^{\rm
cyl}_+ (\mathit{\Omega})$ were introduced in (\ref{MA1}) and in the
proof of Lemma \ref{MAlm}.

\begin{proposition} \label{fkg}
For all $\Lambda \Subset \mathbf{L}$, $\xi \in \mathit{\Omega}^{\rm
t}$ and any  $f, g \in K_+ (\mathit{\Omega})$, it follows that
\begin{equation} \label{w2}
\pi_\Lambda ( f \cdot g |\xi) \geq \pi_\Lambda ( f |\xi) \cdot
\pi_\Lambda ( g |\xi).
\end{equation}
This inequality holds also for any continuous increasing functions,
for which the corresponding integrals exist. This yields in
particular that for all such functions,
\begin{equation} \label{w3}
 \xi \leq \tilde{\xi} \ \
\Longrightarrow \pi_\Lambda ( f |\xi) \leq \pi_\Lambda (
f|\tilde{\xi}).
\end{equation}
\end{proposition}
Next, there follow the GKS inequalities, Theorem 6.2 in \cite{AKKR}.
\begin{proposition} \label{gks}
Let the anharmonic potentials have the form
\begin{equation} \label{w3a}
V_\ell (x) = v_\ell (x^2) - h_\ell x, \quad h_\ell \geq 0 \ \ {\rm
for \ all} \ \ell\in \mathbf{L},
\end{equation}
with $v_\ell$ being continuous. Let also the continuous functions
$f_1, \dots , f_{n+m} : \mathbf{R}\rightarrow \mathbf{R}$ be
polynomially bounded and such that every $f_i$ is either an odd
increasing function on $\mathbf{R}$ or an even positive function,
increasing on $[0, +\infty)$. Then the following inequalities hold
for all $\tau_1 , \dots , \tau_{n+m} \in [0, \beta]$, and all
$\ell_1 , \dots , \ell_{n+m}\in \Lambda$,
\begin{equation} \label{w3b}
\int_{\mathit{\Omega}} \left(\prod_{i=1}^n f_i (\omega_{\ell_i}
(\tau_i))\right) \pi_\Lambda \left({\rm d}\omega |0\right) \geq 0;
\end{equation}
\begin{eqnarray} \label{w3c}
& & \int_{\mathit{\Omega}}\left(\prod_{i=1}^n f_i (\omega_{\ell_i}
(\tau_i))\right) \cdot \left(\prod_{i=n+1}^{n+m}f_i
(\omega_{\ell_i}(\tau_i)) \right)\pi_\Lambda \left({\rm d
}\omega|0\right)
\\ & & \qquad \geq  \int_{\mathit{\Omega}}\left(\prod_{i=1}^n f_i (\omega_{\ell_i}
(\tau_i))\right) \pi_\Lambda \left({\rm d }\omega|0\right) \cdot
\int_{\mathit{\Omega}}\left(\prod_{i=n+1}^{n+m}f_i
(\omega_{\ell_i}(\tau_i)) \right)\pi_\Lambda \left({\rm d
}\omega|0\right). \nonumber
\end{eqnarray}
\end{proposition}
 Given $\xi \in \mathit{\Omega}^{\rm t}$, $\Lambda \Subset
\mathbf{L}$, and $\ell, \ell'$, $\tau , \tau' \in [0, \beta]$,  the
pair correlation function is
\begin{eqnarray}\label{w4}
K^\Lambda_{\ell \ell'}(\tau , \tau' |\xi)& = &
\int_{\mathit{\Omega}} \omega_\ell (\tau) \omega_{\ell'} (\tau')
\pi_\Lambda ({\rm d}\omega|\xi) \\ & - & \int_{\mathit{\Omega}}
\omega_\ell (\tau)  \pi_\Lambda ({\rm d}\omega|\xi) \cdot
\int_{\mathit{\Omega}}
 \omega_{\ell'} (\tau') \pi_\Lambda ({\rm
d}\omega|\xi). \nonumber
\end{eqnarray}
Then, by (\ref{w3}),
\begin{equation} \label{w4b}
K^\Lambda_{\ell \ell'}(\tau , \tau' |\xi) \geq 0,
\end{equation}
which holds for all $\ell, \ell'$, $\tau, \tau'$, and
$\xi\in\mathit{\Omega}^{\rm t}$.
 The following result is a version of the estimate (12.129), page 254 of \cite{FFS},
 which for the Euclidean Gibbs measures may be proven by means of
 the lattice approximation.
\begin{proposition} \label{ffs}
Let $V_\ell$ be of the form (\ref{w3a}) with $h_\ell =0$ and the
functions $v_\ell$ being convex. Then for all $\ell, \ell'$, $\tau,
\tau'$ and for any $\xi \in \mathit{\Omega}^{\rm t}$ such that $\xi
\geq 0$, it follows that
\begin{equation} \label{w5}
K^\Lambda_{\ell \ell'}(\tau , \tau'|\xi) \leq K^\Lambda_{\ell
\ell'}(\tau , \tau'|0).
\end{equation}
\end{proposition}
Let us consider
\begin{eqnarray} \label{v1}
& & U_{\ell_1  \ell_2  \ell_3 \ell_4}^\Lambda (\tau_1 , \tau_2 ,
\tau_3 , \tau_4) =
\int_{\mathit{\Omega}}\omega_{\ell_1}(\tau_1)\omega_{\ell_2}(\tau_2)
\omega_{\ell_3}(\tau_3) \omega_{\ell_4}(\tau_4) \pi_\Lambda ({\rm
d}\omega|0)\\ & &  \qquad \quad  \quad \quad - K_{\ell_1
\ell_2}^\Lambda(\tau_1 , \tau_2|0)K_{\ell_3 \ell_4}^\Lambda(\tau_3 ,
\tau_4|0) - K_{\ell_1 \ell_3}^\Lambda(\tau_1 , \tau_3|0)K_{\ell_2
\ell_4}^\Lambda(\tau_2 , \tau_4|0) \nonumber \\ & &  \qquad \quad
\quad \quad - K_{\ell_1 \ell_4}^\Lambda(\tau_1 , \tau_4|0)K_{\ell_2
\ell_3}^\Lambda(\tau_2 , \tau_3|0), \nonumber
\end{eqnarray}
which is the Ursell function for the measure $\pi_\Lambda
(\cdot|0)$. The next statement gives the Gaussian domination and
Lebowitz inequalities, see \cite{AKKR}.
\begin{proposition} \label{le}
Let $V_\ell$ be of the form (\ref{w3a}) with $h_\ell =0$ and
 the functions $v_\ell$ being convex. Then for all $n
\in \mathbf{N}$, $\ell_1, \dots, \ell_{2n}
 \in \Lambda\Subset \mathbf{L}$, $\tau_1, \dots, \tau_{2n} \in [0, \beta]$, it follows
 that
\begin{eqnarray} \label{vv1}
& &
\int_{\mathit{\Omega}}\omega_{\ell_1}(\tau_1)\omega_{\ell_2}(\tau_2)
\cdots \omega_{\ell_{2n}}(\tau_{2n}) \pi_\Lambda ({\rm d}\omega|0)
\nonumber\\ & & \qquad \quad \leq \sum_{\sigma} \prod_{j=1}^n
\int_{\mathit{\Omega}}\omega_{\ell_{\sigma(2j-1)}}
(\tau_{\sigma(2j-1)})\omega_{\ell_{\sigma(2j)}}(\tau_{\sigma(2j)})\pi_\Lambda
({\rm d}\omega|0),
\end{eqnarray}
where the sum runs through the set of all partitions of $\{1, \dots
, 2n\}$ onto unordered pairs. In particular,
\begin{equation}\label{v2}
U_{\ell_1  \ell_2  \ell_3 \ell_4}^\Lambda (\tau_1 , \tau_2 , \tau_3
, \tau_4) \leq 0.
\end{equation}
\end{proposition}
\subsection{More on extreme elements}

\label{sdss} Here we continue to study the properties of
$\mu_{\pm}$, the existence of which was established in Theorem
\ref{MAtm}. In particular, we give an explicit construction of these
measures.

For $\ell_0$ and $b>0$, let $\hat{\xi} =(\hat{\xi}_\ell)_{\ell \in
\mathbf{L}}$ be the following constant (with respect to $\tau \in
S_\beta$) configuration
\begin{equation} \label{v3}
 \hat{\xi}_{\ell} (\tau) = [ b
\log(1 + |\ell - \ell_0|)]^{1/2}.
\end{equation}
Fix $\sigma \in (0, 1/2)$ and $b$ obeying the condition $b>
d/\lambda_\sigma$ (see the proof of Theorem \ref{3tm}). In view of
(\ref{25}), $\hat{\xi}$ belongs to $\mathit{\Omega}^{\rm t}$. It
also belongs to $\mathit{\Xi}(\ell_0 , b, \sigma)$ and for all
$\xi\in \mathit{\Xi}( b, \sigma)$, one finds $\Delta \Subset
\mathbf{L}$ such that $\xi^{(j)}_\ell (\tau) \leq
\hat{\xi}^{(j)}_{\ell} (\tau)$ for all $\tau$, $j$ and $\ell \in
\Delta^c$. Therefore, for any cofinal sequence $\mathcal{L}$ and
$\xi \in \mathit{\Xi}( b, \sigma)$, one finds $\Delta \in
\mathcal{L}$ such that for all $\Lambda \in \mathcal{L}$, $\Delta
\subset \Lambda$, one has $\pi_\Lambda (\cdot |\xi) \leq \pi_\Lambda
(\cdot |\hat{\xi})$, see  (\ref{w3}). As was established in the
proof of Theorem \ref{1tm}, every sequence $\{\pi_\Lambda (\cdot |
\xi)\}_{\Lambda \in \mathcal{L}}$, $\xi \in \mathit{\Xi}(b
,\sigma)\subset \mathit{\Omega}^{\rm t}$, is relatively compact in
any $\mathcal{W}_\alpha$, $\alpha \in \mathcal{I}$, which by Lemmas
\ref{klm}, \ref{tanlm} yields its $\mathcal{W}^{\rm t}$-relative
compactness. For a cofinal sequence $\mathcal{L}$, let $\hat{\mu}$
be any of the accumulating points of $\{\pi_\Lambda (\cdot |
\hat{\xi})\}_{\Lambda \in \mathcal{L}}$. By Lemma \ref{3lm}
$\hat{\mu} \in\mathcal{G}^{\rm t}$ and by Lemma \ref{locco}
$\hat{\mu}$ dominates every element of ${\rm ex}(\mathcal{G}^{\rm
t})$. Hence, $\hat{\mu} = \mu_+$ since the maximal element is
unique. The same is true for the remaining accumulation points of
$\{\pi_\Lambda (\cdot | \xi)\}_{\Lambda \in \mathcal{L}}$; thus, for
every cofinal sequence $\mathcal{L}$ and for every $\ell_0$, we have
\begin{equation} \label{rf41}
\lim_{\mathcal{L}} \pi_\Lambda (\cdot |\pm \hat{\xi}) = \mu_{\pm}.
\end{equation}
\begin{remark} \label{MPRrk}
 As the configuration
(\ref{v3}) is constant with respect to $\tau\in S_\beta$, the kernel
$\pi_\Lambda (\cdot|\hat{\xi})$ may be considered as the one
$\hat{\pi}_\Lambda (\cdot|0)$ corresponding to the Hamiltonian with
the external field $\hat{\xi}$, that is,
\begin{equation} \label{MPR}
H_\Lambda - \sum_{\ell\in \Lambda} (q_\ell,\hat{\xi}_\ell) .
\end{equation}
\end{remark}

\subsection{Reference models}

\label{rfss} We shall prove Theorems \ref{phtm}, \ref{5tm} by
comparing our model with two reference models, defined as follows.
Let $J$ and $V$ be the same as in (\ref{p1}) and (\ref{ub})
respectively. For $\Lambda \Subset \mathbf{L}=\mathbf{Z}^d$, we set
(c.f., (\ref{sch}))
\begin{equation} \label{rf1}
H^{\rm low}_{\Lambda} = \sum_{\ell \in \Lambda}\left[ H_\ell^{\rm
har} + V(x_\ell)\right] - \frac{1}{2} \sum_{\ell, \ell'\in \Lambda}
J \epsilon_{\ell \ell'} x_\ell x_{\ell'}, \quad x_\ell \in
\mathbf{R},
\end{equation}
where $H_\ell^{\rm har}$ is given by (\ref{16e}) and $\epsilon_{\ell
\ell'} = 1$ if $|\ell - \ell'|=1$ and $\epsilon_{\ell \ell'} = 0$
otherwise. The second reference model is defined on an arbitrary
$\mathbf{L}$ satisfying (\ref{april}). For $\Lambda \Subset
\mathbf{L}$, we set
\begin{equation} \label{rf2}
H^{\rm up}_{\Lambda} = \sum_{\ell \in \Lambda} \left[H_\ell^{\rm
har} + v(x_\ell^2) \right]- \frac{1}{2} \sum_{\ell , \ell' \in
\Lambda} J_{\ell \ell'} x_\ell x_{\ell'} = \sum_{\ell \in \Lambda}
\tilde{H}_\ell - \frac{1}{2} \sum_{\ell , \ell' \in \Lambda} J_{\ell
\ell'} x_\ell x_{\ell'},
\end{equation}
where $\tilde{H}_\ell$ is defined by (\ref{2}) and the interaction
intensities $J_{\ell \ell'}$ are the same as in (\ref{sch}). Since
both these models are particular cases of the model we consider,
their sets of Euclidean Gibbs measures have the properties
established by Theorems \ref{1tm} -- \ref{3tm}. By $\mu^{\rm
low}_{\pm}$, $\mu^{\rm up}_{\pm}$ we denote the corresponding
extreme elements.
\begin{remark} \label{convexrm}
The anharmonic potentials of both reference models have the form
(\ref{w3a}) with the zero external field $h_\ell =0$ and the
functions $v_\ell$ being convex. Hence, they obey the conditions of
all the statements of subsection \ref{ss5.1}. The $low$-reference
model is translation invariant. The $up$-reference model is
translation invariant if $\mathbf{L}$ is a lattice and  $J_{\ell
\ell'}$ are translation invariant.
\end{remark}
In the statements below the comparison with the $low$-reference
model relates to the case of $\mathbf{L}=\mathbf{Z}^d$.
\begin{lemma} \label{rflm}
For every $\ell$, it follows that
\begin{equation} \label{rf3}
\mu^{\rm low}_{+} (\omega_\ell (0)) \leq \mu_{+} (\omega_\ell (0))
\leq \mu^{\rm up}_{+} (\omega_\ell (0)).
\end{equation}
\end{lemma}
{\bf Proof:}
By (\ref{rf41}) we have that for any $\mathcal{L}$,
\begin{equation} \label{soc1}
\int_{\mathit{\Omega}}\omega_\ell (\tau)\mu_{\pm}({\rm d}\omega) =
\lim_{ \mathcal{L}}\int_{\mathit{\Omega}}\omega_\ell
(\tau)\pi_\Lambda ({\rm d}\omega|\pm \hat{\xi}), \ \ {\rm for} \
{\rm all} \ \tau.
\end{equation}
Thus, the proof will be done if we show that for all $\Lambda
\Subset \mathbf{L}$ and $\ell \in \Lambda$,
\begin{equation} \label{rf4}
\pi_{\Lambda}^{\rm low} (\omega_\ell (0)|\hat{\xi}) \leq
\pi_{\Lambda} (\omega_\ell (0)|\hat{\xi}) \leq \pi_{\Lambda}^{\rm
up} (\omega_\ell (0)|\hat{\xi}).
\end{equation}
 First we prove the left-hand
inequality in (\ref{rf4}). For given $\Lambda \Subset \mathbf{L}$
and $t, s \in [0, 1]$, we introduce
\begin{eqnarray} \label{rf5}
& & \varpi^{(t, s)}_\Lambda ({\rm d}\omega_\Lambda ) = \frac{1}{Y(t,
s)} \exp\left( \frac{1}{2} \sum_{\ell , \ell' \in \Lambda} J
\epsilon_{\ell \ell'} (\omega_\ell , \omega_{\ell'})_{L^2_\beta} +
\sum_{\ell \in \Lambda}(\omega_\ell, \eta^{\ell_0, s}_\ell)_{L^2_\beta} \right. \\
& & \qquad - \sum_{\ell \in \Lambda} \int_0^\beta V(\omega_\ell
(\tau)){\rm d}\tau + \frac{s}{2} \sum_{\ell , \ell' \in
\Lambda}\left[J_{\ell \ell'} -   J \epsilon_{\ell
\ell'}\right](\omega_\ell , \omega_{\ell'})_{L^2_\beta} \nonumber
\\& & \qquad - \left. t \sum_{\ell \in \Lambda} \int_0^\beta \left[V_\ell ( \omega_\ell
(\tau)) - V ( \omega_\ell (\tau)) \right]{\rm d}\tau \right)
\chi_\Lambda ({\rm d}\omega_\Lambda), \nonumber
\end{eqnarray}
where, see (\ref{v3}),
\begin{eqnarray} \label{rf6}
\eta^{\ell_0, s}_\ell (\tau) & \stackrel{\rm def}{=} & \sum_{\ell'
\in \Lambda^c} J \epsilon_{\ell \ell'} \hat{\xi}_{\ell'} (\tau) \\
& + & s \sum_{\ell' \in \Lambda^c}\left[ J_{\ell \ell'} - J
\epsilon_{\ell \ell'}\right] \hat{\xi}_{\ell'} (\tau) \geq
\sum_{\ell' \in \Lambda^c} J \epsilon_{\ell \ell'} \hat{\xi}_{\ell'}
(\tau)
> 0, \nonumber
\end{eqnarray}
which in fact is independent of $\tau$, and
\begin{eqnarray*}
& & Y(t, s) = \int_{\mathit{\Omega}_\Lambda} \exp\left( \frac{1}{2}
\sum_{\ell , \ell' \in \Lambda} J \epsilon_{\ell \ell'} (\omega_\ell
, \omega_{\ell'})_{L^2_\beta} +
\sum_{\ell \in \Lambda}(\omega_\ell, \eta^{\ell_0, s}_\ell)_{L^2_\beta} \right. \\
& & \qquad - \sum_{\ell \in \Lambda} \int_0^\beta V(\omega_\ell
(\tau)){\rm d}\tau + \frac{s}{2} \sum_{\ell , \ell' \in
\Lambda}\left[J_{\ell \ell'} -   J \epsilon_{\ell
\ell'}\right](\omega_\ell , \omega_{\ell'})_{L^2_\beta} \nonumber
\\& & \qquad - \left. t \sum_{\ell \in \Lambda} \int_0^\beta \left[V_\ell ( \omega_\ell
(\tau)) - V ( \omega_\ell (\tau)) \right]{\rm d}\tau \right)
\chi_\Lambda ({\rm d}\omega_\Lambda). \nonumber
\end{eqnarray*}
Since the site-dependent `external field' (\ref{rf6}) is positive,
the moments of the measure (\ref{rf5}) obey the GKS inequalities.
Therefore, for any $\ell \in \Lambda$, the function
\begin{equation} \label{rf7}
\phi (t, s) = \varpi_\Lambda^{(t, s)} (\omega_\ell (0)), \quad t, s
\in [0, 1],
\end{equation}
is continuous and increasing in  both variables. Indeed, taking into
account (\ref{p1}), (\ref{ub}), and (\ref{ub1}), we get
\begin{eqnarray*}
 \frac{\partial }{\partial s} \phi (t, s) & = &
\sum_{\ell'\in\Lambda} \left[J_{\ell \ell'} - J \epsilon_{\ell
\ell'}  \right]\hat{\xi}_{\ell'} (0) \\ & \times & \int_0^\beta
\left\{ \varpi_\Lambda^{(t, s)} \left[ \omega_\ell (0)
\omega_{\ell'}(\tau)\right] - \varpi_\Lambda^{(t,s)} \left[
\omega_\ell (0) \right] \cdot \varpi_\Lambda^{(t, s)} \left[
\omega_{\ell'}(\tau)\right] \right\} {\rm d}\tau  \\
  & + &
 \frac{1}{2}\sum_{\ell_1 , \ell_2 \in \Lambda} \left[J_{\ell_1 \ell_2} - J
 \epsilon_{\ell_1
\ell_2}  \right]\left\{\varpi_\Lambda^{(t, s)} \left[
\omega_{\ell}(0) (\omega_{\ell_1}, \omega_{\ell_2})_{L^2_\beta}
\right] \right. \\ & - & \left. \varpi_\Lambda^{(t, s)} \left[
\omega_{\ell}(0)  \right] \cdot \varpi_\Lambda^{(t, s)} \left[
(\omega_{\ell_1}, \omega_{\ell_2})_{L^2_\beta} \right] \right\} \geq
0, \\ \frac{\partial }{\partial t} \phi (t, s) & = & \sum_{\ell' \in
\Lambda} \int_0^\beta \left\{\varpi_\Lambda^{(t, s)} \left(
\omega_{\ell}(0) \cdot \left[ V(\omega_{\ell'} (\tau)) - V_{\ell'}
(\omega_{\ell'} (\tau))\right] \right) \right.
\\ & - & \left. \varpi_\Lambda^{(t, s)} \left[
\omega_{\ell}(0)  \right] \cdot \varpi_\Lambda^{(t, s)} \left[
V(\omega_{\ell'} (\tau)) - V_{\ell'} (\omega_{\ell'} (\tau)) \right]
\right\}{\rm d}\tau \geq 0.
\end{eqnarray*}
But by (\ref{rf5}) and (\ref{rf7})
\[ \phi (0,0) = \pi_\Lambda^{\rm low} (\omega_\ell (0)), \quad \phi (1,1)
 = \pi_\Lambda (\omega_\ell (0)),
 \]
 which proves the left-hand inequality in (\ref{rf4}). To prove
 the right-hand one we have to take the measure (\ref{rf5}) with $s=1$ and
 $v(x^2_\ell)$ instead of $V(x_\ell)$ and repeat the above steps
 taking into account
 (\ref{49}).
$\blacksquare$
\vskip.1cm In the next statement we summarize the properties of
the reference models.
\begin{corollary}[Comparison Criterion]  \label{compco}
The model considered undergoes a phase transition if the ${\rm
low}$-reference model does so. The uniqueness of tempered Euclidean
Gibbs measures of the ${\rm up}$-reference model implies that
$|\mathcal{G}^{\rm t}|=1$.
\end{corollary}
{\bf Proof:}
The proof follows immediately from (\ref{rf3}) and Lemma \ref{bhpn}.
$\blacksquare$

\subsection{Estimates for pair correlation functions}
\label{sss5.3}
 For $\Delta \subset \Lambda$, $\ell , \ell'\in \Lambda$,
$\tau , \tau' \in [0, \beta]$, and $t \in [ 0, 1]$, we set
\begin{equation} \label{w4zz}
Q^\Lambda_{\ell \ell'} (\tau , \tau'|\Delta, t) =
\int_{\mathit{\Omega}_\Lambda} \omega_\ell (\tau) \omega_{\ell'}
(\tau') \varpi_{\Lambda, \Delta}^{(t)} ({\rm d}\omega_\Lambda),
\end{equation}
where this time we have denoted
\begin{eqnarray} \label{w4z}
& & \varpi_{\Lambda, \Delta}^{(t)} ({\rm d}\omega_\Lambda)
  = \frac{1}{Y_{\Lambda, \Delta}(t)} \exp \left\{ \frac{1}{2} \sum_{\ell_1 ,
\ell_2 \in \Lambda \setminus \Delta}J_{\ell_1 \ell_2}
(\omega_{\ell_1}, \omega_{\ell_2})_{L^2_\beta}\right.
\\
& & \quad   + t \left( \sum_{\ell_1 \in \Delta} \sum_{\ell_2 \in
\Lambda \setminus \Delta}J_{\ell_1 \ell_2} (\omega_{\ell_1},
\omega_{\ell_2})_{L^2_\beta} +  \frac{1}{2} \sum_{\ell_1 , \ell_2
\in \Delta}J_{\ell_1 \ell_2} (\omega_{\ell_1},
\omega_{\ell_2})_{L^2_\beta} \right) \nonumber \\& & \quad \left.  -
\sum_{\ell \in \Lambda}\int_0^\beta V_\ell (\omega_\ell (\tau)){\rm
d}\tau \right\}\chi_\Lambda
({\rm d}\omega_\Lambda), \nonumber\\
 & &  Y_{\Lambda , \Delta}(t) =
 \int_{\mathit{\Omega}_\Lambda} \exp \left\{ \frac{1}{2} \sum_{\ell_1 ,
\ell_2 \in \Lambda \setminus \Delta}J_{\ell_1 \ell_2}
(\omega_{\ell_1}, \omega_{\ell_2})_{L^2_\beta}\right.
\nonumber \\
& & \quad   + t \left( \sum_{\ell_1 \in \Delta} \sum_{\ell_2 \in
\Lambda \setminus \Delta}J_{\ell_1 \ell_2} (\omega_{\ell_1},
\omega_{\ell_2})_{L^2_\beta} +  \frac{1}{2} \sum_{\ell_1 , \ell_2
\in \Delta}J_{\ell_1 \ell_2} (\omega_{\ell_1},
\omega_{\ell_2})_{L^2_\beta} \right) \nonumber \\& & \quad \left.  -
\sum_{\ell \in \Lambda}\int_0^\beta V_\ell (\omega_\ell (\tau)){\rm
d}\tau \right\}\chi_\Lambda ({\rm d}\omega_\Lambda). \nonumber
\end{eqnarray}
By literal repetition of the arguments used for proving Lemma
\ref{rflm} one proves the following
\begin{proposition} \label{wzlm}
The above $Q^\Lambda_{\ell \ell'} (\tau , \tau'|\Delta , t)$ is an
increasing continuous function of $t\in [0,1]$.
\end{proposition}
\begin{corollary} \label{wzco}
Let the conditions of Proposition \ref{gks} be satisfied. Then for
any pair $\Lambda \subset \Lambda ' \Subset \mathbf{L}$ and for all
$\tau$ and $\ell$, the functions (\ref{w3}) obey the estimate
\begin{equation} \label{w4zy}
K_{\ell \ell'}^\Lambda (\tau , \tau'|0) \leq K_{\ell
\ell'}^{\Lambda'} (\tau , \tau'|0),
\end{equation}
which holds for all $\ell , \ell' \in \Lambda$ and $\tau , \tau' \in
[0,\beta]$.
\end{corollary}
\noindent Now we obtain bounds for the correlation functions of the
reference models for a one-point $\Lambda = \{\ell\}$. Set
\begin{equation} \label{rf8}
{K}^{\rm up}_\ell (\tau , \tau') = \pi_{\ell}^{\rm up} (\omega_\ell
(\tau) \omega_\ell (\tau')|0), \quad {K}^{\rm low}_\ell (\tau ,
\tau') = \pi_{\ell}^{\rm low} (\omega_\ell (\tau) \omega_\ell
(\tau')|0),
\end{equation}
 We recall that the parameter $\mathit{\Delta}$
was defined by (\ref{51}).
\begin{lemma} \label{dtplm}
For every $\beta$, it follows that
\begin{equation} \label{w70}
K^{\rm up}_\ell \ \stackrel{\rm def}{=} \ \int_0^\beta {K}^{\rm
up}_\ell (\tau , \tau') {\rm d}\tau \leq 1/ m \mathit{\Delta}^2.
\end{equation}
\end{lemma}
{\bf Proof:}
In view of (\ref{mats1}) the above integral in independent of
$\tau$. By (\ref{mats}) and (\ref{mul10})
\begin{equation} \label{z4}
{K}^{\rm up}_\ell = \frac{1}{\tilde{Z}_\ell} \int_0^\beta {\rm
trace} \left\{ x_\ell e^{-\tau \tilde{H}_\ell} x_\ell e^{-(\beta
-\tau) \tilde{H}_\ell}\right\}{\rm d}\tau , \quad \tilde{Z}_\ell =
{\rm trace}[ e^{- \beta\tilde{H}_\ell}],
\end{equation}
where the Hamiltonian $\tilde{H}$ was defined in (\ref{2}). Its
spectrum $\{{E}_n\}_{n \in \mathbf{N}}$ determines by (\ref{51}) the
parameter $\mathit{\Delta}$. Integrating in (\ref{z4}) we get
\begin{eqnarray} \label{w8}
{K}^{\rm up}_\ell & = &\frac{1}{\tilde{Z}_\ell} \sum_{n, n'\in
\mathbf{N}_0, \ n\neq n'} \left\vert (\psi_n , x_\ell
\psi_{n'})_{L^2(\mathbf{R})} \right\vert^2 \frac{(E_{n} -
E_{n'})(e^{-\beta E_{n'}}- e^{-\beta E_{n}})}{(E_{n} - E_{n'})^2}
\nonumber \\ &\leq & \frac{1}{\tilde{Z}_\ell}\cdot
\frac{1}{\mathit{\Delta}^2}\sum_{n, n'\in \mathbf{N}_0} \left\vert
(\psi_n , x_\ell \psi_{n'})_{L^2(\mathbf{R})} \right\vert^2(E_{n} -
E_{n'})(e^{-\beta E_{n'}}- e^{-\beta E_{n}})\nonumber
\\ & = &\frac{1}{\mathit{\Delta}^2} \cdot\frac{1}{\tilde{Z}_\ell} {\rm
trace} \left\{\left[x_\ell , \left[\tilde{H}_\ell, x_\ell\right]
\right] e^{-\beta \tilde{H}_\ell}\right\} = \frac{1}{m
\mathit{\Delta}^2} ,
\end{eqnarray}
where $\psi_n$, $n \in \mathbf{N}_0$ are the eigenfunctions of
$\tilde{H}_\ell$ and $[\cdot, \cdot]$ stands for commutator.
$\blacksquare$
\noindent For the functions $K_{\ell}^{\rm low}$, a representation
like (\ref{z4}) is obtained by means of the following Hamiltonian
\begin{equation} \label{z4z}
\hat{H}_\ell =  H_\ell^{\rm har} + V(x_\ell) = - \frac{1}{2m}
\left(\frac{\partial}{\partial x_\ell} \right)^2 + \frac{a}{2}
x_\ell^2 + V(x_\ell),
\end{equation}
where $m$ and $a$ are the same as in (\ref{2}) but $V$ is given by
(\ref{ub}). Thereby,
\begin{equation} \label{z4y}
{K}^{\rm low}_\ell (0,0) = {\rm trace}[ x_\ell^2 \exp(-\beta
\hat{H}_\ell)]/{\rm trace}[\exp(-\beta \hat{H}_\ell)] \
\stackrel{\rm def}{=} \ \hat{\varrho} (x_\ell^2).
\end{equation}
\begin{lemma} \label{lblm}
Let  $t_*$ be the solution of (\ref{p4}). Then ${K}^{\rm low}_\ell
(0,0) \geq t_*$.
\end{lemma}
{\bf Proof:}
By Bogoliubov's inequality (see e.g., \cite{Simonl}), it follows
that
\[
\hat{\varrho}_\ell \left([p_\ell , [\hat{H}_\ell , p_\ell]]\right)
\geq 0, \quad p_\ell = - \sqrt{-1} \frac{\partial}{\partial x_\ell},
\]
which by (\ref{ub}), (\ref{p3}) yields
\begin{eqnarray*}
& & a + 2 b^{(1)} + \sum_{s=2}^r 2 s (2 s-1)b^{(s)}
\hat{\varrho}\left[ x^{2(s-1)}_\ell\right] \\ & & \quad =  a + 2
b^{(1)} + \sum_{s=2}^r 2 s (2 s-1)b^{(s)} \pi^{\rm low}_{\ell}
\left[\left(\omega_\ell (0)\right)^{2(s-1)}\right] \geq 0. \nonumber
\end{eqnarray*}
Now we use  the Gaussian domination inequality (\ref{vv1}) and
obtain  ${K}^{\rm low}_\ell \geq t_*$.
$\blacksquare$

\subsection{Periodic states and proof of Theorem \ref{phtm}} \label{sss5.2}

In view of Corollary \ref{compco} to prove Theorem \ref{phtm} we
show that
\begin{equation} \label{rf10}
\mu_{+}^{\rm low}(\omega_\ell (0)) >0,
\end{equation}
if the conditions of Theorem \ref{phtm} are satisfied. To this end
we employ the translation invariance and reflection positivity of
the $low$-reference model. With this connection we construct
periodic Euclidean Gibbs states by introducing (c.f., (\ref{19}))
\begin{equation} \label{mm7}
{I}^{\rm per}_\Lambda (\omega_\Lambda) = - \frac{J}{2} \sum_{\ell ,
\ell' \in \Lambda}\epsilon_{\ell\ell'}^\Lambda (\omega_\ell ,
\omega_{\ell'})_{L^2_\beta} + \sum_{\ell\in \Lambda} \int_0^\beta V
\left(\omega_\ell (\tau) \right) {\rm d}\tau,
\end{equation}
where
\begin{equation} \label{mm8}
 \Lambda= (-L, L]^d\bigcap
\mathbf{L}, \ \ L\in \mathbf{N},
\end{equation}
and $\epsilon_{\ell\ell'}^\Lambda = 1$ if $|\ell - \ell'|_\Lambda
=1$ and $\epsilon_{\ell\ell'}^\Lambda = 0$ otherwise. Here
\begin{eqnarray*}
|\ell - \ell'|_\Lambda &= & [|\ell_1 - \ell'_1|^2_L + \cdots +
|\ell_d - \ell'_d|^2_L]^{1/2}, \\   |\ell_j - \ell'_j|_L & = &
\min\{|\ell_j -\ell'_j|; L - |\ell_j - \ell'_j| \}, \ \qquad j = 1 ,
\dots , d.
\end{eqnarray*}
Clearly, ${I}^{\rm per}_\Lambda$ is invariant with respect to the
translations of the torus which one obtains by identifying the
opposite walls of the box (\ref{mm8}). The energy functional
${I}^{\rm per}_\Lambda$ corresponds to the following periodic
Hamiltonian
\begin{equation} \label{schp}
{H}^{\rm per}_\Lambda =  \sum_{\ell \in \Lambda} \left[H^{\rm
har}_\ell + V( x_\ell)\right]- \frac{J}{2} \sum_{\ell ,\ell' \in
\Lambda} \epsilon_{\ell\ell'}^\Lambda x_\ell x_{\ell'},
\end{equation}
in the same sense as $I_\Lambda$ given by (\ref{19}) corresponds to
$H_\Lambda$ given by (\ref{sch}). Now we introduce the periodic
kernels (c.f., (\ref{34}))
\begin{equation} \label{mm9}
{\pi}_\Lambda^{\rm per} ({\rm d}\omega) = \frac{1}{Z^{\rm
per}_\Lambda}\exp\left[- {I}^{\rm per}_\Lambda (\omega_\Lambda)
\right] \chi_\Lambda({\rm d}\omega_\Lambda)\prod_{\ell \in
\Lambda^c} \delta({\rm d}\omega_\ell),
\end{equation}
where $\delta$ is the Dirac measure concentrated at $\omega_\ell =
0$ and
\[
Z^{\rm per}_\Lambda = \int_{\mathit{\Omega}_\Lambda}\exp\left[-
{I}^{\rm per}_\Lambda (\omega_\Lambda) \right] \chi_\Lambda({\rm
d}\omega_\Lambda).
\]
Thereby, for every box $\Lambda$, the above ${\pi}_\Lambda^{\rm
per}$ is a probability measure on $\mathit{\Omega}^{\rm t}$. By
$\mathcal{L}_{\rm box}$ we denote the sequence of boxes (\ref{mm8})
indexed by $L\in \mathbf{N}$. For a given $\alpha \in \mathcal{I}$,
let us choose $\vartheta, \varkappa
>0$ such that the estimate (\ref{67d}) holds.
\begin{lemma} \label{mmlm}
For every box $\Lambda$, $\alpha \in \mathcal{I}$, and $\sigma\in
(0, 1/2)$, the measure ${\pi}_\Lambda^{\rm per}$ obeys the estimate
\begin{equation} \label{ees}
\int_{\mathit{\Omega}} \|\omega\|_{\alpha, \sigma}^2
\pi_\Lambda^{\rm per} ({\rm d}\omega) \leq C_{\ref{ees}}.
\end{equation}
Thereby, the sequence $\{{\pi}_\Lambda^{\rm per}\}_{\Lambda \in
\mathcal{L}_{\rm box}}$ is $\mathcal{W}^{\rm t}$-relatively compact.
\end{lemma}
{\bf Proof:}
For $\ell\in \Lambda$ such that $\{\ell' \in \mathbf{L}  \ | \ |\ell
- \ell'|=1\} \subset \Lambda$, we set $\Delta_\ell =
\mathbf{L}\setminus \{\ell \}$. Then let $\nu_\ell^\Lambda$ be the
projection of ${\pi}_\Lambda^{\rm per}$ onto
$\mathcal{B}(\mathit{\Omega}_{\Delta_\ell})$. Let also $\nu_\ell
(\cdot|\xi)$, $\xi\in \mathit{\Omega}$ be the following probability
measure on the single-spin space $\mathit{\Omega}_{\{\ell\}} =
C_\beta$
\begin{equation} \label{sig}
\nu_\ell ({\rm d}\omega_\ell |\xi )  = \frac{1}{N_\ell (\xi)}
\exp\left\{J \sum_{\ell'}\epsilon_{\ell \ell'}(\omega_\ell,
\xi_{\ell'})_{L^2_\beta} -  \int_0^\beta V( \omega_\ell (\tau)){\rm
d}\tau \right\} \chi({\rm d}\omega_\ell).
\end{equation}
Then (c.f., (\ref{35})) desintegrating $\pi_\Lambda^{\rm per}$ we
get
\begin{equation} \label{mm11}
{\pi}_\Lambda^{\rm per} ({\rm d}\omega) =\nu_\ell ({\rm
d}\omega_\ell |\omega_{\Delta_\ell} )\nu_\ell^\Lambda ({\rm
d}\omega_{\Delta_\ell}).
\end{equation}
Like in Lemma \ref{4lm} and Corollary \ref{1co} one proves that the
measure $\nu_\ell(\cdot |\xi)$ obeys
\[
\int_{C_\beta}\exp\left\{\lambda_\sigma |\omega_\ell
|^2_{C_\beta^\sigma} + \varkappa |\omega_\ell |^2_{L_\beta^2}
\right\}\nu_\ell ({\rm d}\omega_\ell|\omega_{\Delta_\ell} ) \leq
\exp\left\{C_{\ref{53}} + \vartheta J \sum_{\ell' } \epsilon_{\ell
\ell'}|\omega_{\ell'}|^2_{L^2_\beta} \right\},
\]
where $\lambda_\sigma$, $\varkappa$, and $\vartheta$ are as in
(\ref{53}), (\ref{55}). Now we integrate both sides of this
inequality with respect to $\nu_\ell^\Lambda$ and get, c.f.,
(\ref{66}), (\ref{67d})
\[
n_\ell^{\rm per}(\Lambda) \ \stackrel{\rm def}{=} \  \log\left\{
\int_{\mathit{\Omega}}\exp [\lambda_\sigma |\omega_\ell
|^2_{C_\beta^\sigma} + \varkappa |\omega_\ell
|^2_{L_\beta^2}]\pi_\Lambda^{\rm per}({\rm d}\omega) \right\} \leq
C_{\ref{66a}}.
\]
Then the estimate (\ref{ees}) is obtained in the same way as
(\ref{61a}) was proven. The relative
$\mathcal{W}_\alpha$-compactness of $\{\pi_\Lambda^{\rm
per}\}_{\Lambda \in \mathcal{L}_{\rm per}}$ follows from (\ref{ees})
and the compactness of the embeddings $\mathit{\Omega}_{\alpha ,
\sigma} \hookrightarrow \mathit{\Omega}_{\alpha'}$, $\alpha <
\alpha'$. The $\mathcal{W}^{\rm t}$-compactness is a consequence of
by Lemma \ref{tanlm}.
$\blacksquare$
\begin{lemma} \label{nmlm} Every $\mathcal{W}^{\rm t}$-accumulation
point $\mu^{\rm per}$ of the sequence $\{{\pi}_\Lambda^{\rm
per}\}_{\Lambda \in \mathcal{L}_{\rm per}}$ is a Euclidean Gibbs
measure of the ${\rm low}$-reference model.
\end{lemma}
{\bf Proof:}
Let $\mathcal{L}\subset \mathcal{L}_{\rm per}$ be the subsequence
along which $\{{\pi}_\Lambda^{\rm per}\}_{\Lambda \in \mathcal{L}}$
converges to $\mu^{\rm per} \in \mathcal{P}(\mathit{\Omega}^{\rm
t})$. Then $\{\nu_\ell^{\Lambda}\}_{\Lambda \in \mathcal{L}}$
converges to the projection of $\mu^{\rm per}$ on
$\mathcal{B}(\mathit{\Omega}_{\Delta_\ell})$. Employing the Feller
property (Lemma \ref{2lm}) we pass in (\ref{mm11}) to the limit
along this $\mathcal{L}$ and apply both its sides to a function $f
\in C_{\rm b}(\mathit{\Omega}^{\rm t})$. This yields that $\mu^{\rm
per}$ has the same one-point conditional distributions as the
Euclidean Gibbs measures of the reference model. But according to
Theorem 1.33 of \cite{Ge}, page 23, every Gibbs measure is uniquely
defined by its conditional distributions corresponding to one-point
sets $\Lambda = \{\ell \}$ only.
$\blacksquare$
Now we are at a position to prove that (\ref{rf10}) holds if $\beta>
\beta_*$. Given a box $\Lambda$, we introduce
\begin{equation} \label{op}
P_\Lambda (\beta) = \int_{\mathit{\Omega}} \left\vert\frac{1}{\beta
|\Lambda|}\sum_{\ell \in \Lambda}\int_0^\beta\omega_\ell (\tau) {\rm
d}\tau \right\vert^2 {\pi}^{\rm per}_\Lambda ({\rm d}\omega).
\end{equation}
For any $\ell$, one can take the box $\Lambda$ such that the
Euclidean distance from this $\ell$ to $\Lambda^c$ be greater than
$1$. Then by Corollary \ref{wzco} and Lemma \ref{lblm} one gets
\begin{equation} \label{sig7}
\int_{\mathit{\Omega}} [\omega_\ell (0)]^2{\pi}^{\rm per}_\Lambda
({\rm d}\omega) \geq {K}^{\rm low}_\ell (0,0)\geq t_*.
\end{equation}
 The infrared estimates based on the reflection positivity
 of the $low$-reference model, together with the Bruch-Falk
 inequality\footnote{See Theorem VI.7.5, page 392 of \cite{Simonl} or Theorem 3.1 in
 \cite{DLS}} and the estimate (\ref{sig7}),
 lead to the following bound
 \begin{equation} \label{may10}
P_\Lambda (\beta) \geq t_* f(\beta/ 4 m t_*) - \theta_d/ 2 \beta
J,
 \end{equation}
which holds for any box $\Lambda$. By means of the Griffiths
theorem, see \cite{DLS}, Theorem 1.1 and the corollaries, one can
prove that
\begin{equation} \label{may12}
\mu^{\rm per}(\omega_\ell (0)) \geq \limsup_{\mathcal{L}_{\rm
per}} \sqrt{P_\Lambda (\beta)}.
\end{equation}
Therefore, the estimate (\ref{rf10}) holds if the right-hand side of
(\ref{may12}) is positive, which can be ensured by taking $\beta
>\beta_*$, see (\ref{may2}) and (\ref{cj1}), (\ref{cja}).

\subsection{Proof of Theorem \ref{5tm}}
\label{ss5.2}

Now we make precise the parameter $\delta$ participating in the
condition (\ref{26a}). In what follows, we set $\delta = m
\mathit{\Delta}^2$, where the parameter $\mathit{\Delta}$ was
defined by (\ref{51}). Then
\begin{equation} \label{w16a}
\hat{J}_0 < \hat{J}_\alpha < m \mathit{\Delta}^2.
\end{equation}
Let us consider the examples following Assumption (B). If
$J_{\ell \ell'}$ obeys (\ref{24a}), the values of $\alpha$ in
question exist in view of
\begin{equation} \label{26m}
\lim_{\alpha \rightarrow 0+} \hat{J}_\alpha = \hat{J}_0,
\end{equation}
which readily follows from (\ref{24a}), (\ref{24b}). If the
weights are chosen as in (\ref{24d}), one can use
 $\varepsilon$ to ensure
 (\ref{w16a}).  Indeed, simple calculations yield
 \[
0 < \hat{J}^{(\varepsilon)}_\alpha - \hat{J}_0 \leq \varepsilon
\alpha d \hat{J}_\alpha^{(1)},
\]
where  to indicate the dependence of $\hat{J}_\alpha$ on
 $\varepsilon$ we write
 $\hat{J}_\alpha^{(\varepsilon)}$. Thereby, we fix $\alpha \in \mathcal{I}$ and choose
$\varepsilon$ to obey $\varepsilon < m \mathit{\Delta}^2 / \alpha
d \hat{J}_\alpha^{(1)}$.

Now let us turn to the proof of Theorem \ref{5tm}.
 By Corollary \ref{compco} it is enough to prove the uniqueness for
 the $up$-reference model, which by Lemma \ref{bhpn} is equivalent
 to
\begin{equation} \label{rf20}
\mu^{\rm up}_{+} (\omega_\ell (0)) = 0, \quad {\rm for} \ {\rm all}
\ \ \beta>0 \ \ {\rm and} \ \ \ell.
\end{equation}
Given $\Lambda \Subset \mathbf{L}$, we introduce the matrix
$(T^\Lambda_{\ell \ell'})_{\ell , \ell'\in \mathbf{L}}$ as follows.
We set $T^\Lambda_{\ell \ell'} = 0$ if either of $\ell , \ell'$
belongs to $\Lambda^c$. For $\ell , \ell' \in \Lambda$,
\begin{equation} \label{w10}
T^\Lambda_{\ell \ell'} = \sum_{\ell_1 \in \Lambda} J_{\ell \ell_1}
\int_0^\beta \pi_\Lambda^{\rm up}\left[\omega_{\ell_1} (\tau)
\omega_{\ell'} (\tau') \left\vert 0 \right. \right]{\rm d}\tau'.
\end{equation}
By (\ref{mats1}) the above integral is independent of $\tau$.

\begin{lemma} \label{newlm}
If (\ref{52}) is satisfied, there exists $\alpha \in \mathcal{I}$,
such that for every $\Lambda \Subset \mathbf{L}$, the matrix
$(T^\Lambda_{\ell \ell'})_{\ell , \ell'\in \mathbf{L}}$ defines a
bounded operator in the Banach space $l^\infty (w_\alpha)$.
\end{lemma}
{\bf Proof:}
The proof will be based on a generalization of the method used in
\cite{AKKR02b} for proving Lemma 4.7. For $t\in [0, 1]$, let
$\varpi^{(t)}_{\Lambda}\in \mathcal{P}(\mathit{\Omega}_\Lambda))$ be
defined by (\ref{w4z}) with $\Delta = \Lambda$ and each $V_\ell
(\omega_\ell (\tau))$ replaced by $v ([\omega_\ell(\tau)]^2)$, where
$v$ is the same as in (\ref{2}). Then by (\ref{rf2})
\begin{equation} \label{rf30}
\varpi_\Lambda^{(0)} = \prod_{\ell \in \Lambda}\pi_{\ell}^{\rm up}
(\cdot|0), \quad
 \varpi_\Lambda^{(1)} = \pi_\Lambda^{\rm up}(\cdot |0), \quad {\rm for} \ {\rm any} \ \ \Lambda \Subset
\mathbf{L}.
\end{equation}
Thereby, we set
\begin{equation} \label{w14a}
T^\Lambda_{\ell \ell'} (t) = \sum_{\ell_1} J_{\ell \ell_1}
\int_0^\beta \varpi_\Lambda^{(t)}\left[\omega_{\ell_1} (\tau)
\omega_{\ell'} (\tau')  \right]{\rm d}\tau' \quad t\in [0, 1].
\end{equation}
One can show that for every fixed $\ell, \ell'$, the above
$T^\Lambda_{\ell \ell'} (t)$ is differentiable on the interval $t\in
(0, 1)$ and continuous at its endpoints where (see (\ref{w70}))
\begin{equation} \label{w15a}
 T^\Lambda_{\ell \ell'} (0) = J_{\ell \ell'}
K^{\rm up}_{\ell'} \leq J_{\ell \ell'}/ m \mathit{\Delta}^2, \quad \
T^\Lambda_{\ell \ell'}(1) = T^\Lambda_{\ell \ell'}.
\end{equation}
 Computing the derivative we get
\begin{eqnarray} \label{w16}
\frac{\partial }{\partial t} T_{\ell \ell'}^\Lambda (t) & = &
\frac{1}{2} \sum_{\ell_1, \ell_2, \ell_3 } J_{\ell \ell_1} J_{\ell_2
\ell_3} \int_0^\beta \int_0^\beta U^\Lambda_{\ell \ell' \ell_2
\ell_3} (t ,
\tau, \tau' , \tau_1 , \tau_1) {\rm d}\tau' {\rm d}\tau_1 \\
& + & \sum _{\ell_1 } T^\Lambda_{\ell \ell_1} (t) T^\Lambda_{\ell_1
\ell'} (t), \nonumber
\end{eqnarray}
where $U^\Lambda_{\ell \ell' \ell_1 \ell_2} (t , \tau, \tau' ,
\tau_1 , \tau_1)$ is  the Ursell function which obeys the estimate
(\ref{v2}) since the function $v$ is convex. Hence, except for the
trivial case $J_{\ell \ell'} \equiv 0$, the first term in
(\ref{w16}) is strictly negative. Let us consider the following
Cauchy problem
\begin{equation} \label{w17}
\frac{\partial }{\partial t} L_{\ell \ell'} (t) = \sum _{\ell_1}
L_{\ell \ell_1} (t)  L_{\ell_1 \ell'} (t), \quad L_{\ell \ell'} (0)
= \lambda J_{\ell \ell'}, \ \ \ell, \ell' \in \mathbf{L},
\end{equation}
where $\lambda \in (1/ m \mathit{\Delta}^2 , 1/\hat{J}_\alpha) $,
with $\alpha \in \mathcal{I}$ chosen to obey (\ref{w16a}). For such
$\alpha$, one can solve the problem (\ref{w17}) in the space
$l^\infty(w_\alpha)$ (see Remark \ref{1rm}) and obtain
\begin{equation} \label{w17a}
L(t) = \lambda J \left[I - \lambda t J \right]^{-1}, \quad
\|L(t)\|_{l^\infty (w_\alpha)} \leq \frac{\lambda \hat{J}_\alpha}{1
- \lambda t \hat{J}_\alpha}.
\end{equation}
where $I$ is the identity operator. Now let us compare (\ref{w16})
and (\ref{w17}) considering the former expression as a differential
equation subject to the initial condition (\ref{w15a}). Since the
first term in (\ref{w16}) is negative, one can apply Theorem V, page
65 of \cite{Walter} and obtain $T_{\ell \ell'}^\Lambda < L_{\ell
\ell'} (1)$, which in view of (\ref{w17a}) yields the proof.
$\blacksquare$ \vskip.1cm \noindent
\textbf{Proof of Theorem \ref{5tm}:} \ For $\ell, \ell_0$, $\Lambda
\Subset \mathbf{L}$, such that $\ell \in \Lambda$, and $t\in [0,
1]$, we set
\begin{equation}\label{w19}
\psi_\Lambda (t) = \int_{\mathit{\Omega}} \omega_{\ell} (0)\pi^{\rm
up}_\Lambda ({\rm d}\omega | t \xi^{\ell_0}) ,
\end{equation}
where $\xi^{\ell_0}$ is the same as in (\ref{v3}). The function
$\psi_\Lambda$ is obviously differentiable on the interval $t \in
(-1, 1)$ and continuous at its endpoints.
 Then
\begin{equation} \label{w20}
0 \leq \psi_\Lambda (1) \leq  \sup_{t \in [0, 1]} \psi_\Lambda' (t).
\end{equation}
The derivative is
\begin{equation} \label{w21} \psi_\Lambda' (t) =
\sum_{\ell_1 \in \Lambda , \  \ell_2 \in \Lambda^c} J_{\ell \ell_1}
\int_0^\beta \pi_\Lambda^{\rm up}\left[\omega_{\ell_1}(0) \omega_{
\ell_2} ( \tau) \left\vert t \xi^{\ell_0} \right. \right]
\eta_{\ell_2} {\rm d}\tau,
\end{equation}
where the `external field' $\eta_{\ell'} = \left[b \log(1 + |\ell' -
\ell_0|)\right]^{1/2}$ is positive at each site. Thus, we may use
(\ref{w5}) and obtain
\begin{equation} \label{w22}
\psi_\Lambda' (t) \leq \sum_{l' \in \Lambda^c} T^\Lambda_{\ell
\ell'} \eta_{\ell'}.
\end{equation}
By Assumption (B) (b), $\eta \in l^\infty (w_\alpha)$ with any
$\alpha >0$, then employing Lemma \ref{newlm}, the estimate
(\ref{w17a}) in particular, we conclude that the right-hand side of
(\ref{w22}) tends to zero as $\Lambda \nearrow \mathbf{L}$, which by
(\ref{soc1}) and (\ref{w19}), (\ref{w20}) yields (\ref{rf20}).
$\square$

\section{Uniqueness at Nonzero External Field}
\label{7s}

In  statistical mechanics phase transitions may be associated with
nonanalyticity of thermodynamic characteristics considered as
functions of the external field $h$. In special cases one can
oversee at which values of $h$ this nonanaliticity can occur. The
Lee-Yang theorem states that the only such value  is $h=0$; hence,
no phase transitions can occur at nonzero $h$. In the theory of
classical lattice models these arguments were applied in
\cite{LML,LPe,LP}. We refer also to sections 4.5, 4.6 in \cite{GJ}
and sections IX.3 -- IX.5 in \cite{Si74} where applications of such
arguments in quantum field theory and classical statistical
mechanics are discussed.

In the case of lattice models with the single-spin space
$\mathbf{R}$ the validity of the Lee-Yang theorem depends on the
properties of the anharmonic potentials. For the polynomials $V(x) =
x^4 + a x^2$, $a\in \mathbf{R}$, the Lee-Yang theorem holds, see
e.g., Theorem IX.15 on page 342 in \cite{Si74}. But no other
examples of this kind were known, see the discussion on page 71 in
\cite{GJ}. Below we give a sufficient condition for the
 potentials $V$ to have the corresponding property and
discuss some examples. Here we use the family $\mathcal{F}_{\rm
Laguerre}$ defined by (\ref{m1}). We also prove a number of lemmas,
which allow us to apply the arguments based on the Lee-Yang theorem
to our quantum model and hence to prove Theorem \ref{6tm}.

Recall that the elements of $\mathcal{F}_{\rm Laguerre}$ can be
continued to entire functions $\varphi:\mathbf{C}\rightarrow
\mathbf{C}$, which have no zeros outside of $(-\infty, 0]$.
\begin{definition} \label{lydf}
A probability measure $\nu$ on the real line is said to have the
Lee-Yang property if there exists $\varphi\in \mathcal{F}_{\rm
Laguerre}$ such that
\[
\int_\mathbf{R} \exp(xy)\nu({\rm d}y) = \varphi (x^2).
\]
\end{definition}
\noindent In \cite{Koz1}, see also Theorem 2.3 in \cite{KOU}, the
following fact was proven.
\begin{proposition} \label{lypn}
Let the function $u:\mathbf{R} \rightarrow \mathbf{R}$ be such that
for a certain $b\geq 0$, its derivative obeys the condition $b + u'
\in \mathcal{F}_{\rm Laguerre}$. Then the probability measure
\begin{equation} \label{w23}
\nu({\rm d}y) = C \exp[- u(y^2)]{\rm d}y,
\end{equation}
has the Lee-Yang property.
\end{proposition}
\vskip.1cm \noindent This statement gives a sufficient condition,
the lack of which was mentioned on page 71 of \cite{GJ}. The example
of a polynomial given there for which the corresponding classical
models undergo phase transitions at nonzero $h$,
 in our
notations is $u (t) = t^3 - 2t^2 + (\alpha + 1)t$, $\alpha >0$. It
certainly does not meet the condition of Proposition \ref{lypn}.
Turning to the model described by Theorem \ref{6tm} we note that,
for $ v(t)
 = t^3 + b^{(2)} t^2 + b^{(1)} t$, the function $u (t) = v (t) + a
 t/2$ obeys the conditions of Proposition \ref{lypn} if and only if
$b^{(2)} \geq 0$ and $b^{(1)} + a/2 \leq [b^{(2)}]^2/3$. Therefore,
 according to Theorem \ref{6tm}
 we have $|\mathcal{G}^{\rm t}|=1$ at $h \neq 0$ and
$2b^{(1)} + a < 0$, $b^{(2)} \geq 0$. On the other hand, for this
model, by Theorem \ref{phtm} one has a phase transition at $h=0$ and
the same coefficients of $v$.

Set
\begin{equation} \label{w24}
f(h^2) = \int_{\mathbf{R}^n} \exp\left[ h \sum_{i=1}^n x_i +
\sum_{i,j=1}^n M_{ij}x_i x_j \right]\prod_{i=1}^n \nu({\rm d}x_i),
\quad h \in \mathbf{R}.
\end{equation}
By Theorem 3.2 of \cite{LS}, we have the following
\begin{proposition} \label{ly1pn}
If in (\ref{w24}) $M_{ij} \geq 0$ for all $i,j = 1, \dots , n$, and
the measure $\nu$ is as in Proposition \ref{lypn}, then the function
$f$, if exists, belongs to $\mathcal{F}_{\rm Laguerre}$. It
certainly exists if $u'$ is not constant.
\end{proposition}
\noindent Now let the potential $V$  obey the conditions of Theorem
\ref{6tm}. Recall that $p_\Lambda(h)$ stands for the pressure
(\ref{c1}) with $\xi=0$. Define
\begin{equation} \label{w25}
 \varphi_\Lambda (h^2) = p_\Lambda (h), \quad h \in \mathbf{R} .
\end{equation}
\begin{lemma} \label{lylm}
If $V$  obeys the conditions of Theorem \ref{6tm}, the function
$\exp\left(|\Lambda|\varphi_\Lambda\right)$ belongs to
$\mathcal{F}_{\rm Laguerre}$.
\end{lemma}
{\bf Proof:}
With the help of the lattice approximation technique the function
$\exp\left(|\Lambda|\varphi_\Lambda\right)$ may be approximated by
$f_N$, $N\in\mathbf{N}$, having the form (\ref{w24}) with the
measures $\nu$ having of the form (\ref{w23}) with $u(t) = v(t) + a
t/2$, $v$ is as in (\ref{m2}), and non-negative $M_{ij}$ (see
Theorem 5.2 in \cite{AKKR}). For every $h\in \mathbf{R}$, $f_N (h^2)
\rightarrow \exp\left(|\Lambda|\varphi_\Lambda (h^2)\right)$ as $N
\rightarrow +\infty$. The entire functions $f_N$ are ridge, with the
ridge being $[0, +\infty)$. For sequences of such functions, their
point-wise convergence on the ridge implies via the Vitali theorem
(see e.g., \cite{Si74}) the uniform convergence on compact subsets
of $\mathbf{C}$, which yields the property stated (for more details,
see \cite{KozHol,KW}).
$\blacksquare$
\vskip.1cm \noindent {\bf Proof of Theorem \ref{6tm}:}  By Lemma
\ref{lylm}, for every $\Lambda \Subset \mathbf{L}$, $p_\Lambda(h)$
can be extended to a function of  $h \in \mathbf{C}$, holomorphic
in the right and left open half-planes. By standard arguments, see
e.g., Lemma 39, page 34
 of \cite{KozHol}, and Lemma \ref{conlm}
it follows that the limit of such extensions $p(h)$ is holomorphic
in certain subsets of those half-planes containing the real line,
except possibly for the point $h =0$. Therefore, $p(h)$ is
differentiable at each $h \neq 0$. Then the proof of the theorem
follows from Corollary \ref{Mco}. $\square$

\end{document}